\documentclass[12pt,a4paper]{article}
\pdfoutput=1

\usepackage{jheppub}
\usepackage{graphicx,amsmath,amssymb}
\usepackage{mathtools}
\usepackage[T1]{fontenc}
\usepackage[utf8]{inputenc}

\usepackage{ucs}
\usepackage{fontenc}

\usepackage{multicol}
\usepackage{framed}
\usepackage[section]{placeins}   % float placement
\usepackage{dcolumn,makecell,booktabs,multirow}   % tables
\usepackage[shortlabels]{enumitem}  % list environments
\usepackage{float}

\usepackage{amsmath,amssymb,mathtools,empheq,slashed,braket,array}
\usepackage{color}
\usepackage{graphicx}
\usepackage{setspace}
\usepackage{subfigure}

\usepackage{textcomp,starfont,bbold,bm}

\usepackage[table]{xcolor}
\usepackage[compat=1.1.0]{tikz-feynman}
\usepackage{environ}    
\usepackage{verbatim}    
\hypersetup{
    colorlinks,
    citecolor=black,
    filecolor=black,
    linkcolor=black,
    urlcolor=blue
}
\newcommand\cmt[1]{}

\usepackage{accents}

\newcommand{\leri}[1]{\left(#1 \right)}

\usepackage{tikz,xcolor,hyperref}
\definecolor{lime}{HTML}{A6CE39}
\DeclareRobustCommand{\orcidicon}{%
	\begin{tikzpicture}
	\draw[lime, fill=lime] (0,0) 
	circle [radius=0.16] 
	node[white] {{\fontfamily{qag}\selectfont \tiny ID}};	\draw[white, fill=white] (-0.0625,0.095) 
	circle [radius=0.007];	\end{tikzpicture}
	\hspace{-2mm}}
\foreach \x in {A,...,Z}{%
	\expandafter\xdef\csname orcid\x\endcsname{\noexpand\href{https://orcid.org/\csname orcidauthor\x\endcsname}{\noexpand\orcidicon}}
}

\newcommand{\bea}{\begin{eqnarray}}
\newcommand{\eea}{\end{eqnarray}}

\newcommand{\hide}[1]{{ }}
\newcommand{\GeV}{\, {\rm GeV}}
\newcommand{\MeV}{\, {\rm MeV}}
\newcommand{\keV}{\, {\rm keV}}
\newcommand{\eV}{\, {\rm eV}}

\newcommand{\Mpl}{M_{\rm Pl}}
\newcommand{\Mpls}{M_{\rm Pl}^2\,}
\newcommand{\cO}{\mathcal{O}}
\newcommand{\css}{{\rm cs}^2}

\global\long\def\L{\mathcal{L}}

\let\oldquote\quote
\renewcommand\quote{\scriptsize\oldquote}
\let\oldquotation\quotation
\renewcommand\quotation{\scriptsize\oldquotation}\usepackage{type1cm}
\usepackage{wasysym}
\usepackage{feyn}
\usepackage{hyperref}
\numberwithin{equation}{section}
\usepackage{bbold}
\usepackage{simplewick}
\usepackage{amsmath}
\usepackage{amsthm}
\usepackage{cancel}
\usepackage{graphicx}

\makeatletter
\AtBeginDocument{%
  \expandafter\renewcommand\expandafter\subsection\expandafter{%
    \expandafter\@fb@secFB\subsection
  }%
}
\makeatother

\usepackage[nolist]{acronym}
\begin{acronym}
\acro{SM}{Standard Model}
\acro{CPV}{CP Violation}
\acro{EH}{Euler-Heisenberg}
\acro{BSM}{Beyond the Standard Model}
\acro{COM}{Center of Mass}
\acro{QED}{Quantum Electrodynamics}
\acro{EFT}{Effective Field Theory}
\acro{ALP}{Axion-Like Particle}
\acro{LV}{Lorentz Violation}
\acro{DM}{Dark Matter}
\acro{EOM}{Equation of Motion}
\acro{EP}{Equivalence Principle}
\acro{AP}{Atomic Physics}
\acro{DDM}{Direct Dark Matter}
\acro{DP}{Dark Photon}
\acro{EM}{Electro-magnetic}
\acro{SSB}{Spontaneous Symmetry Breaking}
\acro{EWSB}{electroweak symmetry breaking}
\acro{EW}{electro-weak}
\acro{VEV}{vacuum expectation value}
\acro{FC}{fundamental constant}
\acro{ULDM}{ultralight dark matter}
\acro{QFT}{quantum field theory}
\acro{NGB}{Nambu Goldstone Boson}
\acro{GB}{Goldstone Boson}
\acro{pNGB}{pseudo-Nambu Goldstone Boson}
\acro{CW}{Coleman-Weinberg}
\acro{RG}{Renormalization Group}
\acro{pNGB}{pseudo Nambu-Goldstone boson}
\acro{CP}{charge-parity}
\end{acronym}

\title{The Phenomenology of Quadratically Coupled Ultra Light Dark Matter}
\author[a]{Abhishek Banerjee\orcidA{},}
\author[a]{Gilad Perez\orcidB{},}
\author[b,c]{Marianna Safronova\orcidC{},}
\author[a]{Inbar Savoray\orcidD{},}
\author[a]{Aviv Shalit\orcidE{}}

\affiliation[a]{Department of Particle Physics and Astrophysics, Weizmann Institute of Science, 234 Herzl Street, Rehovot 7610001, Israel}
\affiliation[b]{Department of Physics and Astronomy, University of Delaware, Newark, Delaware 19716, USA}
\affiliation[c]{Joint Quantum Institute, National Institute of Standards and Technology and the University of Maryland, Gaithersburg, Maryland 20742, USA}

\abstract{ We discuss models of ultralight scalar \ac{DM} with linear and quadratic couplings to the \ac{SM}. In addition to studying the phenomenology of linear and quadratic interactions separately, we examine their interplay. We review the different experiments that can probe such interactions and present the current and expected future bounds on the parameter space. In particular, we discuss the scalar field solution presented in [A. Hees, O. Minazzoli, E. Savalle, Y. V. Stadnik and P. Wolf, Phys.Rev.D 98 (2018) 6, 064051], and extend it to theories that capture both the linear and the quadratic couplings of the \ac{DM} field to the \ac{SM}. 
Furthermore, we discuss the theoretical aspects and the corresponding challenges for natural models in which the quadratic interactions are of phenomenological importance.}

\begin{document}
%\flushbottom
	
\acresetall

\addtocontents{toc}{\protect\thispagestyle{empty}}
\maketitle
%\tableofcontents
\thispagestyle{empty}
\newpage
\setcounter{page}{1}

\acresetall
\section{Introduction}
One possible solution to the ``missing mass'' problem  is that of an ultralight sub-eV bosonic \acf{DM} field, coherently oscillating to account for the observed \ac{DM} density (e.g.~\cite{Catena:2009mf,Graham:2013gfa,Arvanitaki:2014faa,Graham:2015ifn,Banerjee:2018xmn}). Such a light field would oscillate with a frequency proportional to its mass $m_\phi$, and an amplitude which is determined by $m_\phi$ and the \ac{DM} density  $\rho_\text{DM}$, 
\begin{align}
\langle \phi\leri{t,\vec{x}}\rangle &\simeq \frac{\sqrt{2\rho_\text{DM}}}{m_{\phi}}\cos(m_{\phi}\,t+\delta)\,,
\label{eq:DM_VEV}
\end{align}
where $\delta=m_\phi\, \vec\beta\cdot\vec x$ is some random phase with $|\vec\beta|\sim 10^{-3}$ being the virial \ac{DM} velocity. 
Possibly the most simple model of an \ac{ULDM} field is obtained by augmenting the \ac{SM} (of fundamental interactions and elementary particles) with only one degree of freedom. Adding a free light spin-0 field, with its misalignment angle appropriately tuned towards the end of inflation so that its oscillation amplitude yields the right \ac{DM} abundance, fully address the missing mass problem (see for instance~\cite{Kolb:1990vq} for more detail).  
Such a model can be tested solely by its gravitational interactions~\cite{Bar:2018acw, Bar:2019bqz, Arvanitaki:2014wva}, however, adding self-interactions may render the \ac{DM} distribution, and hence the corresponding bounds, non-robust.

More conceptually, models with spin-0 \ac{ULDM} face two main theoretical challenges. The first is associated with the hierarchy problem, namely, the challenge of keeping the scalar light, although in the presence of interactions microscopic quantum fluctuations are generically expected to contribute dramatically to its mass. We shall discuss it further in the following sections.
The second challenge is associated with the fact that all fields interact via gravity. Even in the absence of a direct coupling of the scalar to the \ac{SM} fields, one may argue that if the spin-0 particle is an elementary, point-like, microscopic field, ``gravity-mediated'' interactions will inevitably generate such an effective coupling. 
Below the Planck scale, an \ac{EFT} describing the \ac{SM} elementary fields and the new spin-0 elementary field would consist of (local) interaction terms between the \ac{ULDM} and the \ac{SM}, suppressed by the Planck scale.
Such a coupling may be eliminated if the field is composite or in the presence of additional discrete symmetries (see~\cite{Dine:2022mjw,Banerjee:2022wzk} for recent discussions).\footnote{This can be thought of as a generalized version of the axion quality problem~\cite{Kamionkowski:1992mf, Barr:1992qq, Davidi:2017gir, Davidi:2018sii}, (for a more general discussion see~\cite{Calmet:2021iid, Perez:2020dbw}, and~\cite{Choi:1985cb, Kim:1984pt, Contino:2021ayn, Perez:2020dbw, Banerjee:2022wzk} for models that address the quality problem). In some more detail, the QCD axion quality problem is attributed to the fact that the axion potential resulting of QCD instanton-corrections can be disrupted by the presence of Planck suppressed operators that do not respect the Pecci-Quinn Symmetry, see for instance~\cite{Dine:2012sla,Hook:2018dlk} for reviews on the topic.
}

Let us first consider the case where the \ac{DM} field is elementary and there are no additional discrete symmetries. 
It is interesting to note that for \ac{ULDM} models with scalar-linear couplings between the \ac{DM} and the \ac{SM} fields, any masses below roughly 10$^{-6}$\, eV are excluded by experiments testing the \ac{EP} (corresponding to regions with effective coupling $d^{(1)}_x\gtrsim1$, see text around Fig.~\ref{fig:dil_direction} for more details). 
Along this line, we point out that one can identify a set of models where the \ac{EP} bound is greatly ameliorated or absent (one example for such a model is a pure dilaton, but there are others, see~\cite{Oswald:2021vtc,WRESL2}). In this case, a weaker bound associated with fifth-force searches can be enforced, excluding models with scalar \ac{ULDM} Planck suppressed couplings for \ac{ULDM} masses below 10$^{-10}$\, eV. 

In addition, there is a broad class of well-motivated models where the \ac{ULDM} is predicted to interact with the \ac{SM} fields beyond merely Planck suppressed couplings.
Two prime examples are models where the \ac{ULDM} is spin-0 but a pseudo-scalar, \ac{ALP}, or where it is a CP-even scalar.
\acp{ALP} naturally arise in theories where a global U(1) symmetry is spontaneously broken, for instance, in Froggatt-Nielsen type of models~\cite{Froggatt:1978nt} that address the mass hierarchies (usually denoted as the flavor puzzle), models which account for lepton number conservation~\cite{Gelmini:1980re}, models of QCD axion solution to the strong CP problem~\cite{Kim:1979if, Shifman:1979if, Zhitnitsky:1980tq, Dine:1981rt}, models which solve the hierarchy problem~\cite{Graham:2015cka} or combinations of the above~\cite{Ema:2016ops,Davidi:2018sii, Davidi:2017gir,Banerjee:2022wzk}. 
As we have already mentioned, scalar \ac{ULDM} models are more involved as they are susceptible to naturalness problems, however, two main options were described in the literature. In the first, the \ac{ULDM} mass is protected by either an approximate scale-invariance symmetry~\cite{Arvanitaki:2014faa} or a discrete $\mathbb Z_{N}$ symmetry~\cite{Brzeminski:2020uhm}. In the second, inspired by the relaxion paradigm~\cite{Graham:2015cka}, the \ac{ULDM} is an exotic type of \ac{ALP}~\cite{Banerjee:2018xmn}, with its associated shift symmetry and \ac{CP} invariance broken by two independent sectors \cite{Flacke:2016szy,Choi:2016luu, Banerjee:2020kww}. 
In all of the above models but the last one, the \ac{ULDM} couplings to the \ac{SM} fields are, to leading order, dominated by the \ac{ULDM} derivative couplings to the appropriate \ac{SM} current operators, with the strength of these interactions dictated by the transformations of the \ac{SM} fields under the \ac{ULDM} shift-symmetry. Generically, we expected the dominant interaction to be linear with the \ac{DM} field. 

Most of the theoretical and experimental effort has been put towards studying the linear-\ac{DM}-\ac{SM} interactions (see~\cite{Hees:2018fpg, Lin:2019uvt, Alexander:2016aln} for a recent review and Refs. therein).
However, as mentioned above, both fundamental scalar and axion models suffer from a rather severe quality problem, namely they do not provide sufficient protection against operators that link the \ac{ULDM} field with the \ac{SM} ones (usually denoted as irrelevant operators), even if they are suppressed by super-Planckian cutoffs. 
It motivates us to consider cases where the quadratic interactions dominate over the linear ones, at least when considering the more severely constrained scalar \ac{SM}-operators.
One example of such a scenario is in the case of ultralight QCD axions, where the product of the mass, $m_\phi$, and the decay constant, $f$, satisfies $m_\phi f \ll \Lambda_{\rm QCD}^2$, where $\Lambda_{\rm QCD}$ is the dynamical scale of QCD. This can arise due to fine-tuning, or in a particular type of $\mathbb Z_{N}$ models~\cite{Hook:2018jle, DiLuzio:2021pxd}. In such a case, quantum sensors looking for scalar-quadratic oscillations of constants of nature, would be more sensitive to the presence of \ac{ULDM} QCD axion~\cite{Kim:2022ype}, compared with the well known conventional searches which are based on magnetometer-type of quantum sensors.
From the properties of \ac{QFT}, and its low energy \ac{EFT} perspective, theories where the quadratic interactions of an \ac{ULDM} scalar are experimentally significant, seem to be exotic. First, it implies that the linear coupling that should dominate the phenomenology is, for some reason, is suppressed. Furthermore, the quadratic coupling of the \ac{ULDM}, which is expected to generate large additive contributions to the \ac{ULDM} mass, is thus being pressured by naturalness arguments. 	
We nevertheless find it interesting to consider in depth models where the phenomenology is dominated by the \ac{ULDM} quadratic coupling, in view of the interplay between the direct and indirect experimental searches. Together with presenting the current and expected future bounds on these models, we outline the experimental and theoretical challenges associated with them, and study concrete examples in which these challenges can be ameliorated.

The paper is organized as follows: in Subsections \ref{subsec:Model_linear_DM_couplings} and \ref{subsec:Model_quad_DM_couplings}, we introduce the \ac{ULDM} models of interest and discuss their phenomenology, in particular, the profile of quadratically coupled \ac{DM}. In Section \ref{sec:Updated bounds from EP and Direct DM searches}, we review the bounds from different \ac{ULDM} searches, considering current and future probes. In Section \ref{sec:Quadratic_Interactions_and_Screening}, we study in detail the behavior of a quadratically coupled \ac{DM} field in the presence of a massive source such as the Earth. In addition, we comment on the challenges of \ac{EP} tests and \ac{DDM} searches due to the \ac{DM} field profile, and show how those can be addressed by performing experiments in space. In Sections \ref{sec:Theoretical_Challenges_of_Models_with_Quadratic_Couplings} and \ref{sec:Examples_of_Technically-Natural_Models}, we review the theoretical aspects of models with sizable quadratic \ac{DM} interactions with the \ac{SM}, and provide various examples in which these couplings are technically-natural, and dominate over the linear ones. In Section \ref{sec:relaxed_relaxion_models}, we study two specific models solving the naturalness problem of the \ac{ULDM} and allowing for a hierarchy between the linear and quadratic coupling of the \ac{DM} field. We conclude our results in Section \ref{sec:conclusion}. 

\subsection{Model with linear \ac{DM} couplings}\label{subsec:Model_linear_DM_couplings}
We start by reviewing the case where the \ac{DM} couples linearly to the \ac{SM} fields. Since we choose to focus on CP invariant theories, we distinguish between a CP odd pseudo-scalar, $\phi(x)=a(x)$, and a CP-even scalar, $\phi(x)=\varphi(x)$. The linear interactions can be characterized by the following low-energy effective Lagrangians
\begin{align}\label{eq:Stadnik_Lint_linear}
&\L^\varphi_\text{lin scalar} =  \frac{d^{(1)}_e}{4 \Mpl} \varphi \, F^{\mu\nu}F_{\mu\nu}-\frac{d^{(1)}_g \beta_g}{2 \Mpl \, g} \varphi\, G^{b\mu\nu}G^b_{\mu\nu}
- \frac{d^{(1)}_{m_i}}{M_{\text{Pl}}}\varphi\, m_f\psi_f\psi^c_f +\text{h.c.}\,,	 
\\
\label{eq:Stadnik_Lint_linear2}
& \L^a_\text{lin pseudo-scalar} =  \frac{\tilde{d}^{(1)}_e}{M_{\text{Pl}}} a \, F_{\mu\nu}\tilde{F}^{\mu\nu}+	\frac{\tilde{d}^{(1)}_g }{M_{\text{Pl}}} a \, G^b_{\mu\nu}\tilde G^{b\mu\nu}
- \frac{i\tilde{d}^{(1)}_{m_i}}{M_{\text{Pl}}}a\, m_f\psi_f\psi^c_f +\rm{h.c.}	 \,,
\end{align}
where, $F_{\mu\nu}$ is the \ac{EM} field strength, $G^b_{\mu\nu}$ is the gluon field strength with color index $b$.  
$\beta_g=-\left(\frac{11}{3}-\frac{2}{3}N_f \right) \frac{g^3}{16\pi^2}$ is the QCD beta functions, with $N_f$ being the number of light quarks. $\psi_i$ ($\psi_i^c$) are the \ac{SM} Weyl fermions (anti-fermion) with mass $m_f$ ($f=u,d,e$ being a flavor index), $M_{\text{Pl}} = 2.4\times 10^{18}\GeV$ is the reduced plank mass,
 
$\tilde X^{\mu\nu}=\epsilon^{\mu\nu\rho\sigma}X_{\rho\sigma}$ with $X=F,G^{b}$. 
$d^{(1)}_i$ and $\tilde{d}^{(1)}_i$ are dimensionless scalar and pseudo-scalar linear \ac{DM} couplings, respectively.\footnote{Eq.~\eqref{eq:Stadnik_Lint_linear2} is basis dependent. One can perform pseudo-scalar dependent redefinition of the fermion fields to remove pseudo-scalar coupling from the mass term or from a topological term.}
\\
The analysis and the bounds on the scalar coupling, $d^{(1)}_i$, can be found in~\cite{Arvanitaki:2014faa,Safronova:2017xyt}. The oscillating background of $\phi$, as given in Eq.~\eqref{eq:DM_VEV}, induces a temporal variation of the mass of the \ac{SM} fermions, the fine-structure constant $(\alpha)$, and the strong coupling constant $(\alpha_s)$ as
\begin{align}
Y_i(t)=Y_i(1+d_{i}^{(1)}\left<\phi\right>/\Mpl)
\end{align} 
where $Y\in(\alpha,\alpha_s, m_f)$. 
\ac{DDM} searches are sensitive to the time variation of \aclp{FC}, with their sensitivity at a given frequency $\omega$ given as
\begin{equation}
\label{eq:linear_eta_DDM_first}
\eta^{\rm DDM}\leri{\omega}=\mathcal{F}_\omega\leri{\left(\frac{\delta Y_i}{Y_i}(t)\right)^{\rm DDM-lin}} \simeq \; \Delta\kappa_i\ \frac{d^{(1)}_{i}}{M_{\text{Pl}}} \mathcal{F}_\omega\leri{\left\langle  \phi \right\rangle } \;\simeq \; \Delta\kappa_i\ \frac{d^{(1)}_{i}}{M_{\text{Pl}}}\frac{\sqrt{2\rho_\text{DM}}}{m_{\phi}}\,.
\end{equation}
In equation~\eqref{eq:linear_eta_DDM_first}, $\Delta\kappa_i\equiv \kappa^A_i -\kappa^B_i$ is the difference of the sensitivity coefficients of a specific transition (see {\it e.g.}~\cite{Antypas:2019yvv,Safronova:2017xyt} and refs. therein) and $\mathcal{F}_\omega\leri{f\leri{t}}$ is the root of power spectral density at frequency $\omega$, given by $\mathcal{F}_\omega\leri{f\leri{t}}=\sqrt{\frac{1}{2\pi T}\Big|\int^T_{0} f\leri{t} e^{-i\omega t} dt\Big|^2}$ where $T$ is total duration of the experiment.

\begin{figure}
	\centering
	\includegraphics[width=0.4\linewidth]{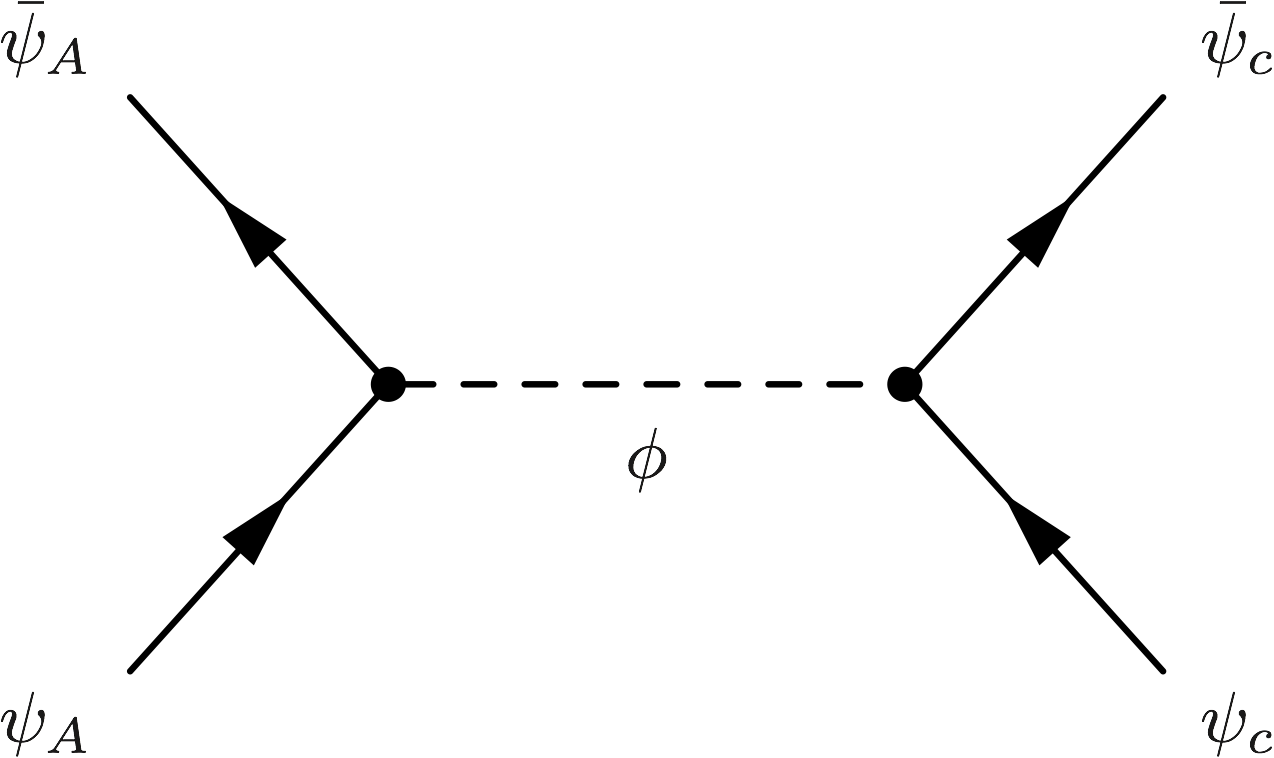}	
	\caption{A Yukawa potential is generated at tree level from the exchange of $\phi$.}
	\label{fig:linear_Yukawa}
\end{figure}

In addition, the scalar field $\phi$ can also mediate long-range forces between two masses. 
Therefore, another constraint on the parameter space of a light scalar \ac{DM} arises from experiments testing the \ac{EP} or deviations from Newtonian gravity~\cite{Fischbach:1996eq}. 
A linear $\phi$ coupling with the \ac{SM} generates a Yukawa force at tree-level, as shown in Figure~\ref{fig:linear_Yukawa}.

The Yukawa poten
tial that affects a test body $A$ in the presence of a massive central body $C$, such as the Earth, has the following form
\begin{equation}\label{eq:linear_Yukawa_potential}
V_{\text{Yukawa}}^{\text{lin}} \simeq -Q_i^A Q_i^C \left( d_i^{(1)}\right)^2 \dfrac{1}{4\pi r}e^{-m_\phi r}\,,
\end{equation}  
where $Q^A$ and $Q^C$ are the dilatonic charges of the test body and the central body respectively,~(see e.g~\cite{Damour:2010rp}) and $r$ is the distance between $A$ and $C$.
The sensitivity of \ac{EP} violation tests is characterized in terms of the differential acceleration $a$ between two test bodies $A$ and $B$ in the presence of a source $C$, and takes the following form
\begin{equation}\label{eq:EP_lin}
\eta^{\rm EP}=\left( \frac{\delta a_{\text{test}}}{a}\right)^{\text{EP-linear}} \propto \;  \left( d^{(1)}_{i}\right)^2\, \Delta Q_{i} \,Q_{i}^{C} \,,
\end{equation}
where $\Delta Q_{i}=Q_{i}^{\rm A}-Q_{i}^{\rm B}$. 

\subsection{Model with quadratic DM couplings}\label{subsec:Model_quad_DM_couplings}
The same analysis for the \ac{DDM} and \ac{EP} bounds on a linear $\phi$ theory can be extended to quadratic interactions. We focus on a CP invariant quadratic $\phi^2$ theory, given by the following effective Lagrangian
\begin{equation}\label{eq:Stadnik_Lint_quad}
\L_\text{quad-int} \!=\frac{\phi^2}{2}\left[ \frac{d^{(2)}_e}{4M_{\text{Pl}}^2 } F^{\mu\nu}F_{\mu\nu}-
\frac{d^{(2)}_g \beta_g}{2\Mpls \,g} G^{b\mu\nu}G^b_{\mu\nu} -\frac{d^{(2)}_{m_i}}{\Mpls} m_i\psi_i\psi^c_i+\text{h.c.} \right],
\end{equation}
where the $d^{(2)}_i$ are the dimensionless quadratic couplings.

Since $\phi$ acquires a time-dependent \ac{VEV} as in Eq.~\eqref{eq:DM_VEV}, oscillations of \aclp{FC} will be generated similarly to the linear theory, but with a background value of $\left\langle  \phi^2  \right\rangle $ instead of $\left\langle  \phi  \right\rangle $. 
For example, the variation of a constant $Y_i$ can be written as,
\begin{equation} \label{eq:AP-quad_electrons}
\frac{\delta Y_i(t)}{Y_i}\simeq \;  \Delta\kappa_i \frac{d^{(2)}_{i}}{2M_{\text{Pl}}^2} \left\langle \phi^2  \right\rangle  \simeq \Delta\kappa_i \frac{d^{(2)}_{i}}{M_{\text{Pl}}^2}\frac{\rho_\text{DM}}{m_{\phi}^2}\cos^2\leri{m_\phi t + \delta}
\end{equation}
One should note that since the temporal modulation of the fundamental constant should follow a $\cos^2\leri{m_\phi t}$ behavior, a bound on $\left(\delta Y_i(t)/Y_i\right)$ at an angular frequency $\nu$ should be interpreted as a bound on the couplings of a \ac{DM} candidate with a mass of $m_\phi=\frac{\hbar\nu}{2c^2}$. Moreover, since the time-averaged value of $\left\langle \cos^2\leri{m_\phi t} \right\rangle  = \frac{1}{2}$, there is an additive constant contribution to the fundamental coupling constants. 
In this work, we are
interested in timescales on which \ac{DM} density variations, and decoherence effects 
can be neglected. For a discussion related to the slow variation of the \ac{DM} amplitude see e.g.~\cite{Stadnik:2014tta,Stadnik:2015kia} (cosmological evolution) and \cite{Masia-Roig:2022net} (velocity dispersion decoherence).

The \ac{EP} bounds on the quadratic interactions are of different nature than the \ac{EP} bounds on the linear theory. In the quadratic theory, there is no Yukawa potential generated at tree level. However the time and space dependent background value of $\left\langle \phi^2 \right\rangle$, in the presence of a massive central object, results with an effective long range force. 
In Appendix~\ref{App:Classical_background_of_phi2}, we analyze the obtained background value of $\left\langle \phi^2 \right\rangle$ and its affect on the acceleration of a test body, and compare it to the 1-loop quantum corrections of $\phi^2$. We find that in the regime of our interest of \ac{ULDM}, the quantum corrections are negligible compared to the classical ones. 
See~\cite{Banks:2020gpu} for a discussion of the Yukawa force for generalised potentials. Below, we take a closer look at the profile of $\phi$ and its implications to \ac{EP} and \ac{DDM} bounds.

\subsubsection{The profile of quadratically coupled \ac{DM}, boundary condition dependence}\label{sec:criticallity}
The dynamics related to quadratic coupling have a strong dependence on the boundary conditions. 
In our analysis, we assume that far away from the Earth, $\phi$ takes its galactic \ac{DM} background, following~\cite{Hees:2018fpg}. 
This assumption is somewhat idealistic since the Earth is moving in the solar system, which consists of the moon, other planets, and the Sun, in addition to our solar system moving within the galactic medium, all perturbing the value of the scalar. 
In that way, it might not be entirely realistic to set the value of $\phi$ to its vacuum solution, assuming it is only be affected by the Earth, as this may depend on other variables such as the relaxation time and the dynamical history of the formation of the scalar background~\cite{Bar:2018acw,Budnik:2020nwz,Balkin:2021zfd,Bar:2021kti}. 

Alternatively, one can assume other boundary conditions and consider the phenomenology of a transient scalar background~\cite{Dailey_2020}. 
In fact the affect of gravitational focusing of ultralight \ac{DM} in the solar system was recently analyzed in~\cite{PhysRevD.105.063032} where it was shown that it leads to some changes to the distribution of the ultralight \ac{DM}, distribution that would have a preferred direction due to the velocity of the Sun in the \ac{DM} galactic halo. Finally, the self interaction of the \ac{DM} in the presence of a central gravitational potential, is expected to modify the \ac{ULDM} distribution, but requires a dedicated focused study beyond the scope of this paper, as reported in~\cite{haloformation}. For the sake of concreteness and despite the fact that the analysis might be incomplete, we shall follow the treatment of~\cite{Hees:2018fpg}, assuming trivial \ac{DM} background at infinity as in Eq.~\eqref{eq:DM_VEV}.

These boundary conditions lead to two important implications; the first is that the field is mediating long-distance forces despite being massive, and the second is a screening behavior near the surface of a massive source. 
Given the boundary conditions discussed above, the solution to the \ac{EOM} extracted from Eq.~\eqref{eq:Stadnik_Lint_quad} near a massive central object yields the following analytical solution to $\phi$
\begin{equation}\label{eq:phi_quadratic_solution_v1}
\phi\left(t,x\right)=\phi_{0}\cos\left(m_{\phi}t+\delta\right)\left[1-s_{C}^{\left(2\right)}[d_i^{(2)}]\,\frac{GM_{C}}{r}\right]\, ,
\end{equation}
where $\phi_0\equiv \sqrt{2\rho_{\rm DM}}/m_\phi$ is the background amplitude of $\phi$ at infinity, $s^{(2)}_{C}[d^{(2)}_i]$ is some function of the quadratic couplings (explained in Appendix~\ref{App:The_solution of the Linear model and a Quadratic}), $G$ is Newton's gravitational constant and $M_C$ the mass the central body, 
see subsection \ref{sec:Screening of the Quadratic potential} for additional details. 

In this Section, we follow two limits: the weakly coupled limit and the strongly coupled limit. In those limits, the solution takes the following form
\begin{equation}
\phi\left(t,x\right)=\phi_{0}\cos\left(m_{\phi}t+\delta\right)	\begin{cases}
\left[1-\frac{R_C}{r}\left(1-\sqrt{\frac{d_i^{\rm crit}}{d^{(2)}_i}}\right)\right]\,{\rm for}\,\,\, d^{(2)}_{i}\gg d_i^{\rm crit} \\
\left[1-\frac{R_C}{r}\frac{d^{(2)}_i}{3\,d_i^{\rm crit}}\right]\,\,{\rm for}\,\,\, d^{(2)}_{i} \ll d_i^{\rm crit}\,,
\end{cases}
\label{eq:phi2sol_v1}
\end{equation}
where the critical value of the coupling is defined as $d_i^{\rm crit} = R_C/(3 Q^C_i GM_C)$, where $R_C$ is the radius of the central body. 
Therefore, for a given \ac{DM} mass $m_\phi$, there exists a critical value of the quadratic coupling $d^{(2)}$, at which the background value of $\left\langle \phi^2 \right\rangle$ is screened, and thus the sensitivity to quadratically coupled \ac{ULDM} is suppressed. This can be seen by taking the limit $r \rightarrow R_C$ in Eq.\eqref{eq:phi2sol_v1}, when the \ac{DDM} sensitivity becomes
\begin{align} \label{eq:AP-quad_electrons_V2}
\eta^{\rm DDM}\leri{\omega} & = \mathcal{F}_\omega\leri{\left( \frac{\delta Y_i}{Y_i}(t)\right)^{\text{DDM-quad}}_{ d^{(2)}_{i} \gg d_i^{\rm crit} ,  r\simeq R_c } } \simeq \;  \frac{\Delta \kappa d^{(2)}_{i}}{2M_{\text{Pl}}^2} \mathcal{F}_\omega\leri{\left\langle \phi^2 \right\rangle} \nonumber
\\ & \simeq \Delta \kappa
\frac{\phi_0^2} {M_{\text{Pl}}^2} \, \left[ d_i^{\rm crit} +\frac{\Delta r}{R_C}d_{i}^{(2)}\sqrt{\frac{d_i^{\rm crit}}{d^{(2)}_i}} +\left( \frac{\Delta r}{R_C}\right)^2 d_{i}^{(2)} \right]\times\mathcal{F}_\omega\leri{\cos^2\left(m_{\phi}t+\delta\right)} \nonumber
\\ & \simeq \Delta \kappa
\frac{\phi_0^2} {M_{\text{Pl}}^2} \, d_i^{\rm crit} \times\mathcal{F}_\omega\leri{\cos^2\left(m_{\phi}t+\delta\right)} \,,
\end{align} 
where $\frac{\Delta r}{r} \equiv \frac{r-R_c}{r}\ll 1$.

Finally, we comment on the negative coupling scenario (when the sign of the \ac{DM} quadratic coupling is opposite relative to the mass term). 
In this case, the loss of control over the system is related to the fact that if inside the Earth the mass is too negative, it would overcome the pressure gradient from the kinetic term of the order of $1/R_C$. Thus, it would lead to tachyonic instabilities (where within the Earth, the mass squared of the field is negative) and to a runaway behavior of the field into large-amplitudes inside and outside~\cite{Budnik:2020nwz,Balkin:2021zfd,Hook:2019pbh}. 

\subsection{Summary of EP and DDM sensitivities}

We end this introduction by summarizing the sensitivities again, $\eta^{\rm DDM}$ and $\eta^{\rm EP}$, and provide their full description as a function of the \ac{DM} scalar field background. The complete derivation of these sensitivities can be found in Appendix~\ref{App:The_solution of the Linear model and a Quadratic}. For the linear couplings \ac{DM} model, we found:
\begin{align}
	& \eta^{\rm EP}_{\text{linear}}  = 2\frac{|\vec a_{A,C}-\vec a_{B,C}|}{|\vec a_{A,C}+\vec a_{B,C}|}\simeq\frac{|\Delta Q_i d^{(1)}_i\vec\nabla V_C\leri{\phi}|}{M_{\text{Pl}}G M_C/r^2}\simeq \frac{\Delta Q_i d^{(1)}_i}{M_{\text{Pl}}G M_C/r^2}\left[ \nabla \phi\left(x,t \right) + \vec{v}\dot{\phi}\left(x,t \right)\right]\, ,
	\label{eq:etq_EP_linear}
	\\
	& \eta^{\rm DDM} _{\text{linear}} \leri{\omega}= \mathcal{F}_\omega\leri{\frac{\delta Y(t) }{Y}}\simeq  \frac{  \Delta\kappa_i^A d^{(1)}_i}{M_{\text{Pl}}} \mathcal{F}_\omega\leri{\left\langle \phi\left(x,t \right)  \right\rangle}\,.	
	\label{eq:etq_DDM_linear}
\end{align}
In both equation~\eqref{eq:etq_EP_linear} and \eqref{eq:etq_DDM_linear}, the letters A and B represent two different test bodies, while C denotes the central heavy object such as the Earth. $d^{(1)}_i$ is the linear \ac{DM} coupling to the $i^{\text{th}}$ \ac{SM} field. For the quadratic \ac{DM} model, we find
\begin{align}
	& \eta^{\rm EP}_{\text{quad}}  = 2\frac{|\vec a_{A,C}-\vec a_{B,C}|}{|\vec a_{A,C}+\vec a_{B,C}|}\simeq 
	\frac{\Delta Q_id^{(2)}_i}{M^2_{\text{Pl}}G M_C/r^2} \phi\left(x,t \right) \left[ \nabla \phi\left(x,t \right) + \vec{v}\dot{\phi}\left(x,t \right)\right] \, ,
	\label{eq:etq_EP_quadratic}
	\\
	& \eta^{\rm DDM} _{\text{quad}}\leri{\omega} = \mathcal{F}_\omega\leri{\frac{\delta Y(t) }{Y}} \simeq  \frac{{ \Delta\kappa_i^A d^{(2)}_i }}{2M_{\text{Pl}}^2} \mathcal{F}_\omega\leri{\left\langle \phi^2\left(x,t \right)  \right\rangle} \,.	
	\label{eq:etq_DDM_quadratic}
\end{align}
Finally, we present the approximate sensitivities given the specific \ac{DM} background solution of Eq.~\eqref{eq:phi_quadratic_solution_v1} with its special boundary conditions. For the linear \ac{DM} model, we get
\begin{align}
	& \eta^{\rm EP}_{\text{linear}}  = 2\frac{|\vec a_{A,C}-\vec a_{B,C}|}{|\vec a_{A,C}+\vec a_{B,C}|}\simeq Q^C_j d^{(1)}_j \Delta Q_i d^{(1)}_ie^{-m_\phi r}\, ,
	\label{eq:etq_EP_linear_Stadnik_background}
	\\
	& \eta^{\rm DDM} _{\text{linear}} \leri{ m_\phi}= \frac{|\delta Y(t) |}{Y}\simeq  \frac{  \Delta\kappa_i^A d^{(1)}_i}{M_{\text{Pl}}} \phi_{0} \,.	
	\label{eq:etq_DDM_linear_Stadnik_background}
\end{align}
Lastly, the results for the quadratic \ac{DM} model are
\begin{align}
	& \eta^{\rm EP}_{\text{quad}}  = 2\frac{|\vec a_{A,C}-\vec a_{B,C}|}{|\vec a_{A,C}+\vec a_{B,C}|}\simeq 
	\Delta Q_id^{(2)}_i  s_{C}^{\left(2\right)}[d^{(2)}_i]\leri{\Delta Q}_j^{AB} d^{(2)}_j \frac{\phi_0^2}{2M^2_{\text{Pl}}}\left[1-s_{C}^{\left(2\right)}\frac{GM_{C}}{r}\right] \, ,
	\label{eq:etq_EP_quadratic_Stadnik_background}
	\\
	& \eta^{\rm DDM} _{\text{quad}}\leri{2 m_\phi} = \frac{|\delta Y(t) |}{Y}\simeq  \frac{{ \Delta\kappa_i^A d^{(2)}_i }}{4M_{\text{Pl}}^2} \phi_0^2\, \left[1-s_{C}^{\left(2\right)}\frac{GM_{C}}{r}\right]^2\,.	
	\label{eq:etq_DDM_quadratic_Stadnik_background}
\end{align}

\section[Updated Bounds from EP and DDM searches]{Updated Bounds from EP and DDM searches } \label{sec:Updated bounds from EP and Direct DM searches}

In this Section we present the bounds on the \ac{DM} models both from \ac{EP} and \ac{DDM} searches. 
We also discuss the interplay between the \ac{EP} and \ac{DDM} searches for the linear and the quadratically coupled \ac{DM} with the \ac{SM}. 
We also discuss two proposed ways to alleviate the \ac{EP} test constraints in details and present the reach of \ac{DDM} searches in those scenarios. 

\subsection{Summary of current and future bounds}\label{sec:Summary of current and future bounds}

\begin{figure}[h]
	\centering
	\includegraphics[scale=0.45]{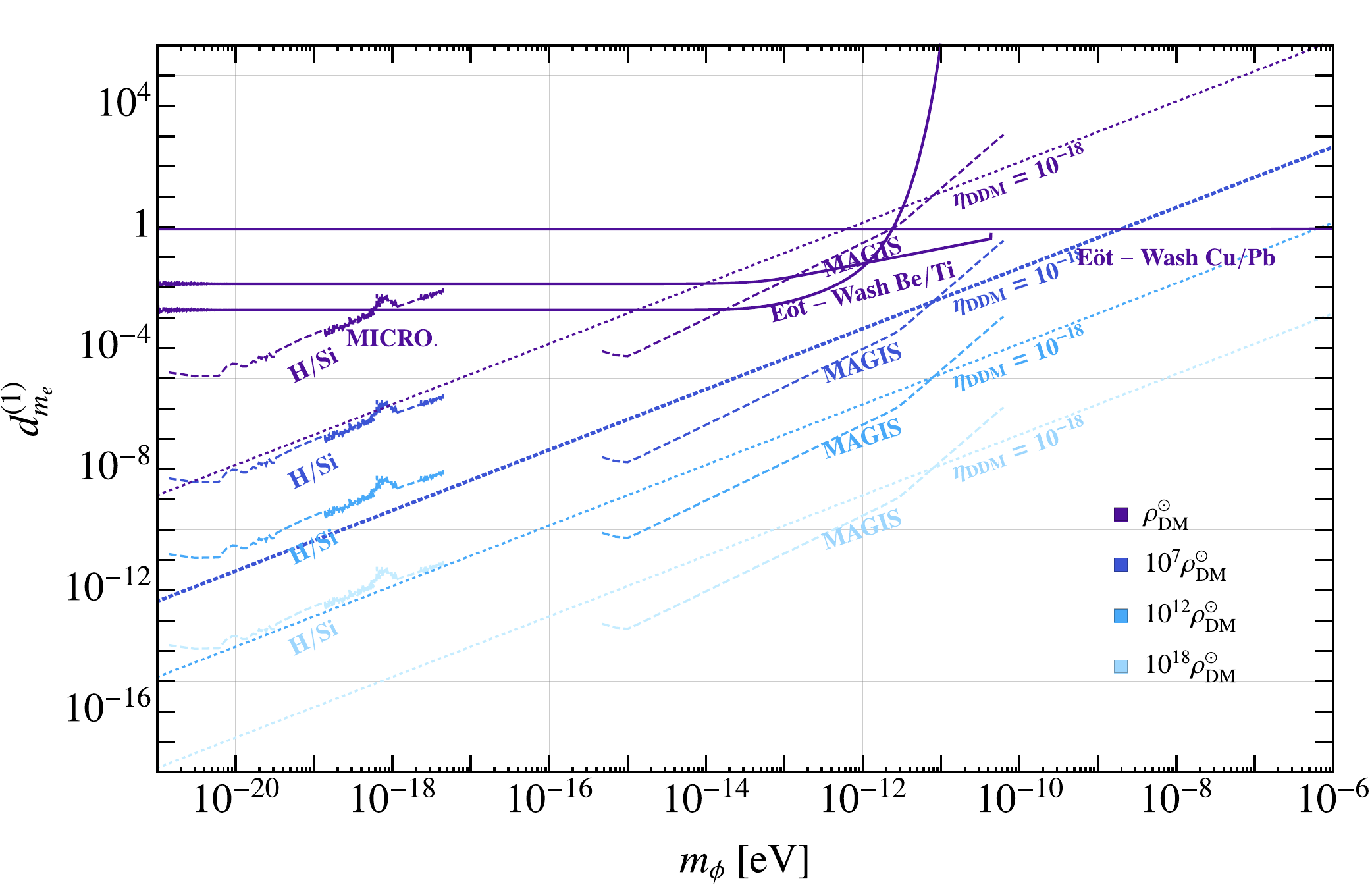}\quad
	\vspace{0.5 cm}
	\includegraphics[scale=0.45]{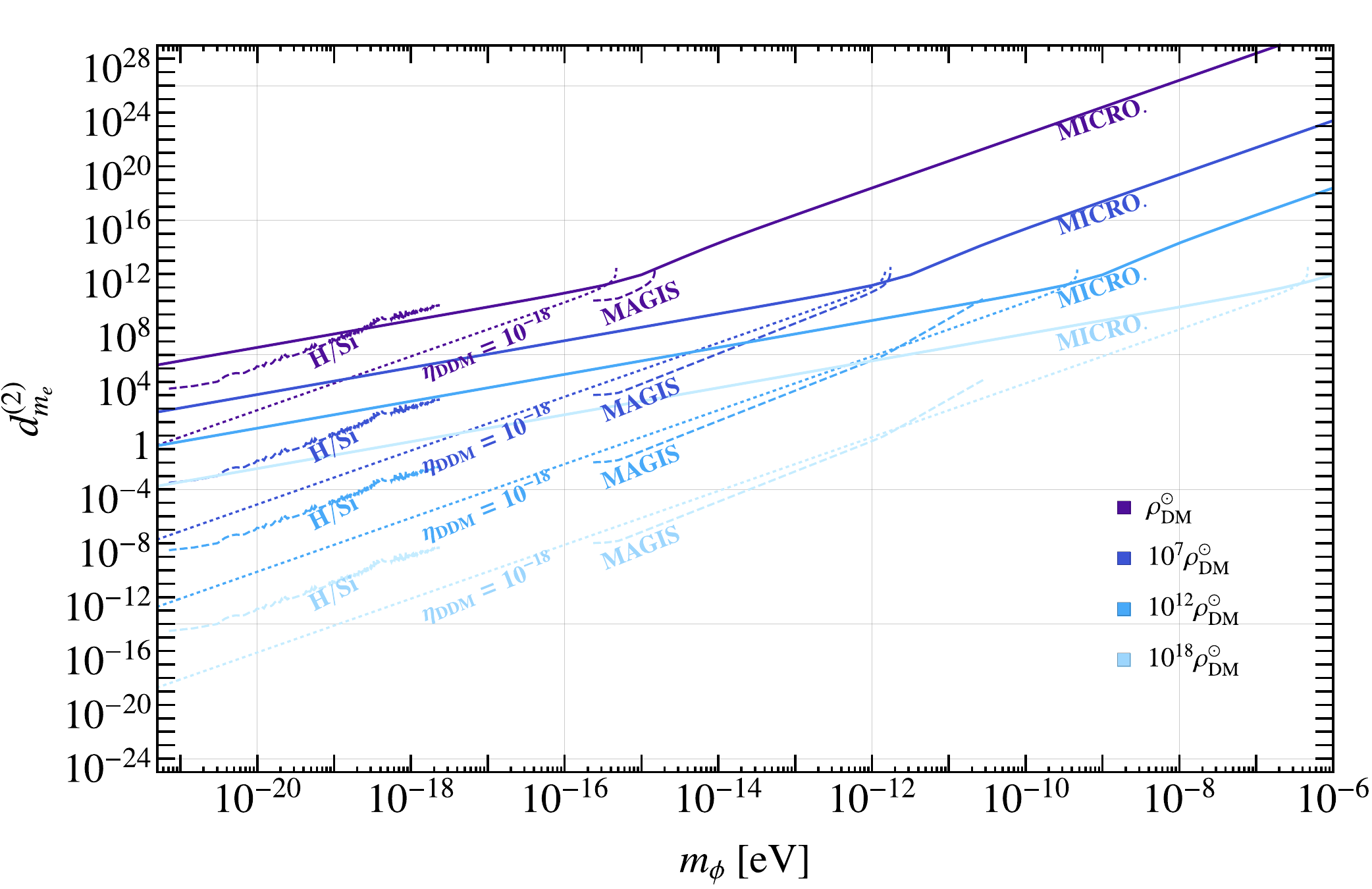}
	\caption{\textbf{Top:} the bounds on $d_{m_e}^{(1)}$ from the linear \ac{DM} electrons couplings.  \textbf{Bottom:} the bounds on $d_{m_e}^{(2)}$ from the quadratic \ac{DM} electron couplings. }
	\label{fig:dme_AP_EP}
\end{figure}

\begin{figure}[h]
	\centering
	\includegraphics[scale=0.45]{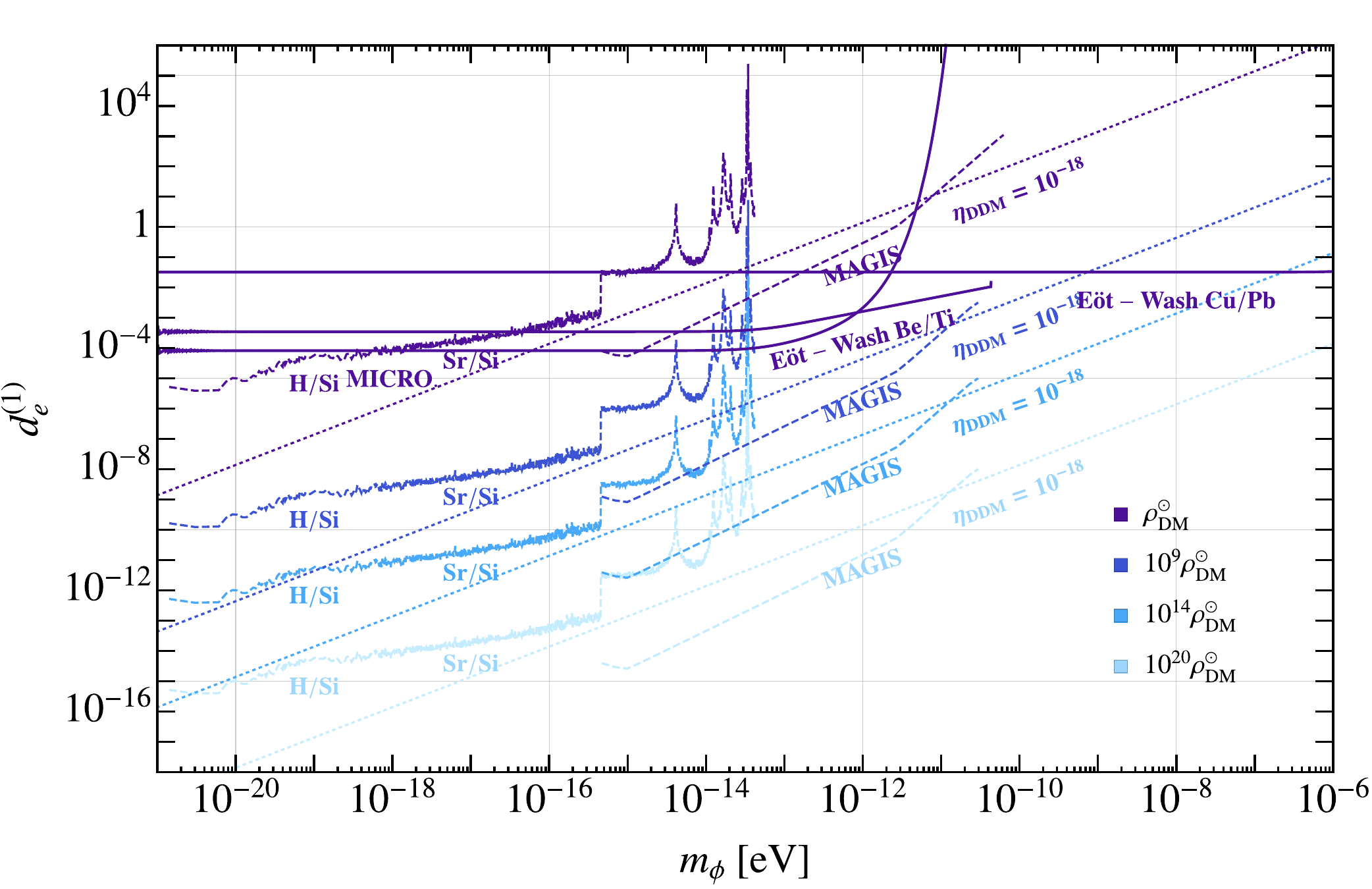}
	\vspace{0.5 cm}
	\includegraphics[scale=0.45]{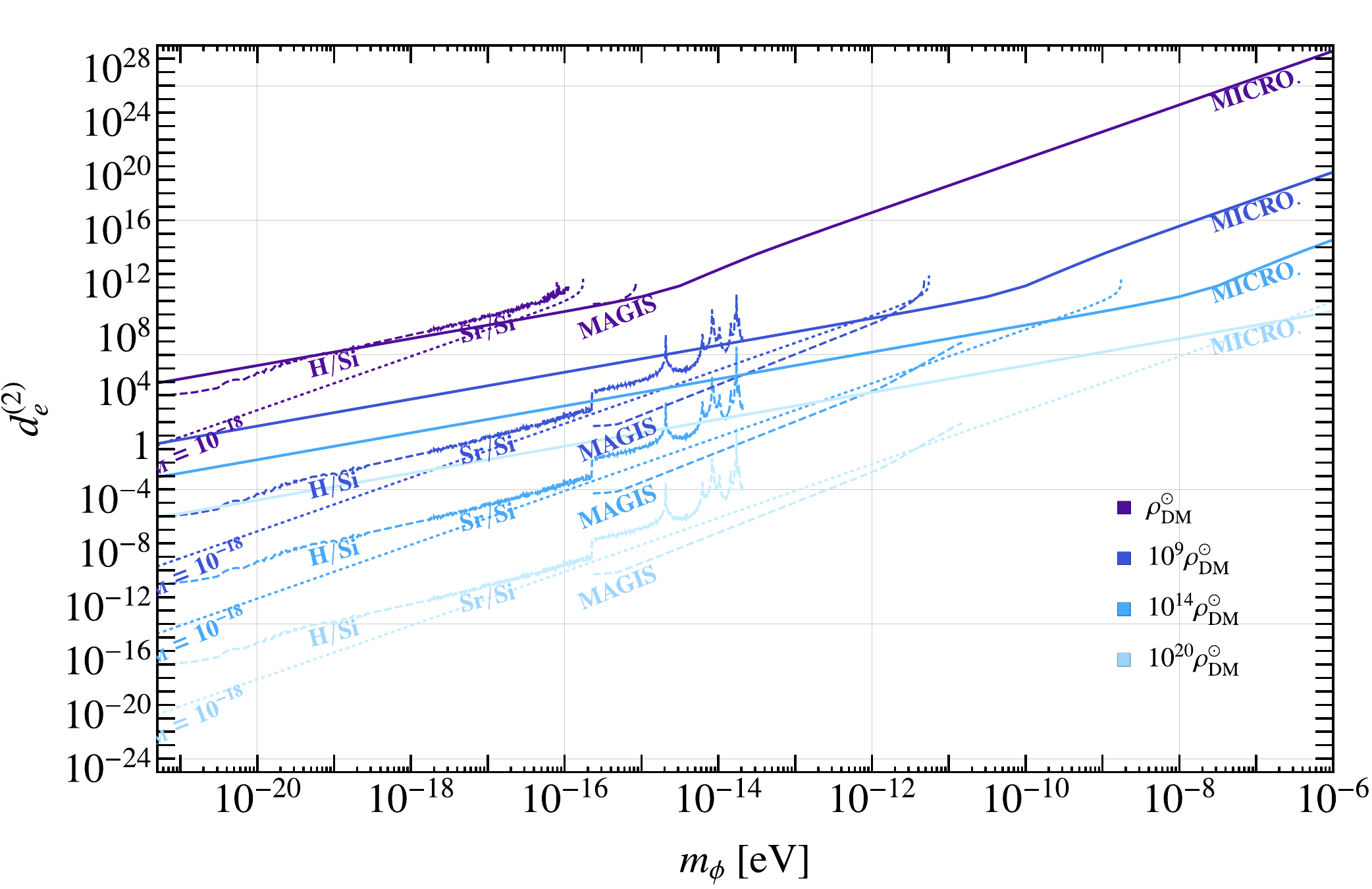}
	\vfill
	\caption{\textbf{Top:} the bounds on $d_{e}^{(1)}$ from the linear \ac{DM} photon couplings.  \textbf{Bottom:} the bounds on $d_{e}^{(2)}$ from the quadratic \ac{DM} photon couplings. Note the bounds in~\cite{Kennedy:2020bac} were modified for the assumption of a stochastic \ac{DM} background.}
	\label{fig:de_AP_EP}
\end{figure}

\begin{figure}[h]
	\centering
	\includegraphics[scale=0.45]{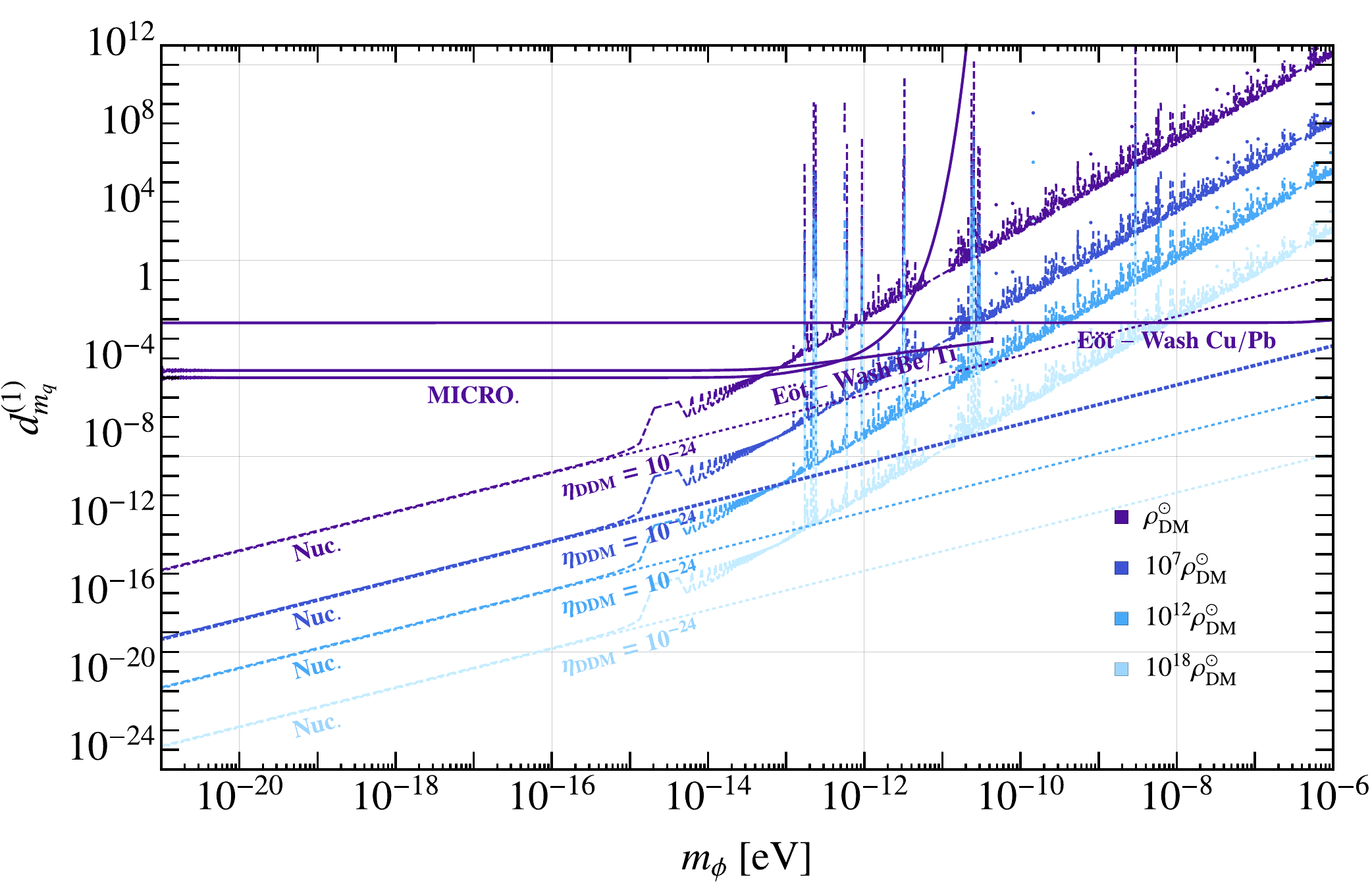}
	\vspace{0.5 cm}
	\includegraphics[scale=0.45]{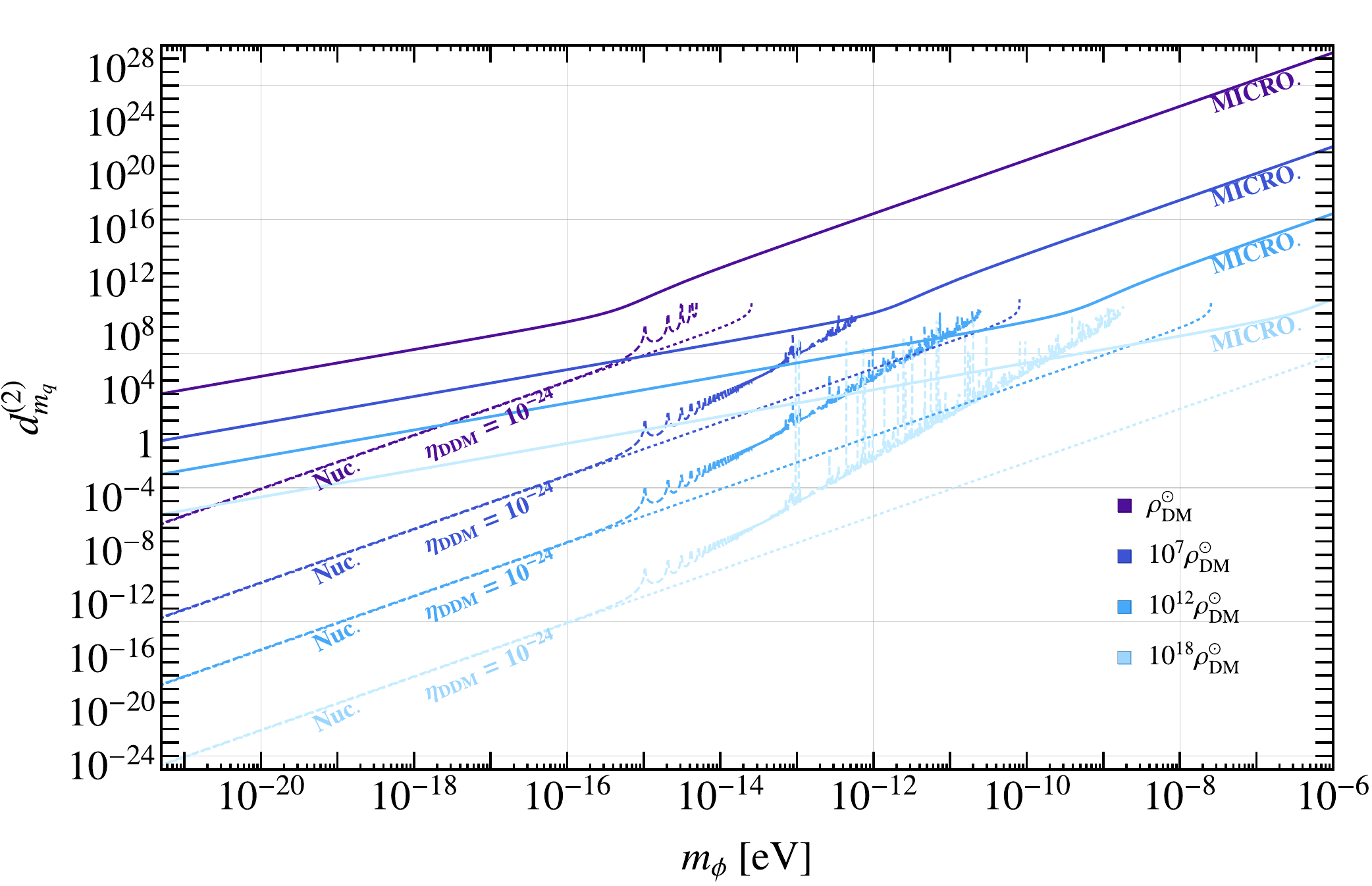}
	\vfill
	\caption{\textbf{Top:} the bounds on $d_{m_q}^{(1)}$ from the linear \ac{DM} light quarks couplings.  \textbf{Bottom:} the bounds on $d_{m_q}^{(2)}$ from the quadratic \ac{DM} light quarks couplings.}
	\label{fig:dmq_AP_EP}
\end{figure}

\begin{figure}[h]
	\centering
	\includegraphics[scale=0.45]{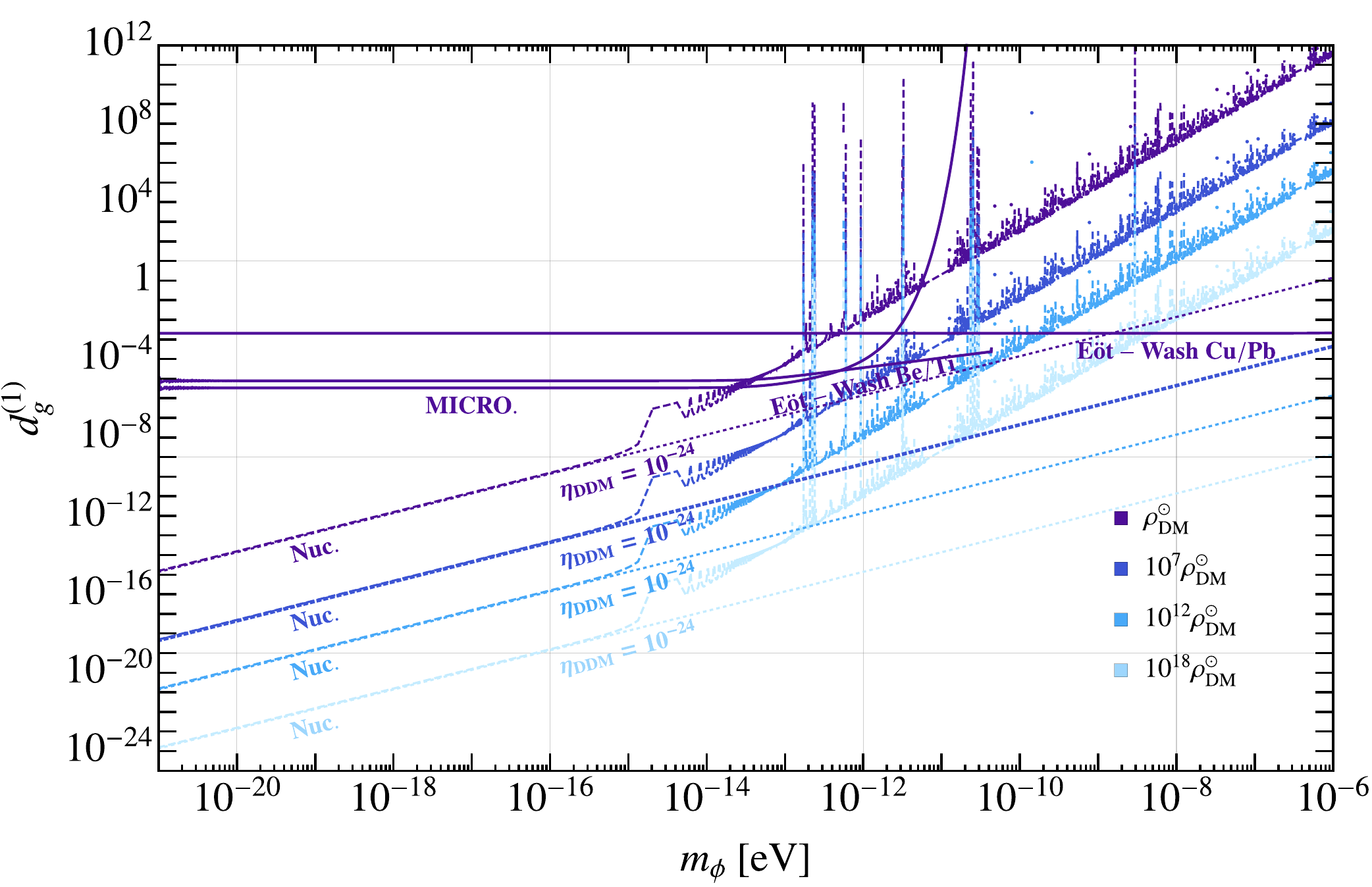}
	\vspace{0.5 cm}
	\includegraphics[scale=0.45]{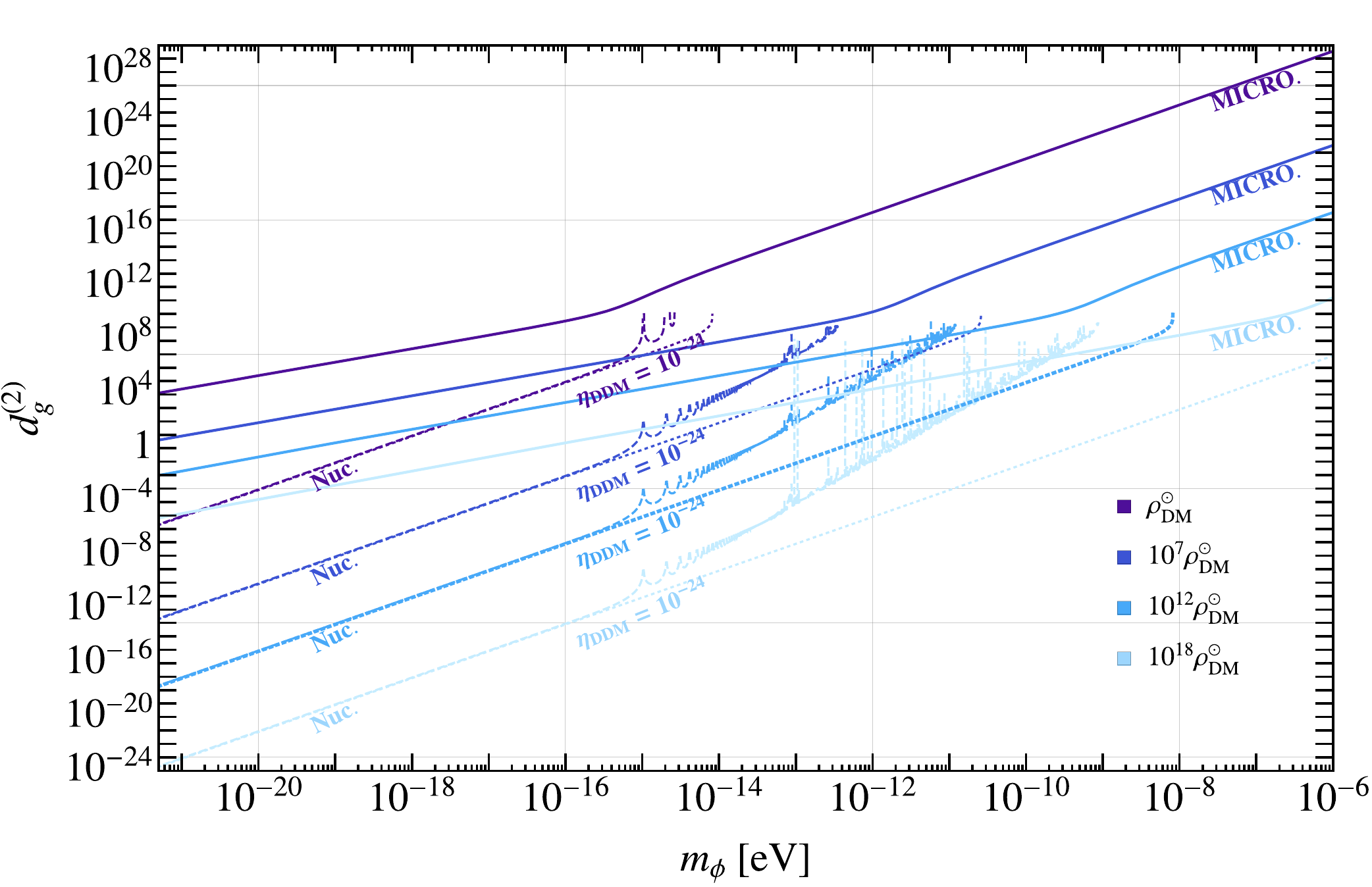}
	\vfill
	\caption{\textbf{Top:} the bounds on $d_{g}^{(1)}$ from the linear \ac{DM} gluon couplings.  \textbf{Bottom:} the bounds on $d_{g}^{(2)}$ from the quadratic \ac{DM} gluon couplings.}
	\label{fig:dg_AP_EP}
\end{figure}

The known bounds on scalar, pseudo-scalar and quadratic \ac{DM} interactions with the \ac{SM} are summarized in Tables~\ref{tab:bounds_summary_10^-8eV}-\ref{tab:bounds_summary_10^-18eV} for various \ac{ULDM} masses. 
In addition, we also present current and future-projected bounds on the linear scalar couplings $d^{(1)}_i$ and on the quadratic couplings $d^{(2)}_i$ as a function of the \ac{DM} mass for various local \ac{DM} densities, up to $10^5$ the \ac{DM} density at the solar position $\rho_{\text{DM}}^{\odot}$~\cite{Salucci:2010qr}, as motivated by~\cite{Pitjev_2013,Tsai:2022jnv,haloformation}. The bounds for the electron couplings, $d_{m_e}$, are shown in Figure~\ref{fig:dme_AP_EP}, the bounds for the photon couplings, $d_e$, are shown in Figure~\ref{fig:de_AP_EP}, the bounds for the quark couplings, $d_{m_q}$, are shown in Figure~\ref{fig:dmq_AP_EP}, and the bounds for the gluon couplings, $d_g$, are shown in Figure~\ref{fig:dg_AP_EP}. 
For all linear couplings, the \ac{EP} test bounds are derived from the terrestrial E\"ot-Wash Be/Ti~\cite{Schlamminger:2007ht} and E\"ot-Wash Cu/Pb~\cite{PhysRevD.61.022001} measurements, as well as from the MICROSCOPE data~\cite{PhysRevLett.129.121102} taken on a satellite orbiting the Earth at an approximate altitude of 700 km. 
For the quadratic couplings, we present only the bounds from MICROSCOPE, which are expected to be the strongest~\cite{Hees:2018fpg}. 
 The current \ac{DDM} bounds for $d_{m_e}$ are given from the H/Si clock-cavity comparison measurements presented in~\cite{Kennedy:2020bac}. 
 For $d_e$, the current \ac{DDM} bounds are given both from H/Si and Sr/Si clock-cavity comparisons~\cite{Kennedy:2020bac}, where for masses larger than $\sim 10^{-16}$~eV an additional measurement with using dynamical decoupling was applied to improve the sensitivity at high frequencies~\cite{Aharony:2019iad}.  
 For both $d_{m_e}$ and $d_e$, we also show a line representing a \ac{DDM} sensitivity of $\eta_{\rm DDM}=10^{-18}$ at all masses, as well as the expected bound from the future \ac{DDM} MAGIS-100 experiment~\cite{Abe_2021}. 
A \ac{DDM} experiment involving hyperfine transitions~\cite{Kennedy:2020bac,Hees:2016gop} and/or vibrational levels of a molecule~\cite{Oswald:2021vtc} can be used to constraint \ac{DM} couplings to nucleus i.e. $d_{m_q}$ and $d_{g}$.
However, here we present the expected bounds from a nuclear clock with a sensitivity of $\eta_{\rm DDM}=10^{-24}$, using a Ramsey sequence with the parameters given in~\cite{Banerjee:2020kww}. 
The astrophysical constraints mentioned in Tables~\ref{tab:bounds_summary_10^-8eV}-\ref{tab:bounds_summary_10^-18eV}, are coming from various stellar cooling processes as mentioned in the given references. 

 \begin{table}[h!]
	\begin{equation}
	\boxed{m_\phi = 10^{-8}\,\eV}\nonumber
	\end{equation}
	\\
	\centering
	\hspace*{-2cm} \begin{tabular}{c|cc}
		operator & current bound & type of experiment \tabularnewline
		\hline 
		$\frac{d^{(1)}_e}{4\,M_\text{Pl}} \phi \, F^{\mu\nu}F_{\mu\nu}$  & $d^{(1)}_e \lesssim 10^{-2}$~\cite{PhysRevD.61.022001}   & \ac{EP} test: E\"ot-Wash Cu/Pb
		\tabularnewline	
		$\frac{\tilde{d}^{(1)}_e}{M_\text{Pl}} \phi \, F^{\mu\nu}\tilde{F}_{\mu\nu}$  & $\tilde{d}^{(1)}_e\lesssim 2\times 10^8 $~\cite{Anastassopoulos:2017ftl}  & axion/ALP searches: CAST
		\tabularnewline
		$\frac{\left| d^{(1)}_{m_e} \right| }{M_\text{Pl}} \phi \, m_e \psi_e \psi_e^c$  & $\left| d^{(1)}_{m_e} \right|  \lesssim 1$~\cite{PhysRevD.61.022001} & \ac{EP} test: E\"ot-Wash Cu/Pb
		\tabularnewline
		$i\frac{ \left| \tilde{d}^{(1)}_{m_e} \right| }{M_\text{Pl}} \phi \, m_e \psi_e \psi_e^c$  & $\left| \tilde{d}^{(1)}_{m_e} \right|\lesssim 8 \times 10^8$~\cite{Capozzi:2020cbu} & Astrophysics
		\tabularnewline
		& & 
		\tabularnewline
		$\frac{d^{(1)}_g\beta(g)}{2M_\text{Pl}\, g} \phi \, G^{\mu\nu}G_{\mu\nu}$  & $d^{(1)}_g \lesssim  10^{-3}$~\cite{PhysRevD.61.022001} &  \ac{EP} test: E\"ot-Wash Cu/Pb
		\tabularnewline
		$\frac{\tilde{d}^{(1)}_g}{M_\text{Pl}} \phi \, G^{\mu\nu}\tilde{G}_{\mu\nu}$  & $\tilde{d}^{(1)}_g \lesssim 10^{8}$~\cite{Raffelt:2006cw, Buschmann:2021juv} & Astrophysics
		\tabularnewline
		$\frac{\left| d^{(1)}_{m_N} \right| }{M_\text{Pl}} \phi \, m_N \psi_N \psi_N^c$  & $\left| d^{(1)}_{m_N} \right|\lesssim 6\times 10^{-3}$~\cite{Schlamminger:2007ht} & \small{\ac{EP} test: 
			E\"ot-wash 2008}
		\tabularnewline
		$i\frac{ \left| \tilde{d}^{(1)}_{m_N} \right| }{M_\text{Pl}} \phi \, m_N \psi_N \psi_N^c$  & $\left| \tilde{d}^{(1)}_{m_N} \right|\lesssim 3\times 10^8$~\cite{Beznogov:2018fda} & Astrophysics
		\tabularnewline
		& & 
		\tabularnewline
		& & 
		\tabularnewline
		$\frac{d^{(2)}_e}{8M^2_\text{Pl}} \phi^2 \, F^{\mu\nu}F_{\mu\nu}$  & $d^{(2)}_e  \lesssim 10^{25}$~\cite{PhysRevLett.129.121102}   & \ac{EP} test: MICROSCOPE
		\tabularnewline
		$\frac{\left| d^{(2)}_{m_e} \right| }{2M^2_\text{Pl}} \phi^2 \, m_e \psi_e \psi_e^c$  & $\left| d^{(2)}_{m_e} \right|   \lesssim 10^{27}$~\cite{PhysRevLett.129.121102}   & \ac{EP} test: MICROSCOPE
		\tabularnewline
		& & 
		\tabularnewline
		$\frac{d^{(2)}_g \beta_g}{4\Mpls \,g}\phi^2 \, G^{\mu\nu}G_{\mu\nu}$  & $d^{(2)}_g \lesssim 10^{25}$~\cite{PhysRevLett.129.121102} & \ac{EP} test: MICROSCOPE
		\tabularnewline
		$\frac{\left| d^{(2)}_{m_N} \right| }{2M^2_\text{Pl}} \phi^2 \, m_N \psi_N \psi_N^c$  & $\left| d^{(2)}_{m_N} \right| \lesssim 10^{25}$~\cite{PhysRevLett.129.121102} & \ac{EP} test: MICROSCOPE
		\tabularnewline
		
	\end{tabular}
	\caption{Strongest existing bounds on various \ac{DM} couplings for a mass of the order of $m_\phi = 10^{-8}\,\eV$.}
  \label{tab:bounds_summary_10^-8eV}
\end{table}

\begin{table}[h!]
	\begin{equation}
	\boxed{m_\phi = 10^{-15}\,\eV}\nonumber
	\end{equation}
	\\
	\centering
	\hspace*{-2cm} \begin{tabular}{c|cc}
		operator & current bound & type of experiment \tabularnewline
		\hline 
		$\frac{d^{(1)}_e}{4\,M_\text{Pl}} \phi \, F^{\mu\nu}F_{\mu\nu}$  & $d^{(1)}_e \lesssim  10^{-4}$~\cite{PhysRevLett.129.121102}   & \ac{EP} test: MICROSCOPE
		\tabularnewline	
		$\frac{\tilde{d}^{(1)}_e}{M_\text{Pl}} \phi \, F^{\mu\nu}\tilde{F}_{\mu\nu}$  & $\tilde{d}^{(1)}_e\lesssim 2\times 10^6 $~\cite{Reynolds:2019uqt}  & Astrophysics
		\tabularnewline
		$\frac{\left| d^{(1)}_{m_e} \right| }{M_\text{Pl}} \phi \, m_e \psi_e \psi_e^c$  & $\left| d^{(1)}_{m_e} \right|  \lesssim  10^{-3}$~\cite{PhysRevLett.129.121102}   & \small{\ac{EP} test: MICROSCOPE}
		\tabularnewline
		$i\frac{ \left| \tilde{d}^{(1)}_{m_e} \right| }{M_\text{Pl}} \phi \, m_e \psi_e \psi_e^c$  & $\left| \tilde{d}^{(1)}_{m_e} \right|\lesssim 8 \times 10^8$~\cite{Capozzi:2020cbu} & Astrophysics
		\tabularnewline
		& & 
		\tabularnewline
		$\frac{d^{(1)}_g\beta(g)}{2M_\text{Pl}\, g} \phi \, G^{\mu\nu}G_{\mu\nu}$  & $d^{(1)}_g \lesssim 6\times 10^{-6}$~\cite{PhysRevLett.129.121102} & \ac{EP} test: MICROSCOPE
		\tabularnewline	
		$\frac{\tilde{d}^{(1)}_g}{M_\text{Pl}} \phi \, G^{\mu\nu}\tilde{G}_{\mu\nu}$  & $\tilde{d}^{(1)}_g \lesssim 10^{8}$~\cite{Raffelt:2006cw, Buschmann:2021juv} & SN1987A, NS 
		\tabularnewline
		$\frac{\left| d^{(1)}_{m_N} \right| }{M_\text{Pl}} \phi \, m_N \psi_N \psi_N^c$  & $\left| d^{(1)}_{m_N} \right| \lesssim 6 \times 10^{-6}$~\cite{PhysRevLett.129.121102} &  \ac{EP} test: MICROSCOPE
		\tabularnewline
		$i\frac{ \left| \tilde{d}^{(1)}_{m_N} \right| }{M_\text{Pl}} \phi \, m_N \psi_N \psi_N^c$  & $\left| \tilde{d}^{(1)}_{m_N} \right|\lesssim 3\times 10^8$~\cite{Beznogov:2018fda} & Astrophysics
		\tabularnewline
		& & 
		\tabularnewline
		& & 
		\tabularnewline
		$\frac{d^{(2)}_e}{8M^2_\text{Pl}} \phi^2 \, F^{\mu\nu}F_{\mu\nu}$  & $d^{(2)}_e  \lesssim 10^{11}$~\cite{PhysRevLett.129.121102}   & \ac{EP} test: MICROSCOPE
		\tabularnewline
		$\frac{\left| d^{(2)}_{m_e} \right| }{2M^2_\text{Pl}} \phi^2 \, m_e \psi_e \psi_e^c$  & $\left| d^{(2)}_{m_e} \right|  \lesssim 10^{12}\,$~\cite{PhysRevLett.129.121102}   & \ac{EP} test: MICROSCOPE
		\tabularnewline
		$\frac{d^{(2)}_g \beta_g}{4\Mpls \,g}\phi^2 \, G^{\mu\nu}G_{\mu\nu}$  & $d^{(2)}_g  \lesssim 10^{11} $~\cite{PhysRevLett.129.121102}& \ac{EP} test: MICROSCOPE.
		\tabularnewline	
		$\frac{\left| d^{(2)}_{m_N} \right| }{2M^2_\text{Pl}} \phi^2 \, m_N \psi_N \psi_N^c$  & $\left| d^{(2)}_{m_N} \right|  \lesssim 10^{11} $~\cite{PhysRevLett.129.121102}& \ac{EP} test: MICROSCOPE
		\tabularnewline
	\end{tabular}
	\caption{Strongest existing bounds on various \ac{DM} couplings for a mass of the order of $m_\phi = 10^{-15}\,\eV$.}
	\label{tab:bounds_summary_10^-15eV}
\end{table}
\begin{table}[h!]
	\begin{equation}
	\boxed{m_\phi = 10^{-18}\,\eV}\nonumber
	\end{equation}
	\\
	\centering
	\hspace*{-2cm} 	\begin{tabular}{c|cc}
		operator & current bound & type of experiment \tabularnewline
		\hline 
		$\frac{d^{(1)}_e}{4\,M_\text{Pl}} \phi \, F^{\mu\nu}F_{\mu\nu}$  & $d^{(1)}_e \lesssim 10^{-4}\,$~\cite{Kennedy:2020bac}& \ac{DDM} oscillations
		\tabularnewline	
		$\frac{\tilde{d}^{(1)}_e}{M_\text{Pl}} \phi \, F^{\mu\nu}\tilde{F}_{\mu\nu}$  & $\tilde{d}^{(1)}_e\lesssim 2\times 10^6 $~\cite{Reynolds:2019uqt}  & Astrophysics
		\tabularnewline
		$\frac{\left| d^{(1)}_{m_e} \right| }{M_\text{Pl}} \phi \, m_e \psi_e \psi_e^c$  & $\left| d^{(1)}_{m_e} \right|  \lesssim 2\times10^{-3}$ \cite{Kennedy:2020bac}  &   \ac{DDM} Oscillations 
		\tabularnewline
		$i\frac{ \left| \tilde{d}^{(1)}_{m_e} \right| }{M_\text{Pl}} \phi \, m_e \psi_e \psi_e^c$  & $\left| \tilde{d}^{(1)}_{m_e} \right|\lesssim 7\times 10^8$~\cite{Capozzi:2020cbu} & Astrophysics
		\tabularnewline
		& & 
		\tabularnewline
		$\frac{d^{(1)}_g\beta(g)}{2M_\text{Pl}\, g} \phi \, G^{\mu\nu}G_{\mu\nu}$  & $d^{(1)}_g \lesssim 6\times 10^{-6}$~\cite{PhysRevLett.129.121102} & \ac{EP} test: MICROSCOPE
		\tabularnewline	
		$\frac{\tilde{d}^{(1)}_g}{M_\text{Pl}} \phi \, G^{\mu\nu}\tilde{G}_{\mu\nu}$  & $\tilde{d}^{(1)}_g \lesssim 4$~\cite{Abel:2017rtm} & Oscillating neutron EDM
		\tabularnewline
		$\frac{\left| d^{(1)}_{m_N} \right| }{M_\text{Pl}} \phi \, m_N \psi_N \psi_N^c$  & $\left| d^{(1)}_{m_N} \right| \lesssim 2 \times 10^{-6}$~\cite{PhysRevLett.129.121102} & \ac{EP} test: MICROSCOPE
		\tabularnewline
		$i\frac{ \left| \tilde{d}^{(1)}_{m_N} \right| }{M_\text{Pl}} \phi \, m_N \psi_N \psi_N^c$  & $\left| \tilde{d}^{(1)}_{m_N} \right| \lesssim4$~\cite{Abel:2017rtm} & Oscillating neutron EDM
		\tabularnewline
		& & 
		\tabularnewline
		& & 
		\tabularnewline
		$\frac{d^{(2)}_e}{8M^2_\text{Pl}} \phi^2 \, F^{\mu\nu}F_{\mu\nu}$  & $d^{(2)}_e   \lesssim10^{6}$~\cite{Kennedy:2020bac}& \ac{DDM} oscillations
		\tabularnewline
		$\frac{\left| d^{(2)}_{m_e} \right| }{2M^2_\text{Pl}} \phi^2 \, m_e \psi_e \psi_e^c$  & $\left| d^{(2)}_{m_e} \right| \lesssim 10^{6}\,$~\cite{Kennedy:2020bac}   & \ac{DDM} oscillations 
		\tabularnewline
		& & 
		\tabularnewline
		$\frac{d^{(2)}_g \beta_g}{4\Mpls \,g}\phi^2 \, G^{\mu\nu}G_{\mu\nu}$  & $d^{(2)}_g  \lesssim 10^{7} $~\cite{PhysRevLett.129.121102}& \ac{EP} test: MICROSCOPE.
		\tabularnewline	
		$\frac{\left| d^{(2)}_{m_N} \right| }{2M^2_\text{Pl}} \phi^2 \, m_N \psi_N \psi_N^c$  & $\left| d^{(2)}_{m_N} \right|  \lesssim 10^{7} $~\cite{PhysRevLett.129.121102}& \ac{EP} test: MICROSCOPE
		\tabularnewline
		
	\end{tabular}
	\caption{Strongest existing bounds on various \ac{DM} couplings for a mass of the order of $m_\phi = 10^{-18}\,\eV$.}
	\label{tab:bounds_summary_10^-18eV}
\end{table}

\subsection{Complementarity of EP tests and DDM searches}
We can compare the bounds on the \ac{DM} couplings coming from \ac{DDM} experiments to the ones coming from \ac{EP} tests, both for linear and quadratic interactions. We begin by summarizing the scaling of the \ac{DDM} and \ac{EP} bounds, both in the linear and quadratic theories, as presented in Table~\ref{tab:scaling_summary}. 
\begin{table}
	\centering
	\begin{tabular}{c|cc} 
		& {\footnotesize{}linear theory: $\frac{d_{i}^{(1)}}{M_{\text{Pl}}}\phi\,\mathcal{O}^i_{\rm SM}$ } & {\footnotesize{}quadratic theory: $\frac{d_{i}^{(2)}}{2M_{\text{Pl}}^{2}}\phi^{2}\,\mathcal{O}^i_{\rm SM}$ }\tabularnewline
		\hline 
		\ac{DDM} bounds: $\left(\frac{\delta Y_i}{Y}(t)\right)<\eta_{\text{DDM}}$ & $\frac{d_{i}^{(1)}}{M_{\text{Pl}}}\frac{\sqrt{2\rho_{\text{DM}}}}{m_{\phi}}$ & $\frac{d_{i}^{(2)}}{2M_{\text{Pl}}^{2}}\frac{\rho_{\text{DM}}}{m_{\phi}^{2}}$\tabularnewline
		\ac{EP} bounds: $\left(\frac{\delta a_{\text{test}}}{a}\right)<\eta_{\text{EP}}$ & $\left(d_{i}^{(1)}d_{j}^{(1)}\right)\Delta Q_{i}^{\text{test}}Q_{j}^{\text{Earth}}$ & $\frac{1}{M_{\text{Pl}}^{2}}\left(d_{i}^{(2)}d_{j}^{(2)}\right)\Delta Q_{i}^{\text{test}}Q_{j}^{\text{Earth}}\,\frac{\rho_{\text{DM}}}{m_{\phi}^{2}}$\tabularnewline
		&  & \tabularnewline
		Ratio: $\frac{\left(d\right)_{\text{DDM}}}{\left(d\right)_{\text{EP}}}$ & $\frac{M_{\text{Pl}}m_{\phi}\sqrt{\Delta Q_{i}^{\text{test}}Q_{i}^{\text{Earth}}}}{\sqrt{2\rho_{\text{DM}}}}\,\frac{\eta_{\text{DDM}}}{\sqrt{\eta_{\text{EP}}}}$ & $\frac{2M_{\text{Pl}}m_{\phi}\sqrt{\Delta Q_{i}^{\text{test}}Q_{i}^{\text{Earth}}}}{\sqrt{\rho_{\text{DM}}}}\,\frac{\eta_{\text{DDM}}}{\sqrt{\eta_{\text{EP}}}}$\tabularnewline
	\end{tabular}
	\caption{Theoretical estimation of the bounds on the linear and quadratic couplings between the light scalar \ac{DM} and the \ac{SM} light fields. $\eta_{\text{DDM}}$ and $\eta_{\text{EP}}$ are defined as the sensitivity of the \ac{DDM} and \ac{EP} experiments, respectively. $Y$ is a fundamental constant $Y\in(\alpha,\alpha_s, m_f)$ and $\mathcal{O}_{\rm SM}$ is the appropriate \ac{SM} operator $\mathcal{O}_{\rm SM}\in \leri{\frac14 F^{\mu\nu} F_{\mu\nu}\,, \frac{\beta(g)}{2g}G_a^{\mu\nu} G^a_{\mu\nu} \,,m_{i}\psi_{i}\psi_{i}^{c}}$. The spatial dependence of the bounds is disregarded (equivalent to taking the $m_\phi \ll 1/R_C$ limit for the linear case and the $d^{(2)}_i\ll d_i^{\rm crit}$ limit for the quadratic case). }
	\label{tab:scaling_summary}
\end{table}
As one can easily read from Table~\ref{tab:scaling_summary}, the ratio of the bounded couplings from different types of experiments, i.e., the ratio: $\left(d\right)_{\text{DDM}} / \left(d\right)_{\text{EP}}$ has the same parametric dependence in both the linear and the quadratic theory, as long as the spatial dependence of the bounds may be neglected (namely, away from the Yukawa decoupling of the linear bounds and in the sub-critical region for the quadratic bounds). Therefore, under these conditions, if one type of experiment dominates the bounds on the linear interaction in some region of the parameter space, we expect it to also dominate the bounds on the quadratic couplings and vice versa. As is further shown in the table, in agreement with the plots above, \ac{DDM} experiments tend to be more powerful at lower masses. Their corresponding constraints improve linearly with the experimental sensitivity, whereas \ac{EP} tests are expected to take over at higher masses while scaling only with the square root of the experimental sensitivity.

While one of these searches usually dominates the bounds for specific masses, we would like to argue that EP-tests and the \ac{DDM} searches are complementary to each other and provide independent information. Below we point out two engaging scenarios in which the naive ratio between \ac{EP} and \ac{DDM} bounds is violated, demonstrating their complementary.

\subsubsection{Enhanced DM Density }
\begin{figure}
	\centering
	\includegraphics[width=0.8\linewidth]{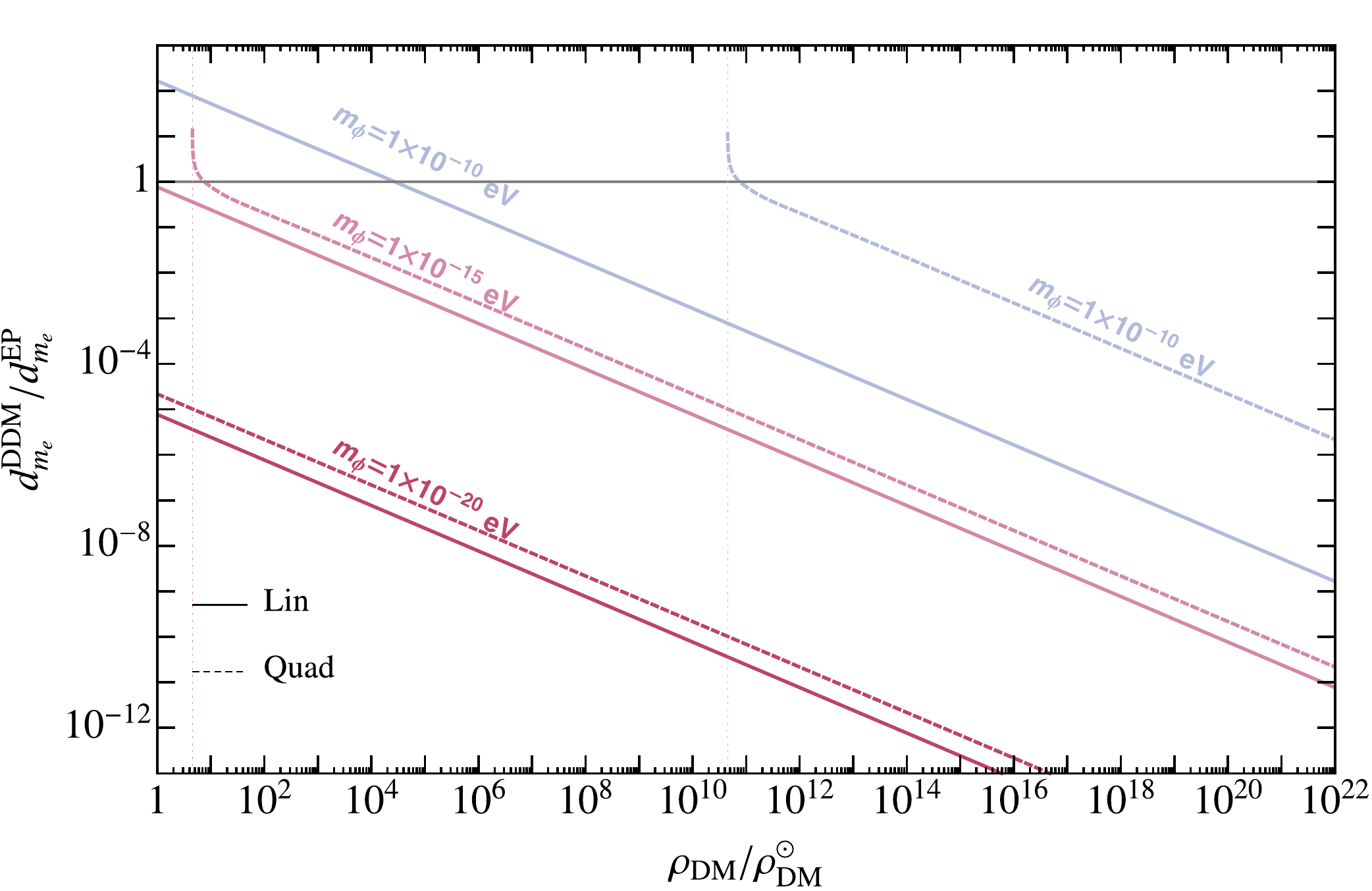}
	\includegraphics[width=0.8\linewidth]{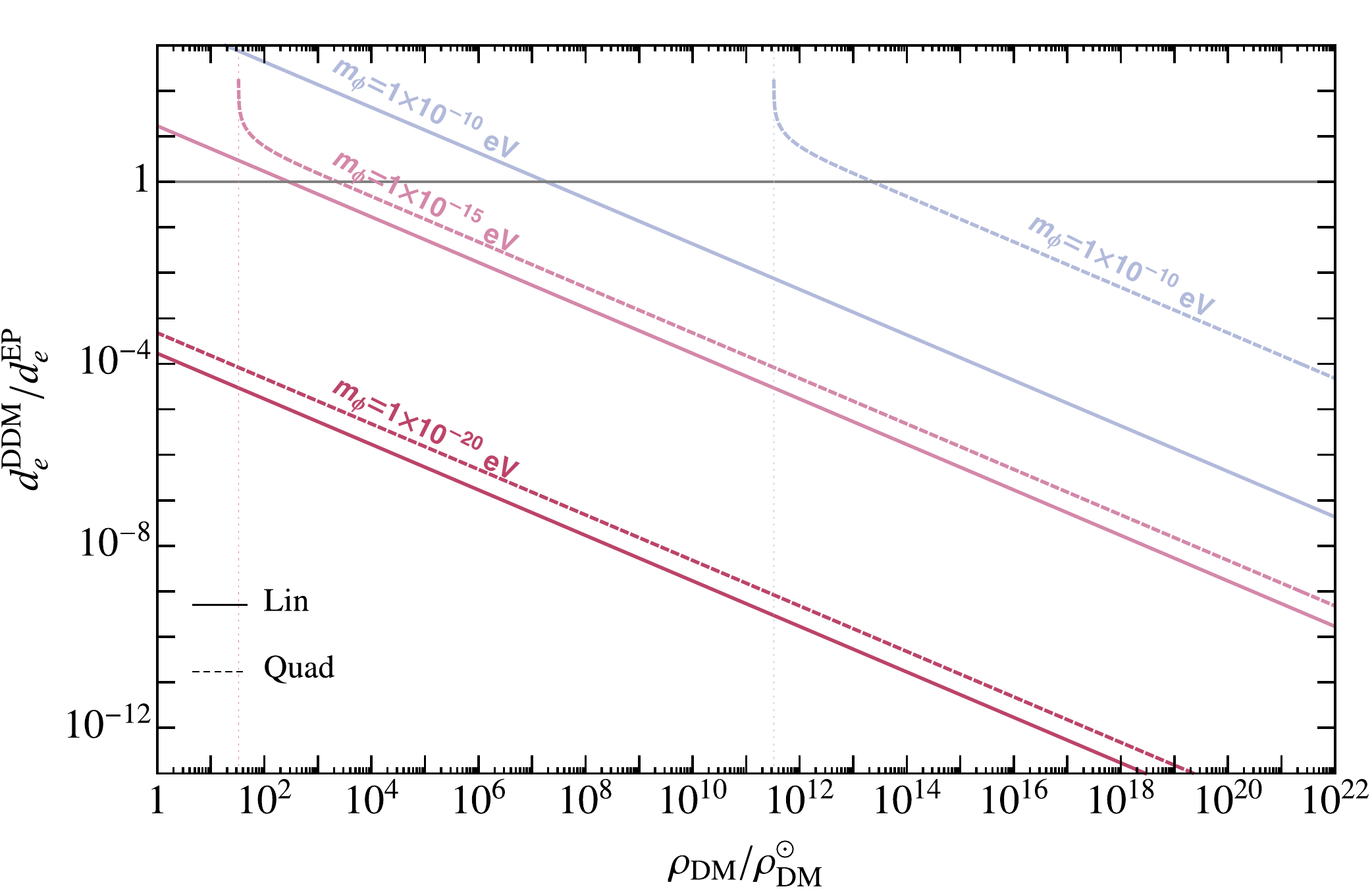}
	\caption{Ratio between \ac{DDM} bounds and \ac{EP} bounds on the \ac{DM}-\ac{SM} electron coupling $d_{m_e}$ (top) and photon coupling $d_{e}$ (bottom) as a function of the on-Earth \ac{DM} density enhancement, relative to the \ac{DM} density at the solar position $\rho_{\text{DM}}^{\odot}$. \textit{Solid} -- ratio for the linear coupling $d^{(1)}$, \textit{dashed} -- ratio for the quadratic coupling $d^{(2)}$. The vertical lines mark the minimal density enhancement required to probe sub-critical quadratic couplings by \ac{DDM} tests. The \ac{DDM} sensitivity is taken to be $\eta_{\text{DDM}} = 10^{-18}$ for all \ac{DM} masses $m_\phi$. The \ac{EP} bound is taken as from existing \ac{EP}-tests results.}
	\label{fig:ratio_dme}
\end{figure}

\begin{figure}
	\centering
	\includegraphics[width=0.8\linewidth]{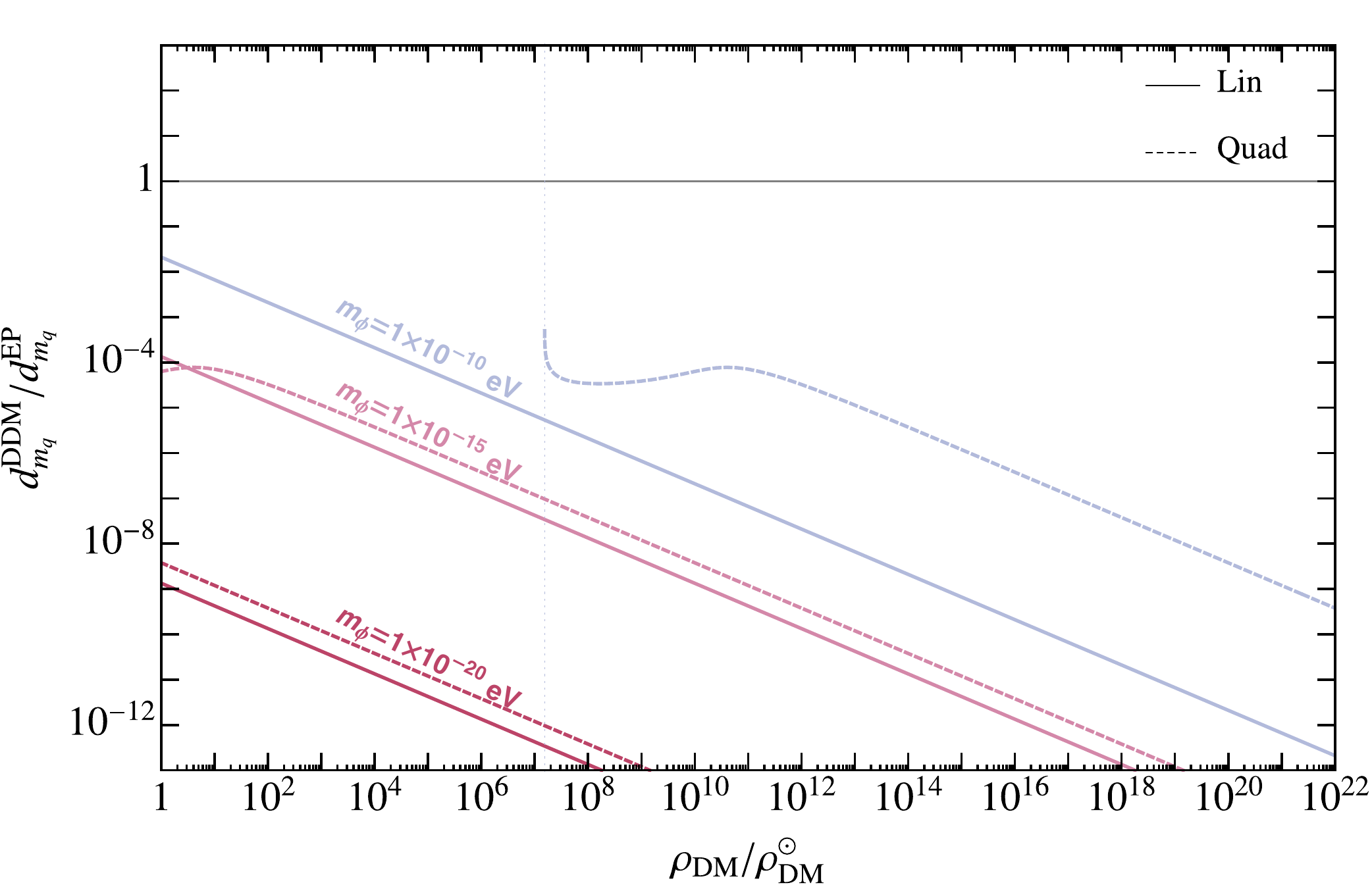}
	\includegraphics[width=0.8\linewidth]{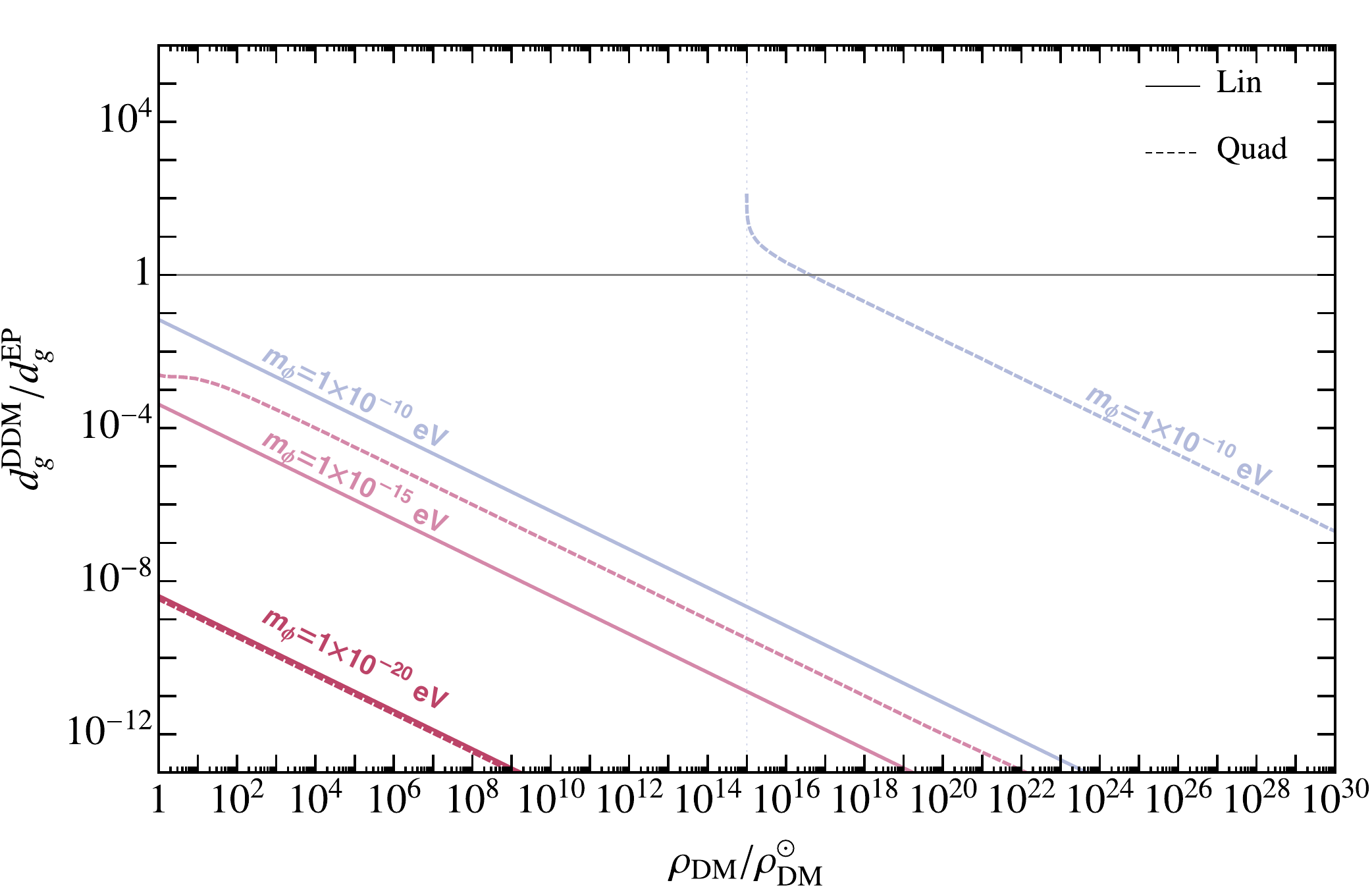}
	\caption{Ratio between \ac{DDM} bounds and \ac{EP} bounds on the \ac{DM}-\ac{SM} quark coupling $d_{m_q}$ (top) and the gluon coupling $d_{g}$ (bottom) as a function of the on-Earth \ac{DM} density enhancement, relative to the \ac{DM} density at the solar position $\rho_{\text{DM}}^{\odot}$. \textit{Solid} -- ratio for the linear coupling $d^{(1)}$, \textit{dashed} -- ratio for the quadratic coupling $d^{(2)}$. The vertical lines mark the minimal density enhancement required to probe sub-critical quadratic couplings by \ac{DDM} tests. The \ac{DDM} sensitivity is taken to be $\eta_{\text{DDM}} = 10^{-24}$ for all \ac{DM} masses $m_\phi$. The \ac{EP} bound is taken as from existing \ac{EP}-tests results.}
	\label{fig:ratio_dmq}
\end{figure}

The current \ac{EP} bounds for the quadratic theory and the \ac{DDM} bounds for both the linear and the quadratic couplings strongly depend on the local \ac{DM} density. These bounds become more stringent if the on-Earth \ac{DM} density is enhanced compared to the \ac{DM} density at the solar position $\rho_{\text{DM}}^{\odot}$~\cite{Salucci:2010qr}, as would be the case if a compact boson star consisting of $\phi$ is formed in the early universe, and is gravitationally bounded to the Sun or the Earth~\cite{Banerjee:2019epw,Banerjee:2019xuy,haloformation}. Importantly, note that \ac{DDM} searches are more sensitive to the local \ac{DM} density than \ac{EP} tests, and thus the ratio of their corresponding bounds $d_i^{\text{DDM}}/d_i^{\text{EP}}$ would vary with the density. The ratios $d_i^{\text{DDM}}/d_i^{\text{EP}}$, are presented as a function of the \ac{DM} on-Earth density enhancement $\rho_{\text{DM}}/\rho_{\text{DM}}^{\odot}$ for a few different benchmark \ac{DM} masses in Figs.~\ref{fig:ratio_dme}-\ref{fig:ratio_dmq}. Although a density enhancement factor much larger than $10^5$ is currently not motivated by theoretical or experimental considerations~\cite{Pitjev_2013,Tsai:2022jnv,haloformation}, higher densities are included for completeness. For $\left(d_{m_e}\right)^{\text{DDM}} / \left(d_{m_e}\right)^{\text{EP}}$ and $\left(d_{e}\right)^{\text{DDM}} / \left(d_{e}\right)^{\text{EP}}$ in Fig~\ref{fig:ratio_dme}, the atomic/molecular clock sensitivity is taken to be $\eta_{\text{DDM}} = 10^{-18}$ for all values of $m_\phi$. For $\left(d_{m_q}\right)^{\text{DDM}} / \left(d_{m_q}\right)^{\text{EP}}$ and $\left(d_{g}\right)^{\text{DDM}} / \left(d_{g}\right)^{\text{EP}}$ in Fig~\ref{fig:ratio_dmq}, the nuclear clock sensitivity is taken to be $\eta_{\text{DDM}} = 10^{-24}$ for all values of $m_\phi$. The \ac{EP} sensitivity is taken from current experiments and depends on the mass of the \ac{DM}.

As expected, an enhanced \ac{DM} density would make the ratio between the \ac{DDM} bounds and the \ac{EP} bounds smaller. In particular, for the electron coupling and for the photon coupling, the hierarchy between the two searches may be flipped for masses greater than $\sim 10^{-15}\eV$. 
In addition, for the quadratic interactions, the \ac{DM} density enhancement could also effectively shift the onset of the critical behavior to higher masses, making \ac{DDM} searches sensitive to the quadratic couplings at these masses, as opposed to the $\rho_{\text{DM}}=\rho_{\text{DM}}^{\odot}$ case. Therefore, when considering the possibility of a larger \ac{DM} density, the \ac{DDM} and \ac{EP} searches may have competing sensitivities, making them complimentary.

\subsubsection{Non-generic couplings}
\label{direction}
Let us discuss the bounds from the \ac{EP} test experiments, which are generically stronger than the constraints arising from the \ac{DDM} searches for individual coupling in the region $10^{-18}\,{\rm eV}\lesssim m_\phi\lesssim \cO(\eV)$~\cite{Kennedy:2020bac,Hees:2018fpg}.
As discussed, the \ac{EP} tests compare the dilatonic charges of two test bodies. 
To calculate the ``dilatonic charge" of an atom {\bf a} with $Z$($N$) being the number of protons (neutrons), one can write the mass of an atom $m^{\bf a}$ as,  
$ m^{\bf a}(Z,N) = m^{\bf a}_{\rm nuc}(Z,N) + Z m_e\,,$
where, $m^{\bf a}_{\rm nuc}$ is the mass of the nucleus of {\bf a}. 
Furthermore, the nucleus mass contribution can be decomposed in terms of the proton ($m_p$) and the neutron ($m_n$) masses, and the binding energy of the strong ($E_3$) and electromagnetic ($E_1$) interaction as,
$m^{\bf a}_{\rm nuc}(Z,N) = Z m_p + N m_n + E_3 + E_1\,.$
Note that $E_1$ is dominated by the \ac{EM} force within the nucleus, and thus we will ignore the electrons' effect on it~\cite{Damour:2010rp}.  
 For a generic atom {\bf a}, the dilatonic charges, $\vec Q^{\bf a}$, can be written as~\cite{Damour:2010rp},
\begin{equation*}\label{Qatom}
\!\! \vec{Q}^{\mathbf a} \!\approx \! F^{\mathbf a}\!\!\left(\! 3\!\times\! 10^{-4} -  
4\, r_I +8\, r_Z\,,\,   3\!\times\!10^{-4} - 3\,r_I\,,\,0.9\,, 0.09 - \frac{0.04}{A^{1/3}}- 2 \times 10^6 r_I^2 - r_Z ,0.002\, r_I\!\right)\! .
\end{equation*}
In what follows we use the following notation for a vector $\vec X \equiv X_{e,m_e,g,\hat m,\delta m}$\,, with $\hat{m}\equiv (m_d+ m_u)/2$ , $\delta m \equiv (m_d- m_u),$ 
$10^4\,r_{I;Z}\equiv 1-2Z/A;Z(Z-1)/A^{4/3}\,,$ and $F^{\mathbf a} =931\, A^{\mathbf a}/(m^{\mathbf a}/{\rm MeV})$ with $A^{\mathbf a}$ being the atomic number of the atom {\bf a}. 
The MICROSCOPE experiment~\cite{Touboul:2017grn,PhysRevLett.129.121102,Berge:2017ovy}, which provides the strongest \ac{EP} bounds for masses below $10^{-12}\,$eV, is sensitive to the difference between the dilatonic charges of Platinum/Rhodium alloy (90/10) and Titanium/Aluminum/Vanadium (90/6/4) which is given by 
\begin{equation*}
	(\stackrel{\longrightarrow}{\Delta Q \ })^{\rm Mic}\simeq 10^{-3} (-1.94\,,\, 0.03\,,\, 0.8\,,\, -2.61\,,\, -0.19)\,.
\end{equation*}
$(\stackrel{\longrightarrow}{\Delta Q\ \!})$ for other experiments looking for \ac{EP} violation are discussed in~\cite{Oswald:2021vtc}. 
We find that these sensitivities map to directions in the five-dimensional \ac{ULDM} coupling space that are very different from that of \ac{DDM} searches, denoted by $(\stackrel{\longrightarrow}{\Delta \kappa \ })^{\rm DDM}$, which usually have $\mathcal{O}(1)$ sensitives for variations in $\alpha$ and $m_e$ ({\it e.g.}~\cite{Safronova:2019lex, Antypas:2019yvv,Antypas:2020rtg} and refs. therein). 
Examining the four best \ac{EP} bounds, Cu-Pb~\cite{PhysRevD.61.022001}, Be-Ti~\cite{Schlamminger:2007ht}, and Be-Al~\cite{Wagner:2012ui} along with the previously mentioned MICROSCOPE experiment, in the five-dimensional vector space of coupling, we can construct a combination that would be orthogonal to all of them, approximately given by,
\bea
\hat{Q}^\perp\sim\big(0.01\,, \,0.99\,, \,0.01\,,\, -0.01\,, \,0.13\,\big)\,.\eea
It implies that models of light scalar \ac{DM} with coupling direction defined according to $\hat{Q}^\perp\cdot \vec d$ would not be subject to these four leading \ac{EP} bounds.

\begin{figure}
	\centering
	\includegraphics[width=0.9\linewidth]{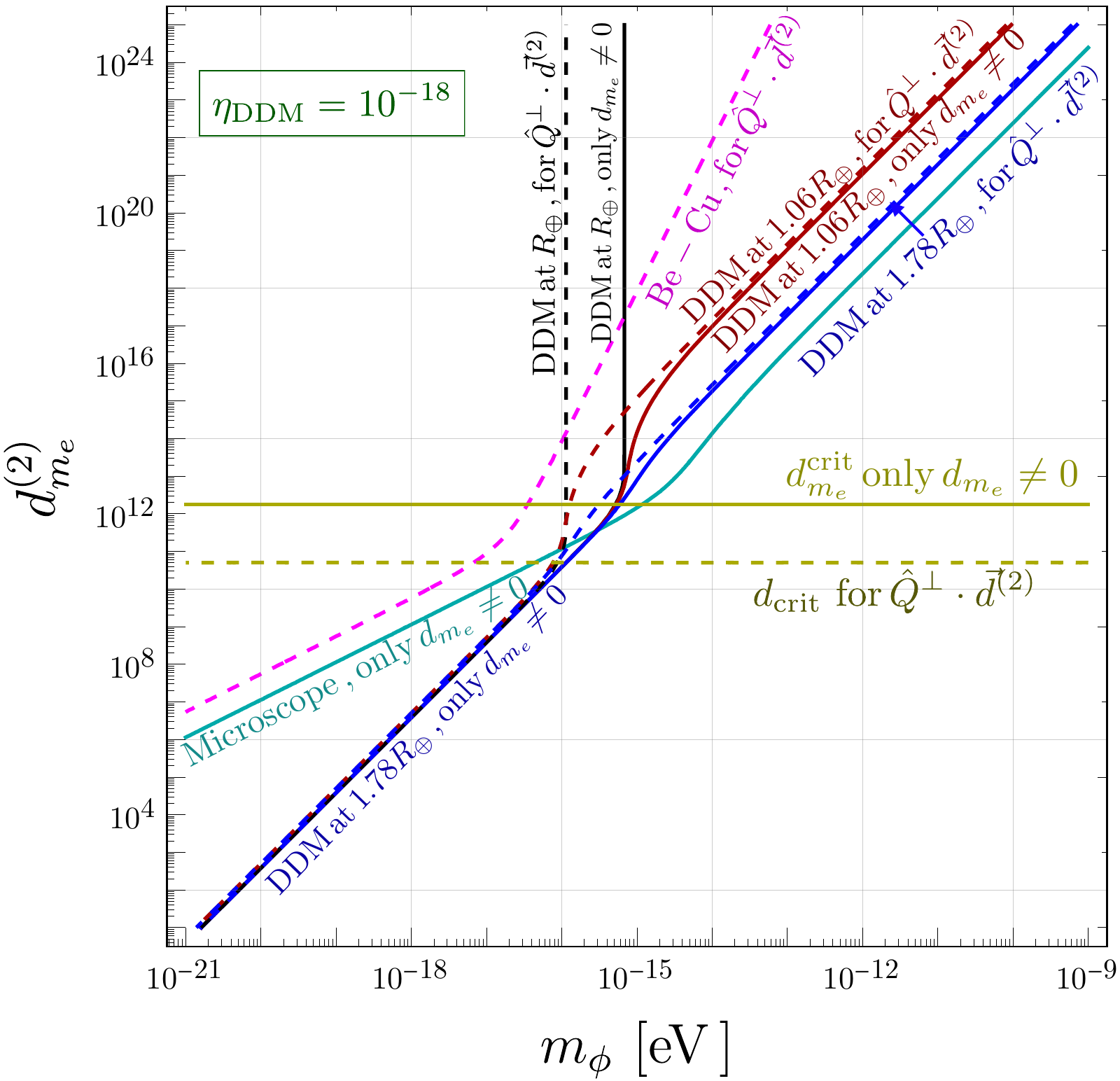}
	\caption{Exclusion plot for $d^{(2)}_{m_e}$; the solid lines assume a model where only $d^{(2)}_{m_e}\neq 0$. 
		The dashed lines depict the bounds for a model defined by a vector of sensitivities, $\hat Q^\perp\cdot\,\vec{d}^{(2)}$, that is orthogonal to the sensitivities of four leading \ac{EP} test experiments projected onto $d^{(2)}_{m_e}$. 
		The black lines depict the bounds from \ac{DDM} searches located on the surface of the Earth, whereas the red and the blue lines represent bounds from \ac{DDM} searches located at 400 km and 5000 km above the surface of the Earth, respectively. 
		We have assumed the sensitivity of the \ac{DDM} searches is $\eta_{\rm DDM}=10^{-18}$. 
		The magenta line depicts the strongest bound from the \ac{EP} tests (which is coming from MICROSCOPE experiment~\cite{PhysRevLett.129.121102}) for only $d^{(2)}_{m_e}\neq 0$ models. 
		The pink line represents the strongest \ac{EP} bound for a model defined by a vector of sensitivities $\hat Q^\perp\cdot\,\vec{d}^{(2)}$(which is coming from Be-Cu test~\cite{Su:1994gu}).
		The yellow lines represent the critical value of the couplings.}
	\label{fig:dme_direction_phi2}
\end{figure}

Here for simplicity, we consider an \ac{ULDM} quadratically coupled to the \ac{SM} and assume two scenarios:
\begin{enumerate}
	\item A model where only $d^{(2)}_{m_e}\neq 0$,
	\item A model defined by a vector of sensitivities, $\hat Q^\perp\cdot\,\vec{d}^{(2)}$, that is orthogonal to the sensitivities of the four leading \ac{EP} test experiments. 
\end{enumerate}	
We present the bounds on these two models in Figure~\ref{fig:dme_direction_phi2} by the solid and the dashed lines, respectively. 
For the second model, we have projected the bounds on $d^{(2)}_{m_e}$ as $\hat{Q}^\perp$ has a relatively significant overlap with the direction corresponding to $m_e$ (the second entry of it, which is the largest).  
Also for simplicity we have considered a  \ac{DDM} search experiment, which only depends on $m_e$ and thus the sensitivity vector can be written as $(\stackrel{\longrightarrow}{\Delta \kappa \ })^{\rm DDM} = (\,0,1,0,0,0)\,$. 
Note that, due to the large overlap of $\hat{Q}^\perp$ with the $m_e$ direction, the specific choice of $(\stackrel{\longrightarrow}{\Delta \kappa \ })^{\rm DDM}$ has a negligible effect on the final conclusion. 
In our case, $\hat{Q}^\perp\cdot(\stackrel{\longrightarrow}{\Delta \kappa\ })^{\rm DDM}\simeq 0.99$, which is approximately the sensitivity coefficients corresponding to $m_e$. 
In addition, the sensitivity of both the \ac{EP} tests and the \ac{DDM} searches depend on the geometry of the source body as discussed in Eqs.~\eqref{eq:etq_EP_quadratic_Stadnik_background} and \eqref{eq:etq_DDM_quadratic_Stadnik_background} respectively. 
We also assume a homogeneous spherically symmetric Earth as the source body, which is made of $32\%$ Iron and $68\%$ silicon oxide. 
With the above assumption, we get the dilatonic charge of the source body as
\bea
Q^{\rm source} \simeq 10^{-3} \,(\,1.87\,,\, 0.27\,,\, 1000.19\,, \,80.51\,,\, 0.04\,)\,. 
\eea   

As mentioned below Eq.~\eqref{eq:scdcrit}, the critical value of a coupling is inversely proportional to the corresponding dilatonic charge of the source. 
In Figure~\ref{fig:dme_direction_phi2}, we see that $d_{m_e}^{\rm crit}$ (shown by the solid yellow line) is $35$ times larger than the critical value of the second model which is defined by a vector of sensitivities, $\hat Q^\perp\cdot\,\vec{d}^{(2)}$ (shown by the dashed yellow line) as $Q^{\rm source}\cdot \hat Q^\perp=9.49\times 10^{-3}$, whereas $Q^{\rm source}_{m_e}=0.27\times 10^{-3}$.  Unlike the first model, where only $d_{m_e}^{(2)}\neq 0$, the second model is not constrained by the four leading \ac{EP} experiments. 
In the $\hat Q^\perp$ direction, the strongest \ac{EP} bound is coming from the Be-Cu test~\cite{Su:1994gu} (the fifth best one), which is more than five orders of magnitude weaker than the MICROSCOPE~\cite{PhysRevLett.129.121102} experiment, which provides the most stringent bound for any model with only one non-zero coupling~\cite{Hees:2018fpg}. 
The turquoise and the pink lines in Figure~\ref{fig:dme_direction_phi2} depict the strongest bounds from the \ac{EP} tests on the first and the second models, respectively. 
The black lines depict the bounds from \ac{DDM} searches located on the surface of the Earth, whereas the red and the blue lines represent bounds from \ac{DDM} searches located 400 km (average altitude  of the International Space Station) and 5000 km above the surface of the Earth (see  \cite{FOCOS} clock proposal), respectively. See Section~\ref{subsec:space_DDM} for more details. We have assumed the sensitivity of the \ac{DDM} searches to be $\eta_{\rm DDM}=10^{-18}$. 
Note that, below criticality, even the terrestrial \ac{DDM} searches provide stronger bounds than the \ac{EP} tests for both the scenarios discussed here. 
These bounds are $\mathcal{O}(10^5)$ stronger than that of the \ac{EP} test in the lower mass region. 
As explained before, for a given sensitivity, the reach of the space-based \ac{DDM} searches is better than the Earth-based ones due to the screening effect of theories with quadratic couplings. 
Above $m_\phi\gtrsim 2 \times 10^{-16}\eV$, the bound from the MICROSCOPE experiment is slightly stronger than that of the space-based \ac{DDM} searches for the first scenario, where only $d_{m_e}^{(2)}\neq 0$. 
However, for the second scenario, due to the considerable overlap of $\hat Q^\perp$ with the $m_e$ direction,
the reach of the \ac{DDM} searches is not reduced, unlike the \ac{EP} tests. 
This allows the space-based \ac{DDM} searches to provide the strongest bound even for the higher masses and above the critical value of the coupling. 
Around $m_\phi\sim 10^{-13}\eV$, the bounds from \ac{DDM} searches are $\mathcal{O}(10^9)$ stronger than the best \ac{EP} bound (coming from the Be-Cu test shown by the dashed magenta line in Figure~\ref{fig:dme_direction_phi2}).  

Let us consider a case where the five-dimensional coupling is universal, thus does not violate \ac{EP}. As discussed in~\cite{Oswald:2021vtc}, if a scalar-field coupling to the \ac{SM} is defined according to $\vec Q_{\rm dil}\cdot\vec{d}^{(i)}$ where $i=1,2$ with $
\vec Q_{\rm dil} \simeq (-0.01,1,1,1,0)\,,   
$ then it will not be subjected to the \ac{EP} tests bounds. However, it will still give rise to deviations from the inverse square law and, thus, will be constrained by fifth-force search experiments. 
To simplify our discussion, we will consider the case of a linearly coupled scalar to the \ac{SM}; however, our main result also applies to a quadratically coupled theory.  

To briefly see how the \ac{EP} non-violation works, we know that gravity couples to the Ricci scalar $R$, and using Einstein's equation, one can write $R\propto T^{\mu}_{\mu}$
where $T^{\mu}_{\mu}$ is the trace of the energy-momentum tensor. 
Thus, if a scalar-field coupling to the \ac{SM} is proportional to $T^\mu_\mu$, it will not generate any \ac{EP} violation. 
This is an idealistic limit and is realized only in pure dilaton models, where the dilaton $(\phi)$ couplings are precisely given by
\begin{equation}
\mathcal{L}\supset \frac{\phi}{f_{\rm dil}}T^{\mu}_{\mu}\,,
\label{eq:L_dilaton}
\end{equation} 
where $f_{\rm dil}$ is the conformal invariance breaking scale~\cite{Dymarsky:2013pqa}. 
As discussed before and in~\cite{Damour:2010rp,Hees:2018fpg}, above interaction would induce a Yukawa interaction between two bodies. The interaction strength can be written as 
\begin{equation}
\alpha = \frac{1}{\sqrt{4\pi G_{\rm N}}}\frac{\partial \ln m^{\rm}(\phi)}{\partial \phi}\propto \frac{1}{f_{\rm dil}}\,.
\label{eq:dilaton_yukawa}
\end{equation}
This shows that the dilaton coupling is universal, and the conformal invariance breaking scale determines the coupling strength. 
The differential acceleration between two test bodies is proportional to the difference between their Yukawa interaction strength, as shown in Eq.~\eqref{eq:Sensitivity_EP_linear}, and it vanishes due to the universality of the dilaton coupling. 
Thus, a pure dilaton does not generate \ac{EP} violation. 
However, it is easy to see using Eq.~\eqref{eq:dilaton_yukawa} that in the presence of a source with mass $m^{\rm s}$, the acceleration of a test body can be written as
\begin{equation}
	\vec{a}_{\rm A} = - \hat{r}\frac{G_{\rm N}\, m^{\rm S}}{r^2}\left[1+ \frac{2\Mpl^2}{f_{\rm dil}^2}\left(1+m_{\rm dil}\,  r\right) e^{- m_{\rm dil}\, r}\right],
\end{equation}
where $\Mpl^2 = 1/(8\pi G_N)$ and $m_{\rm dil}$ is the mass of the dilaton. 
The above equation shows that the presence of a dilaton causes deviation from $1/r^2$ force and thus can be
constrained by various experiments that test for deviations from Newtonian gravity (fifth-force searches) (see~\cite{Fischbach:1996eq} and Refs. therein). 
We are also assuming that the dilaton would acquire a small mass from a sector other than the \ac{SM}~\cite{Coradeschi:2013gda} in order to be a viable \ac{DM} candidate~\cite{Arvanitaki:2014faa}. 

\begin{figure}[h!]
	\begin{subfigure}
	\centering
	\includegraphics[width=0.48\linewidth]{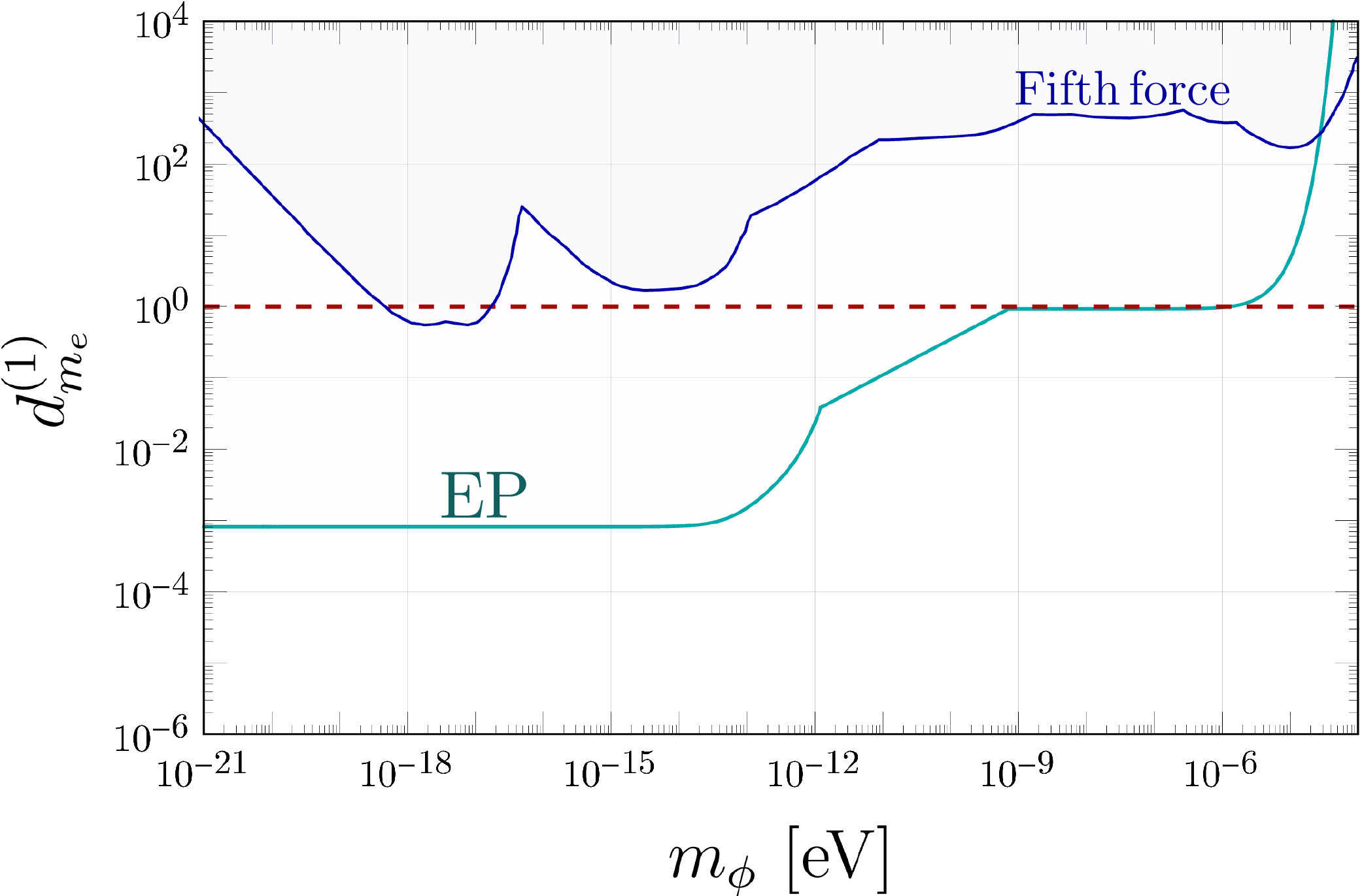}
	\includegraphics[width=0.48\linewidth]{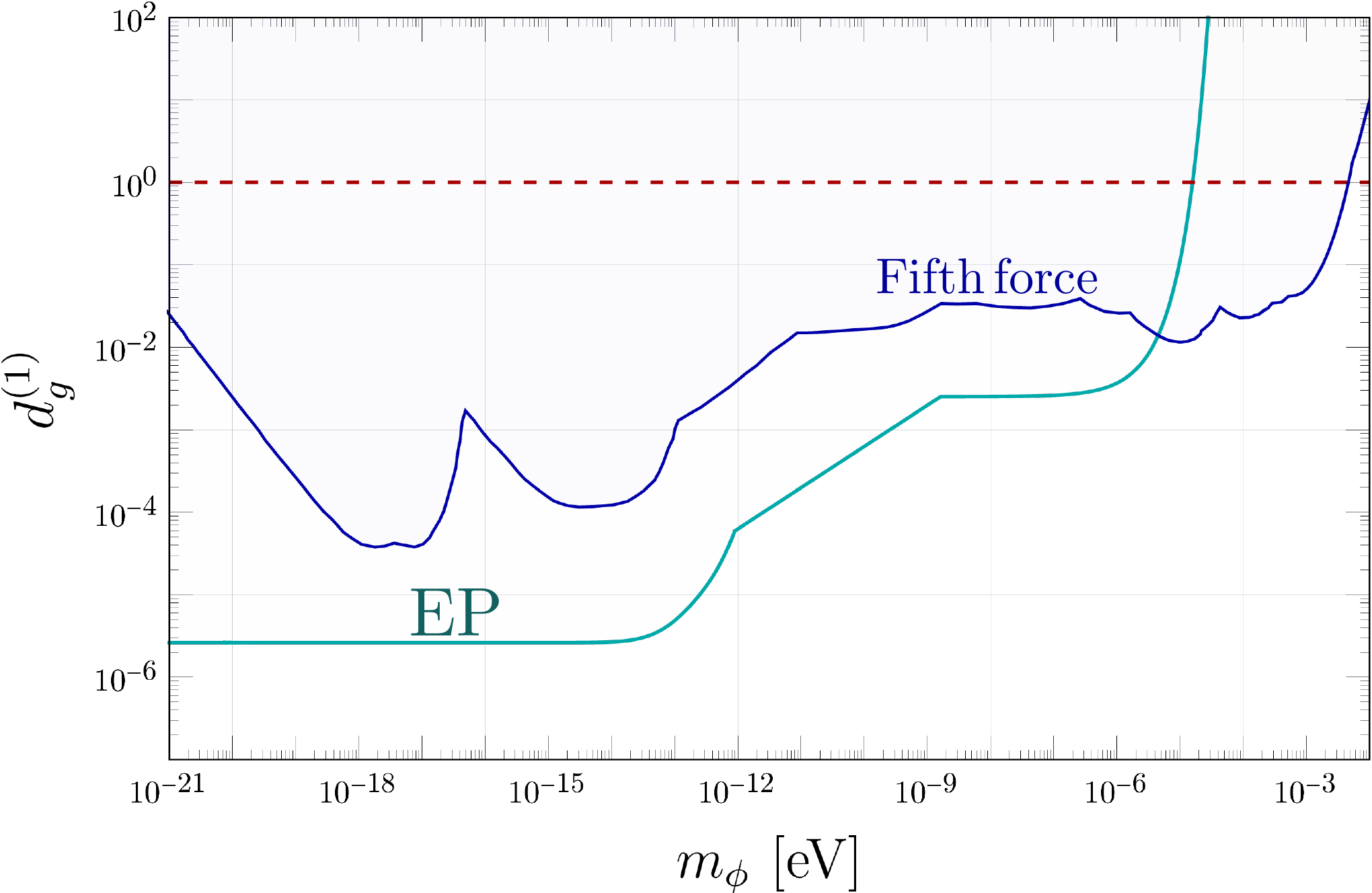}
	\end{subfigure}
	\caption{Bounds from various experiments which are looking for \ac{EP}-violation and/or deviation from Newtonian gravity (fifth force searches) on $d^{(1)}_{m_e}$ (left) and $d^{(1)}_g$ (right). 
		The turquoise lines show the strongest constraints from various \ac{EP} violation searches~\cite{PhysRevLett.129.121102,Wagner:2012ui,Schlamminger:2007ht,PhysRevD.61.022001} assuming a model where only $d^{(1)}_{i}\neq 0$ where $i=m_e,g$. 
		The blue lines depict the strongest bound coming from various fifth force experiments (see~\cite{Fischbach:1996eq,Lee:2020zjt,Tan:2020vpf} and Refs. therein). 
		The red dashed lines indicate $d^{(1)}_{i}=1$.}
	\label{fig:dil_direction}
\end{figure}

So far, we have argued a scalar that interacts with the \ac{SM} as given in Eq.~\eqref{eq:L_dilaton} would not violate \ac{EP} but would give rise to deviations from Newtonian gravity. Now to get the direction in the five-dimensional coupling space, we need to write the expression for $T^\mu_\mu$. 
Assuming that the \ac{SM} is valid up to the scale $f_{\rm dil}$, $T^{\mu}_{\mu}$ can be written as 
\begin{equation}
T^{\mu}_{\mu}=\frac{\beta(g)}{2 g}G^2+ \frac{\beta(g_2)}{2 g_2}W^2+\frac{\beta(y)}{2y}B^2 +(1-\gamma)\sum_{\psi}m_\psi\bar \psi \psi\,,
\end{equation}
where $g$, $g_2$ are the coupling strength of $SU(3)$ and $SU(2)$ gauge groups, respectively, $y$ represents the hypercharge corresponding to $U(1)_y$, and $G$, $W$ and $B$ are the corresponding gauge fields respectively. Also, $\psi$ denotes the \ac{SM} fermions with mass $m_\psi$.  
For simplicity, in the above formula, we assume that the conformal breaking scale $f_{\rm dil}$ is far below the Landau pole of $U(1)_y$ and above the \ac{EW} scale. 
As $T^\mu_\mu$ is invariant under the evolution of the \ac{RG} equation (manifested in the above equation), the dilaton always couples through anomaly matching to the same quantity at any scale $\mu$. 
For our purpose, we consider our theory at $\mu=1\GeV$ and $T^\mu_\mu$ can be expressed as, 
\begin{equation}
	T^{\mu}_{\mu}=\frac{\beta(g_s)}{2 g_s}G^2+\frac{\beta(e)}{2e}F^2 +\sum_{\psi}m_\psi\bar \psi \psi\,.
\end{equation}
In the above equation, we have redefined the fermion masses in terms of their pole masses. 
Combining this with Eq.~\eqref{eq:L_dilaton} and along with our convention of defining a vector in the five-dimensional coupling space, we can write the dilaton coupling vector, $\vec Q_{\rm dil}$ as 
\begin{equation}
	\vec Q_{\rm dil} = (-\frac{2\beta(e)}{e},1,1,1,0)\,,
\end{equation} 
where $e$ is the electric charge and $\beta(e) = e^3/(12\pi^2)$. 
As below $m_e\sim \MeV$, the theory essentially becomes free, we find $2 \beta(e)/e\sim e^2/(6\pi^2) \times 10\sim 0.015$ as $\log(\rm GeV/\rm MeV)\sim 10$. 
Thus, we get $ \vec Q_{\rm dil} = (-0.01,1,1,1,0)\,.$
Thus, we have argued that a scalar field (the dilaton), whose coupling is defined according to $\vec Q_{\rm dil}\cdot\vec d^{(1)}$, will not generate an \ac{EP}-violating acceleration, as the direction is indistinguishable from that of gravity. 
In Fig.~\ref{fig:dil_direction} we show the bounds from various fifth-force searches projected on the $d^{(1)}_{m_e}$ and $d^{(1)}_g$ directions on a pure dilaton model by the blue solid line. 
For comparison, we also show the bounds from various \ac{EP} tests on a model which only couples to either $m_e$ or gluon field strength linearly i.e. $\mathcal{L}\supset d^{(1)}_{m_e} \phi\, m_e\bar e e/\Mpl$ or $\mathcal{L}\supset d^{(1)}_g \beta(g)\,\phi \, G^2/(2g\,\Mpl)$ by the turquoise line. 
Various constraints on these models are shown in details in Fig.~\ref{fig:dme_AP_EP} and Fig.~\ref{fig:dg_AP_EP} respectively.

We, finally, note that if the dilaton couplings are not perfectly aligned with that obtained from the trace of the energy-momentum tensor, it will generate \ac{EP}-violating acceleration as discussed in~\cite{Taylor:1988nw,Kaplan:2000hh}.  

\section{Quadratic Interactions and Screening}\label{sec:Quadratic_Interactions_and_Screening}

In the previous section, we observed that the bounds on the quadratic couplings are weaker than the bounds on the linear couplings. One reason is the cutoff-suppression, attenuating the effects of the quadratic coupling by a factor of $\phi/\Lambda_{\rm UV}$ compared to those of the linear coupling, where $\Lambda_{\rm UV}$ is the UV scale that characterizes the \ac{EFT}. This happens as a quadratic couplings represents a higher-dimensional effective operator than the linear one. 

The other reason is the fact that the quadratic coupling might be screened at the surface of a central body as discussed in Section~\ref{sec:criticallity}, as well as in Appendix~\ref{App:The_solution of the Linear model and a Quadratic}, and previously in~\cite{Hees:2018fpg}. 
Below we discuss two important scenarios that alter the effects of the screening behavior. The first is a theoretical one - a model in which both linear and quadratic couplings are present simultaneously, and the second is an experimental one - positioning \ac{DDM} experiments in space. 

\subsection{Screening and criticality in a model with both linear and quadratic couplings}~\label{subsec:lin_and_quad_model}

Let us discuss the sensitivity of \ac{EP} tests and \ac{DDM} searches in the presence of both linear and quadratic couplings between the \ac{ULDM} and the \ac{SM}. While the interplay of linear and quadratic interactions has been previously discussed in the context of non-\ac{DM} models~\cite{Olive:2007aj}, the \ac{DM} boundary condition plays a crucial role in setting the field's profile, and thus leads to qualitatively and quantitatively different results. As we will show here, a theory with linear and quadratic couplings has more severely constrained linear and quadratic couplings than a theory with only one of the couplings turned on. Also, in the presence of both the couplings, there is a small region of the parameter space where the \ac{DDM} bounds are stronger than those of the \ac{EP} tests. 

The Lagrangian of that model can be written as 
\begin{align}
	\L=\L_\text{lin-int}+\L_\text{ quad-int}\,,
\end{align}
where $\L_\text{lin-int}$ and $\L_\text{ quad-int}$
are defined in Eq.~\eqref{eq:Stadnik_Lint_linear} and Eq.~\eqref{eq:Stadnik_Lint_quad} respectively. 
As we are interested in calculating the sensitivities of \ac{DDM} searches and \ac{EP} tests, we would like to solve for the profile of $\phi$. 
As discussed above, we assume a homogeneous boundary condition at large distances for $\phi$ i.e.
\begin{equation}
\label{eq:bcs_static}
	\phi(r\to\infty,t)=\phi_0 \cos(m_\phi t+\delta)\,,
\end{equation}
with $\phi_0=\sqrt{2\rho_{\rm DM}}/m_\phi$. 
The \ac{EOM} of this combined model is
\begin{equation}
	\left(\frac{\partial^{2}}{\partial t^{2}}-\nabla\cdot\nabla+\tilde{m}_{\phi}^{2}\left(r\right) \right)\phi= J_{\text{source}}\left(r\right)\,,
	\label{eq:EOM_lin_quad}
\end{equation} 
where we define  
\begin{align}
	\tilde{m}_{\phi}^{2}\left(r\right) \equiv m_{\phi}^{2}+\frac{Q_i^C d_i^{(2)}}{M_{\text{Pl}}^{2}}\rho_{C}\left(r\right) \,\,{\rm and}\,\,\,\, J_{\text{source}}\left(r\right)\equiv -\frac{Q^C_id_i^{\left(1\right)}}{M_{\text{Pl}}}\rho_{C}\left(r\right).
\end{align} 
Notice that the linear coupling, $d_i^{(1)}$, provides a source term for the \ac{EOM} whereas the quadratic coupling $d_i^{(2)}$ modifies the mass term of $\phi$. 
As discussed in~\cite{Hees:2018fpg}, in the presence of a source body $C$, the \ac{SM} fields can be replaced by the density of the source body $\rho_{C}\left(r\right)$ with corresponding dilatonic charge $Q_i^C$, where $i$ runs over the \ac{SM} species coupled to the \ac{ULDM}. 
If we model the source body $C$ as a uniform density sphere of radius $R_C$ and mass $M_C$, then one can write 
\begin{equation}
	\rho_{C}\left(r\right)=\begin{cases}
		\frac{3M_{C}}{4\pi R_{C}^{3}} & r\leq R_{C}\\
		0 & r>R_{C}\,.
	\end{cases}
\end{equation}
See the discussion around Eq.~\eqref{eq:EOM_quadratic} for more details. 
The solution of the \ac{EOM} given in Eq.~\eqref{eq:EOM_lin_quad} with the boundary condition of Eq.~\eqref{eq:bcs_static} can be written as
\begin{align}
	\phi\left(r,t\right)= & \begin{cases}
		\phi_{0} \,{\rm cs}\, \frac{\text{sinhc}\left(\frac{r}{R_C}  \sqrt{d_i^{(2)}/d_i^{\rm crit}}\right)}{\cosh\left[\sqrt{d_i^{(2)}/d_i^{\rm crit} }\right]}	- M_{C}\frac{r^{2}}{R_{C}^{3}}\frac{Q^C_jd_j^{\left(1\right)}}{M_{\text{Pl}}}I\left( \tilde{m}_{\phi}r\right)\frac{e^{-\tilde{m}_{\phi}r}}{4\pi} & r<R_{C}\\
		\phi_{0}\,{\rm cs}\,\left(1-s_{C}[d_{i}^{(2)}]\frac{GM_{C}}{r}\right) - M_{C}\frac{Q^C_jd_j^{\left(1\right)}}{M_{\text{Pl}}}I\left( m_{\phi}R_c\right)\frac{e^{-m_{\phi}r}}{4\pi r} & r\geq R_{C}\,,
	\end{cases}
\end{align}
where we define ${\rm cs}= \cos\left(m_{\phi}t+\delta\right)$ and the functions $I(x)$ and $s_C[d_i^{(2)}]$ are given by
\begin{align}
	I(x) = 3 \, \dfrac{x \cosh x - \sinh x}{x^3} \,,\,\,
	s^{(2)}_{C}[d^{(2)}_i] = Q^C_i d^{(2)}_i J_{+}\leri{\sqrt{d^{(2)}_i/d_i^{\rm crit} }}\,. 
\end{align}
In addition, $d_i^{\rm crit} = R_C/(3 Q^C_i GM_C)$ as defined before, and the function $J_+(x)$ is defined as $J_+(x) = 3(x-\tanh x)/x^3$~\cite{Hees:2018fpg}. 
$G$ and $\Mpl$ can be used interchangeably with $4\pi G=1/\Mpls$. 
Using the solution of the \ac{EOM} for $r\geq R_C$, and assuming the signal will be time averaged, we get the sensitivity of the \ac{EP} tests as		
\bea 
\!\!\!\!\!\!\!\!\!\!\!\!\!\!\!\!\!\!\!\!\!\!\!\!\!\!\!\!\!\!\!\!\!\!\!\!
\eta^\text{EP}(r) &=& 2 \frac{|\vec a_{A,C}-\vec a_{B,C}|}{|\vec a_{A,C}+\vec a_{B,C}|} \simeq \frac{\leri{\Delta Q}_i^{AB}}{G M_C/r^2}\leri{\frac{d^{(1)}_i}{M_{\text{Pl}}}+\frac{d^{(2)}_i}{M^2_{\text{Pl}}}\phi} |\vec\nabla\phi|\nonumber\\
	&\simeq& \, \leri{\Delta Q}_i^{AB}\Bigg(s_{C}^{\left(2\right)}[d_{j}^{(2)}] d^{(2)}_i \frac{ \phi_0^2}{2M^2_{\text{Pl}}}\left[1-s_{C}^{\left(2\right)}[d_{j}^{(2)}]\frac{GM_{C}}{r}\right] \nonumber\\
	&& + Q^C_jd_j^{(1)} d^{(1)}_i\, I(m_\phi R_C)\, (1+m_\phi r)\, e^{-m_\phi r}  \nonumber\\ 
	&& - \left( Q^C_jd_j^{(1)}\right)^2  d^{(2)}_i \frac{G\, M_C }{r} I(m_\phi R_C)^2\, (1+m_\phi r)\, e^{-2 m_\phi r}\Bigg)\,, 
	\label{eq:Sensitivity_EP_lin_quad}
\eea
and the sensitivity of \ac{DDM} searches as
\begin{align} \label{eq:Sensitivity_AP_lin_quad}
	\eta^\text{DDM}_\omega(r) = & \; \frac{|\delta Y(t) |}{Y}\approx 
	\frac{\Delta \kappa_id_{i}^{(2)}}{2 M_{\text{Pl}}^2}\, \times\phi_0^2 \left(1-s_{C}[d_{j}^{(2)}]\frac{GM_{C}}{r}\right)^2 \mathcal{F}_\omega\leri{\css} \nonumber
	\\ - \;& \nonumber
	\frac{\Delta \kappa_i d_{i}^{(2)}}{M_{\text{Pl}}^2} \frac{M_{C} Q^C_j d_j^{\left(1\right)}}{M_{\text{Pl}}}I\left( m_{\phi}R_c\right)\frac{e^{-m_{\phi}r}}{4\pi r}\times \phi_0\, \left[1-s_{C}[d_{k}^{(2)}]\frac{GM_{C}}{r}\right]\mathcal{F}_\omega\leri{\rm cs}
	\\ + \;&
	\frac{\Delta \kappa_i d_{i}^{(1)}}{M_{\text{Pl}}}\,\times \phi_0\, \left(1-s^{(2)}_{C}[d_{j}^{(2)}]\frac{GM_{C}}{r}\right) \mathcal{F}_\omega\leri{{\rm cs}}\,.
\end{align} 

We want to describe the screening effect in this model. 
As most of the \ac{DDM} searches are terrestrial, in this section we consider them to be performed very close to the surface of the source body, i.e., at $r\simeq R_C$. In Section~\ref{subsec:space_DDM} we discuss the space based \ac{DDM} searches where $r\gtrsim R_C$. 

As discussed before, in a model where the \ac{DM} interacts only quadratically with the \ac{SM}, if the coupling is larger than the critical value, the \ac{DDM} sensitivity is screened, and the dependence on the quadratic couplings is suppressed. 
However, in a model with both linear and quadratic couplings, due to the mixed $d^{(2)} d^{(1)}$ term, there is no such criticality for the \ac{DDM} searches. 

To see how it works, let us start with the case when $d^{(2)}_{i} \gg d_i^{\rm crit}$.  
In this limit $s^{(2)}_{C}[d^{(2)}_i]$ can be written as Eq.~\eqref{eq:scdcrit} and we get,
\bea
1-s^{(2)}_{C}[d_{i}^{(2)}]\frac{GM_{C}}{r}= \sqrt{\frac{d_i^{\rm crit}}{d^{(2)}_i}} +\mathcal{O}\left(\frac{\Delta r}{R_C}\right)\,, 
\eea
and Eq.~\eqref{eq:Sensitivity_AP_lin_quad} becomes
\begin{align} \label{eq:Sensitivity_AP_lin_quad_criticallity}
	\left. \eta^\text{DDM}(\omega)  \right|_{\left( d^{(2)}_{i} \gg d_i^{\rm crit} , \:  r\simeq R_c\right) }&  \approx 
	\frac{\Delta \kappa_i}{M_{\text{Pl}}^2}  \phi_0^2\,\, d_i^{\rm crit}\times\mathcal{F}_\omega\leri{{\rm cs}^2}\nonumber \\ 
	& - \frac{\Delta \kappa_i}{M_{\text{Pl}}}\phi_0\, \left[\frac{Q^C_j}{3 Q^C_i} d_{j}^{(1)}\sqrt{  \frac{d_i^{(2)}}{d_i^{\rm crit}} } \right] I\left( m_{\phi}R_c\right)e^{-m_{\phi}R_c}\times\mathcal{F}_\omega\leri{{\rm cs}}\nonumber \\ 
	& + \frac{\Delta \kappa_j}{M_{\text{Pl}}} \phi_0\, \left[ d_{j}^{(1)}\sqrt{  \frac{d_i^{\rm crit}}{d_{i}^{(2)}} } \right]  \times\mathcal{F}_\omega\leri{{\rm cs}} +\mathcal{O}\left(\frac{\Delta r}{R_C}\right)\,.
\end{align}
As expected, the above equation shows that if we have only quadratic couplings, i.e., $d_i^{(1)}=0$, then \ac{DDM} searches become insensitive to the quadratic couplings, as the sensitivity does not depend on $d^{(2)}_{i}$ in the supercritical limit. 
However when $d_i^{(1)}\neq 0$, it is sensitive to $\sqrt{d_i^{(2)}}$. 
We can further simplify the above expression for $m_\phi\ll 1/R_C$ as
\begin{align} \label{eq:Sensitivity_AP_lin_quad_criticallity_low_mass}
	\!\!
	\left. \eta^\text{DDM}\leri{\omega}  \right|_{\left( d^{(2)}_{i} \gg d_i^{\rm crit} , \:  r\simeq R_c\ll1/m_\phi\right) } \approx -\frac{\Delta \kappa_j}{3 M_{\text{Pl}}}\phi_0\,  d_{j}^{(1)}\sqrt{\frac{d_i^{(2)}}{d_i^{\rm crit}}}\times \mathcal{F}_\omega\leri{{\rm cs}}+ \mathcal{O}\left(\frac{\Delta r}{R_C},\frac{d_i^{\rm crit}}{d_i^{(2)}}\right)\,,
\end{align}
and for $m_\phi\gg 1/R_C$ as
\begin{align}
	\label{eq:Sensitivity_AP_lin_quad_criticallity_high_mass}
	\left. \eta^\text{DDM}  \right|_{\left( d^{(2)}_{i} \gg d_i^{\rm crit} , \:  r\simeq R_c\gg1/m_\phi\right)}\approx  \frac{\Delta \kappa_j}{M_{\text{Pl}}} \phi_0\, d_{j}^{(1)}\sqrt{\frac{d_{i}^{(2)}}{d_i^{\rm crit}}} \left(\frac{d_i^{\rm crit}}{d_i^{(2)}}- \frac{1}{2 m_\phi^2R_C^2}  \right) \times\mathcal{F}_\omega\leri{{\rm cs}} + \mathcal{O}\left(\frac{\Delta r}{R_C}\right)\,,
\end{align}
by noting the limiting case of the function $I(x)$
\bea
\lim_{x\to 0} I(x)\to 1\,\,{\rm and}\,\,\lim_{x\to \infty} I(x)e^{-x}\to \frac{1}{2 x^2}\,.
\eea
For completeness we also discuss the case of small quadratic coupling, $d^{(2)}_{i} \ll d_i^{\rm crit}$. 
In this case $s_{C}[d^{(2)}_i]\simeq Q^C_i d^{(2)}_i$ and  Eq.~\eqref{eq:Sensitivity_AP_lin_quad} becomes 
\begin{align} \label{eq:Sensitivity_AP_lin_quad_below_criticallity}
	&\left. \eta^\text{DDM}  \right|_{\left( d^{(2)}_{i} \ll d_i^{\rm crit} , \:  r\simeq R_c\right) } \simeq  
	\frac{\Delta \kappa_id_{i}^{(2)}}{2 M_{\text{Pl}}^2}\, \times\phi_0^2  \, \css + \frac{\Delta \kappa_i d_{i}^{(1)}}{M_{\text{Pl}}}\,\times \phi_0\,  {\rm cs}+ \mathcal{O}\left(\frac{d_{i}^{(2)}}{d_i^{\rm crit}}\right)\!,\!\!
\end{align}
for all masses.

\begin{figure}
\centering
	\includegraphics[scale=0.35]{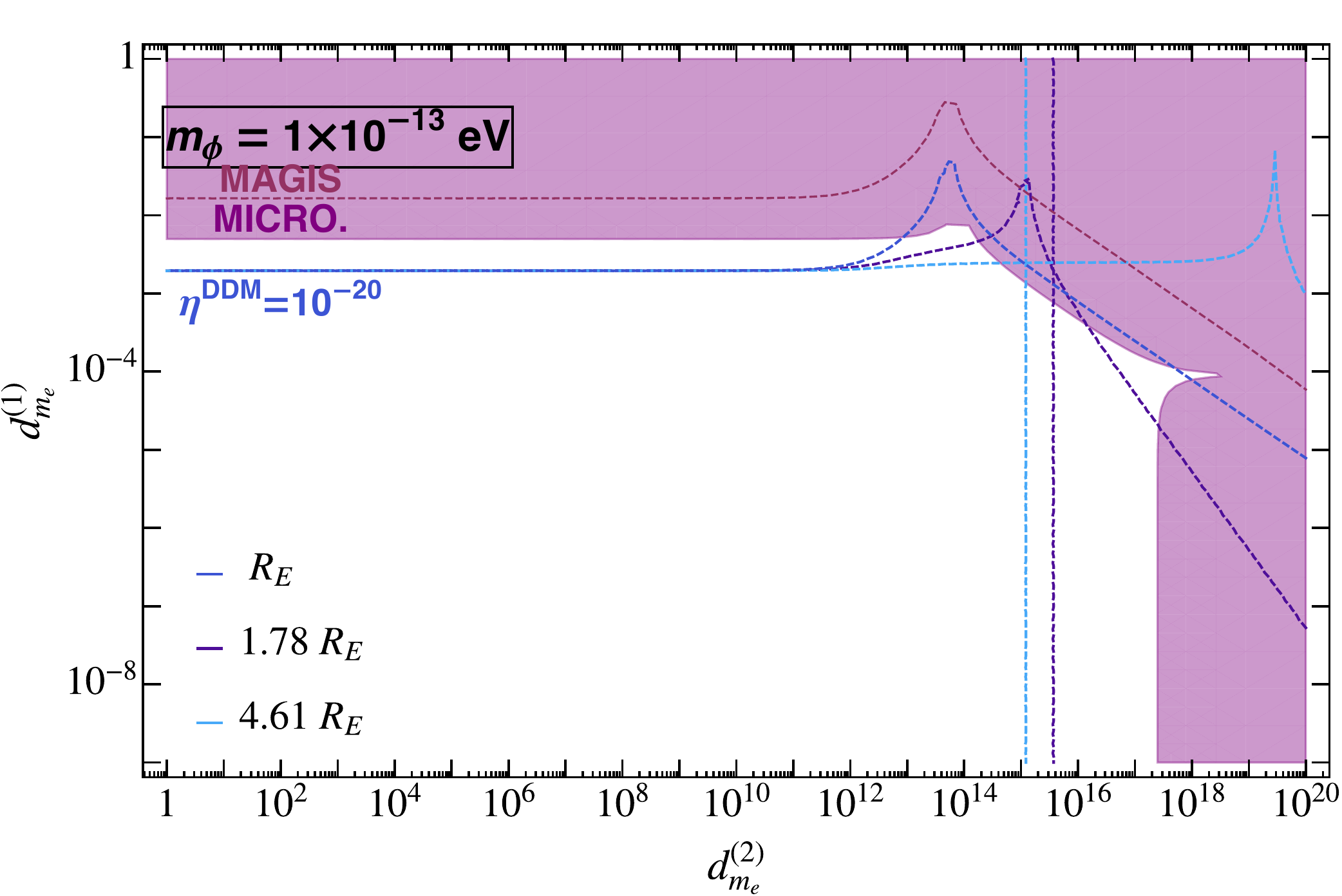}
	\vspace{0.5 cm}
	\includegraphics[scale=0.35]{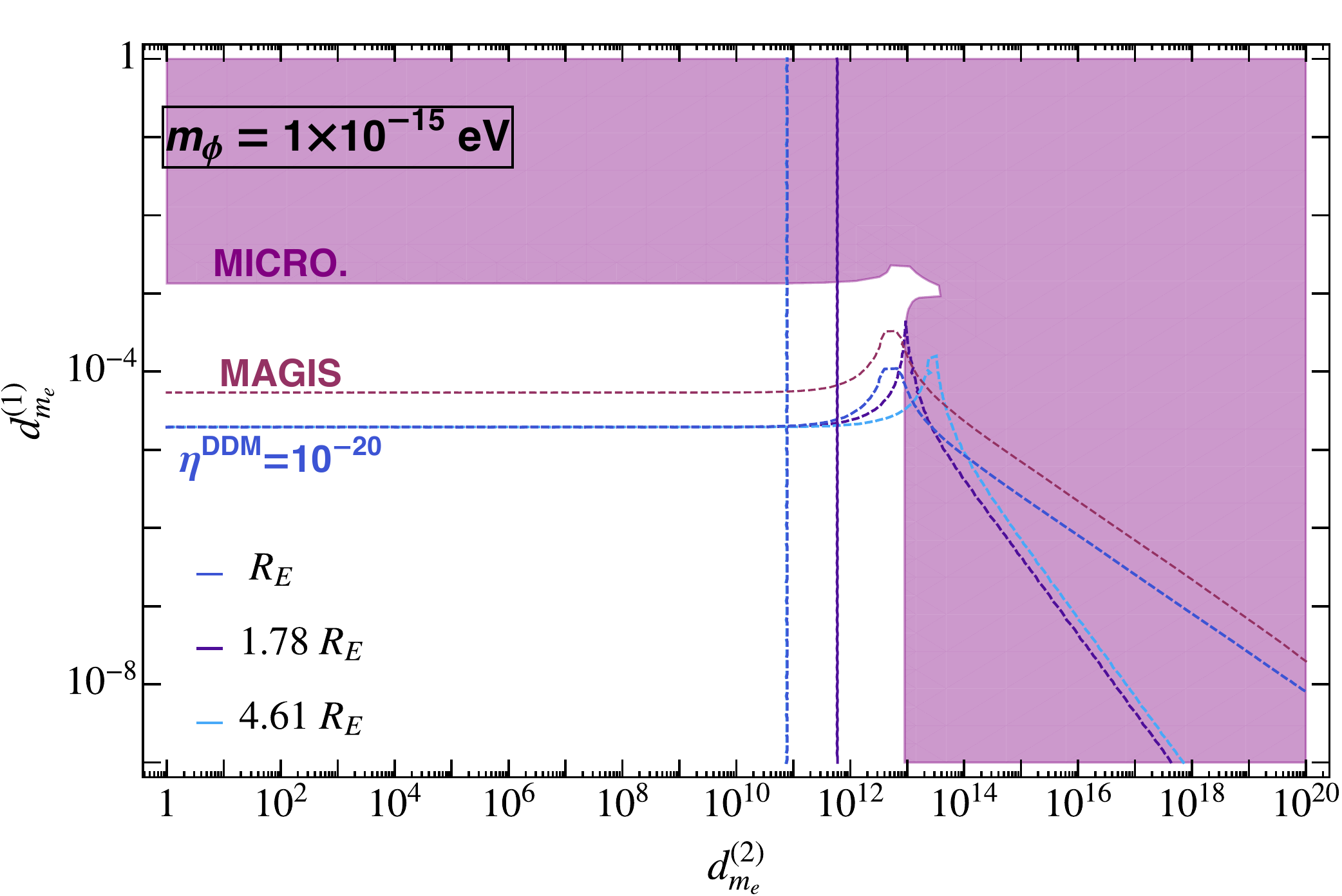}
	\vfill
	\caption{Allowed parameter space for the linear and quadratic electron coupling for masses of $m_{\phi} = 10^{-13}\,\eV\, \text{\textbf{(top)}}, \:10^{-15}\,\eV\, \text{\textbf{(bottom)}} $.}
	\label{fig:Allowed_parameter_space_lin+quad_dme_dalpha}
\end{figure}

\begin{figure}
	\centering
	\includegraphics[scale=0.35]{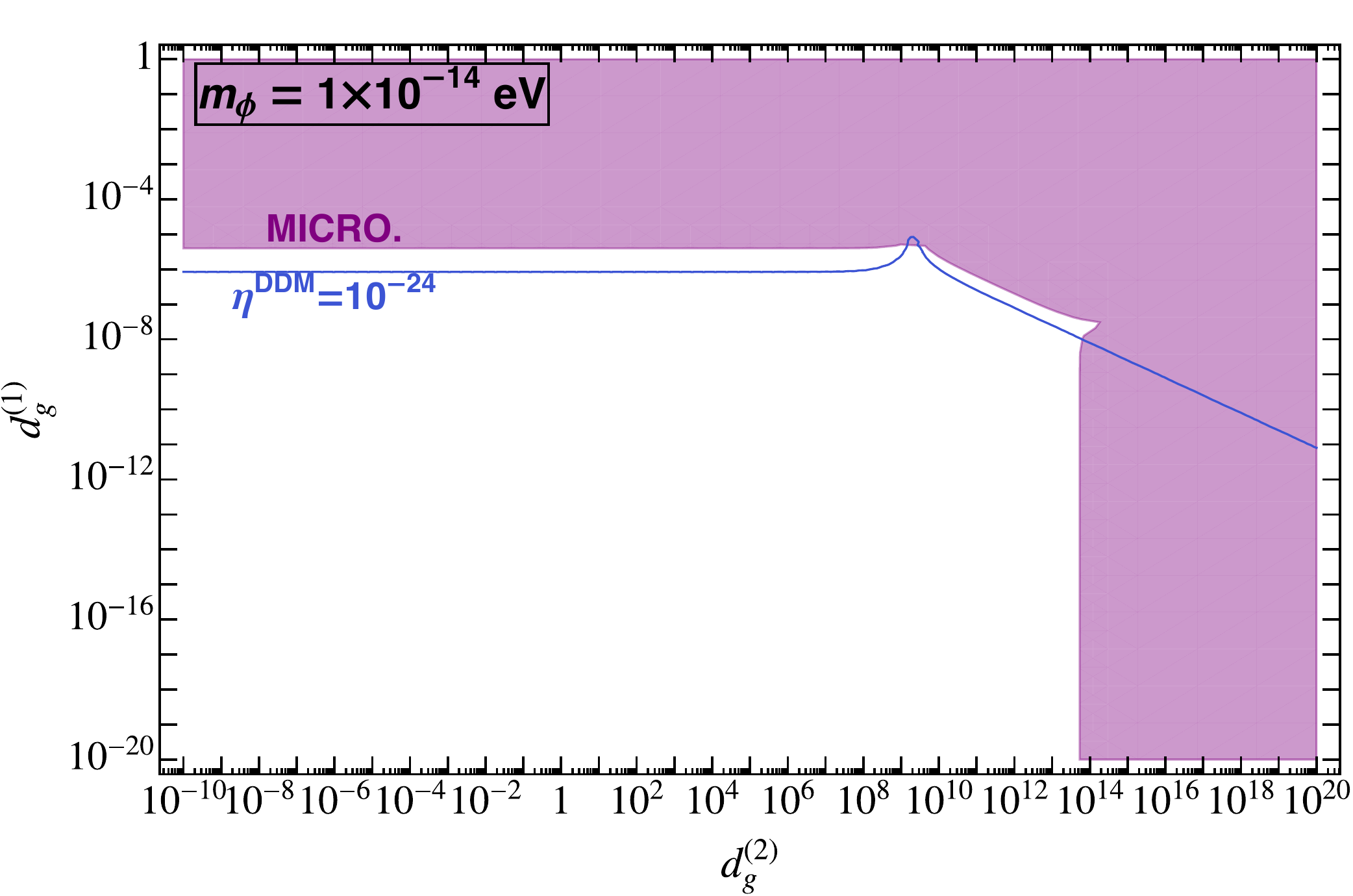}
	\vspace{0.5 cm}
	\includegraphics[scale=0.35]{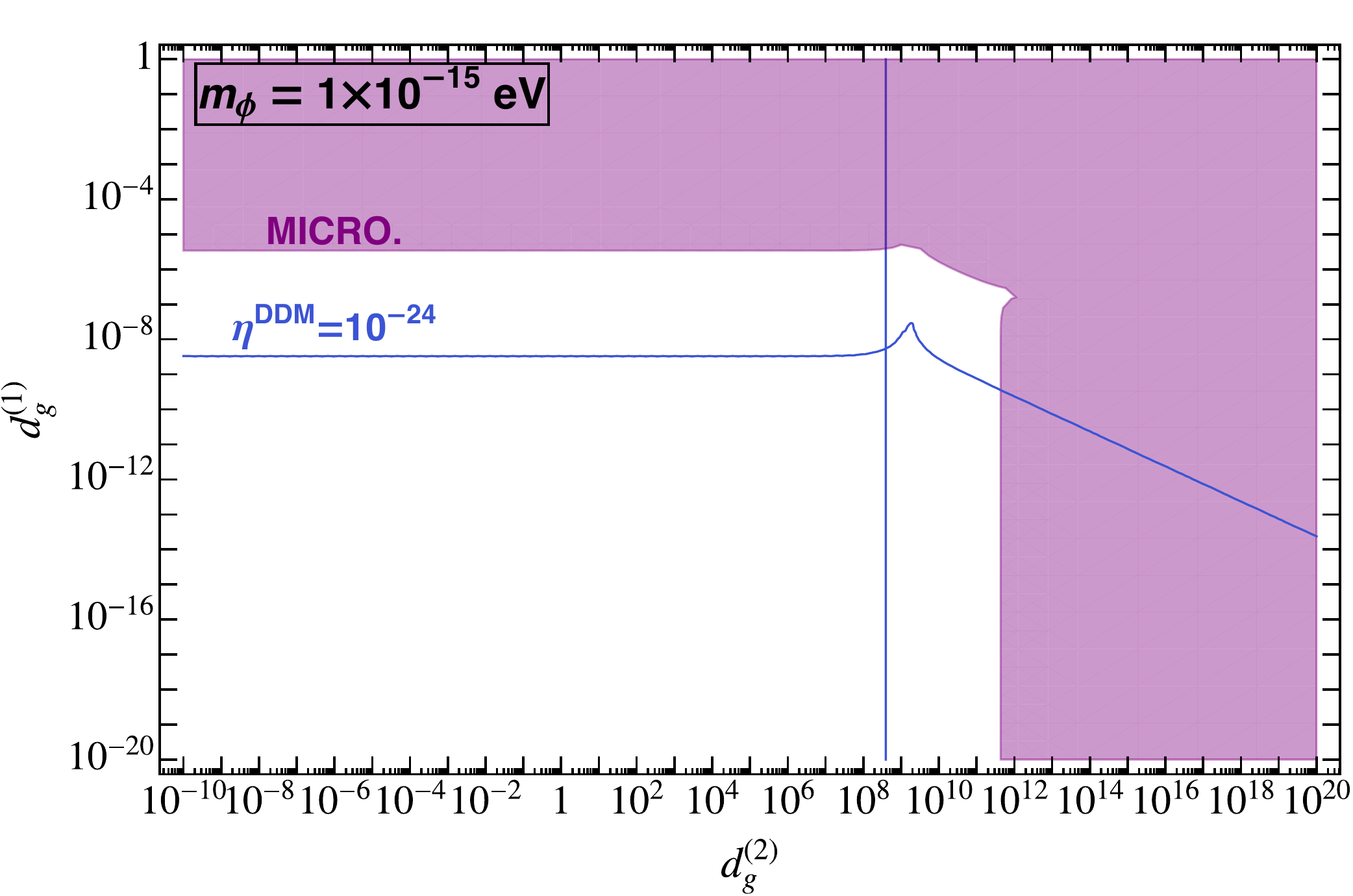}
	\vfill
	\caption{Allowed parameter space for the linear and quadratic gluon coupling for masses of $m_{\phi} = 10^{-14}\,\eV \,\text{\textbf{(top)}}, \:10^{-15}\,\eV\, \text{\textbf{(bottom)}}$.} 
	\label{fig:Allowed_parameter_space_lin+quad_dg_dmq}
\end{figure}

In Figures.~\ref{fig:Allowed_parameter_space_lin+quad_dme_dalpha}-\ref{fig:Allowed_parameter_space_lin+quad_dg_dmq}, we present the allowed parameter space of a model with both the linear and quadratic couplings for different masses of  the \ac{DM}. 
We notice that introducing a non-zero quadratic coupling changes the bounds on the linear coupling and vice-versa. 
This means that a theory that has both linear and quadratic couplings has a stronger constraint on each compared to a theory with only one of the couplings. 
Despite that, we see that in most of the parameter space of interest, the \ac{EP} bounds from the linear coupling are still the dominant ones. 
Due to the scalar field profile in the presence of both linear and quadratic couplings, there is a small region of the parameter space where the \ac{DDM} bounds are stronger than the \ac{EP} bounds, as can be seen in all the figures below.  
This could be of potential interest to \ac{DDM} searches whose accuracy has improved vastly in the last few years.

The above observation 
motivates us to present a few models where the linear couplings are suppressed compared to the quadratic ones. 
We require that these theories are natural in the sense that there is no fine-tuning in order to achieve such hierarchies between the small values of the linear coupling compared to the relatively large values of the quadratic coupling. We present two such models within the Clockwork framework in Section~\ref{subsec:Clockwork}, and another model of within the Relaxed Relaxion framework in Section~\ref{sec:Relaxed_Relaxion_hierarchis_lin_vs_quad}.


\subsection{Screening and criticality in space}\label{subsec:space_DDM}

\begin{figure}
	\centering
	\includegraphics[scale=0.4]{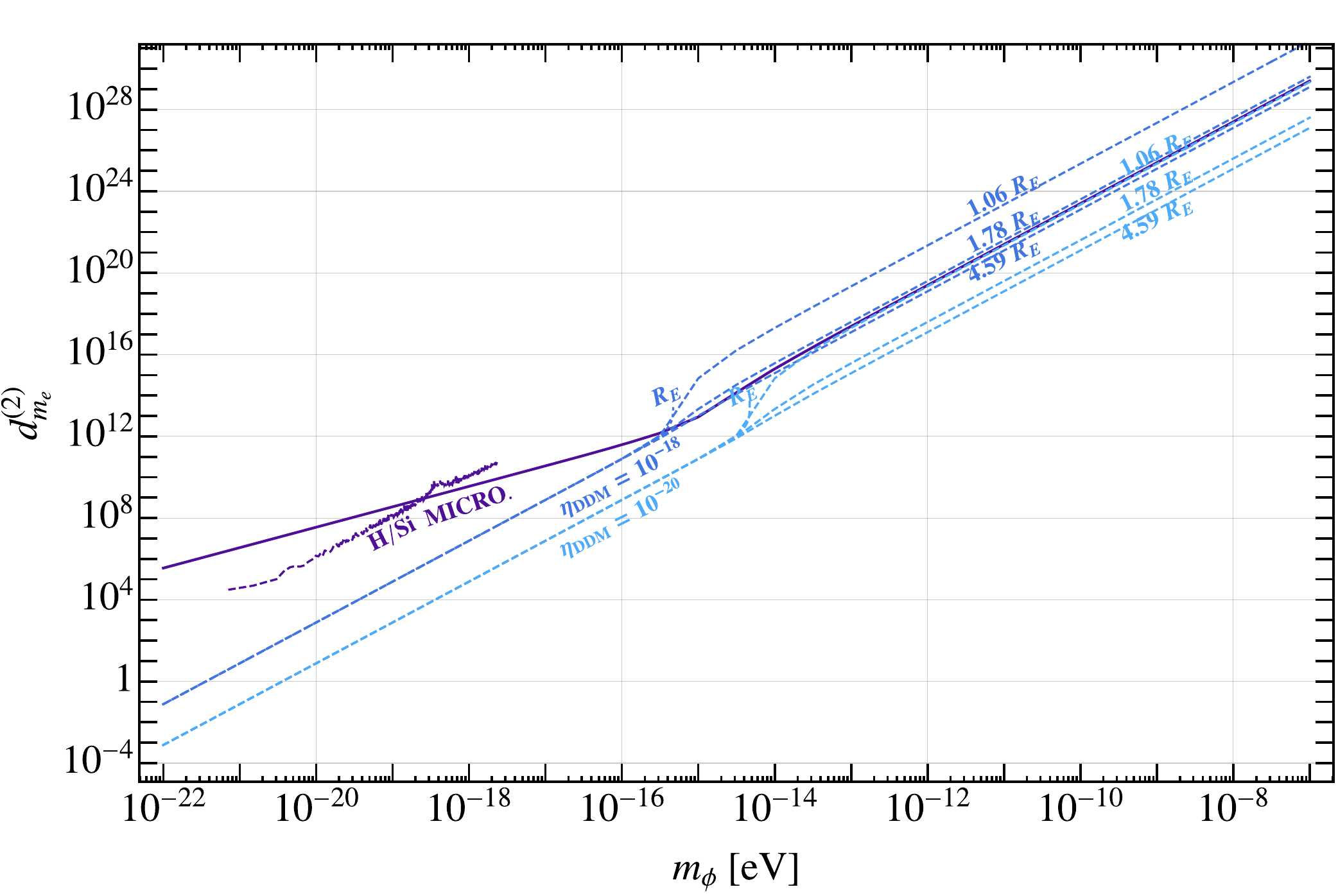}
	\includegraphics[scale=0.4]{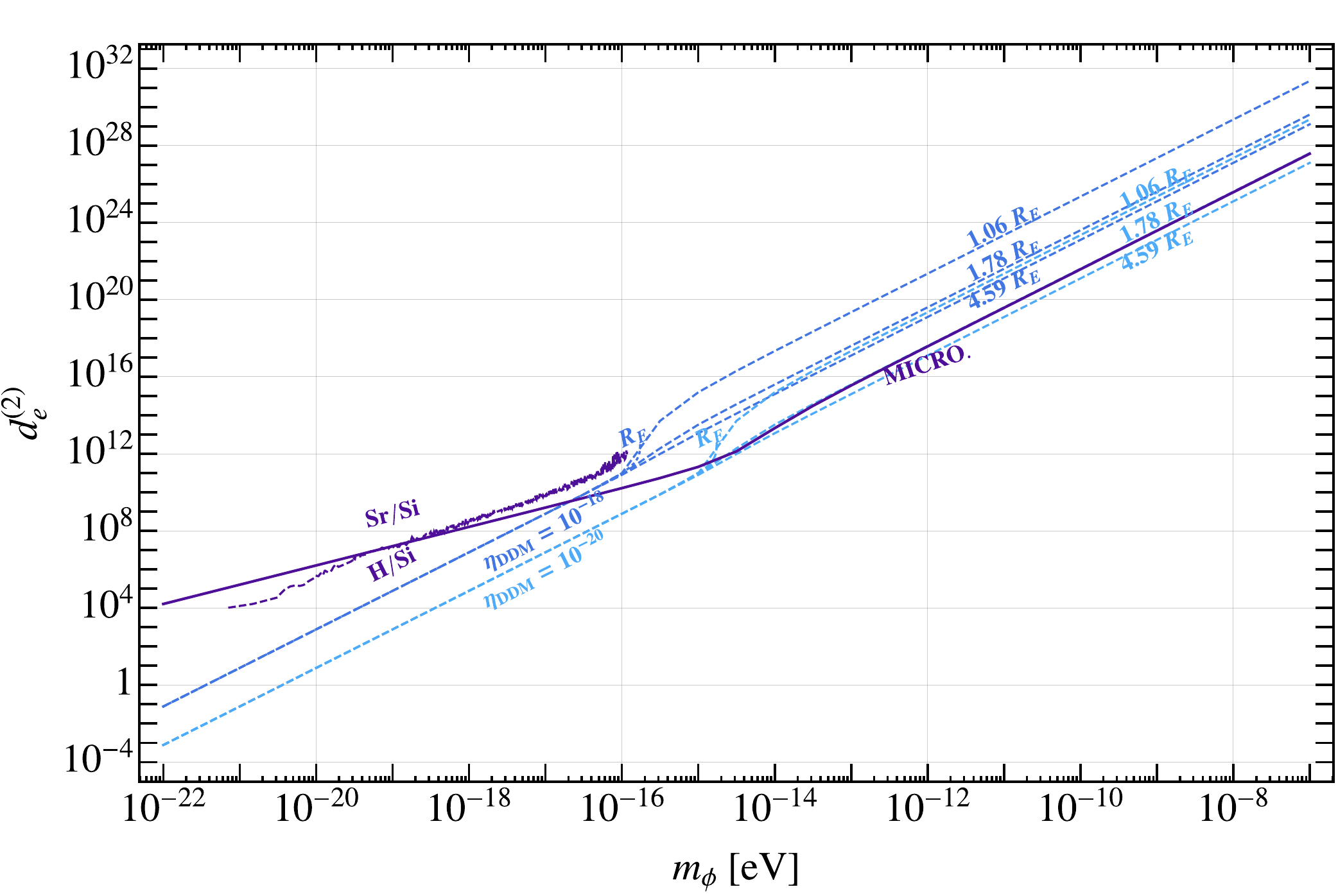}
	\caption{The bounds on the quadratic interactions with electrons (\textbf{top}) and photons (\textbf{bottom}) from a hypothetical \ac{DDM} experiment done on satellites orbiting at different radii.}
	\label{fig:sat_dme_de}
\end{figure}

Since the screening of the \ac{ULDM} is most dominant at the surface of the Earth, experiments done further away are less affected by it, as can be seen by Eq.~\eqref{eq:phi2sol_v1}. This is the key to the dominance of the MICROSCOPE \ac{EP} test, positioned at an altitude of roughly $700$~km, over the bounds on the quadratic coupling. If, however, \ac{DDM} searches are also performed in geocentric orbits, they too would become sensitive to the \ac{ULDM} quadratic couplings. To demonstrate this point, we show in Fig.~\ref{fig:sat_dme_de} the bounds on the electron coupling and photon coupling, $d^{(2)}_{m_e}$ and $d^{(2)}_{e}$ respectively, as expected for \ac{DDM} experiments with sensitivities of $10^{-18}$, and $10^{-20}$, located at 400 km, 5000 km, and 23000 km above the surface of the Earth.  Below we survey some of the recent proposals with the potential to launch highly sensitive \ac{DDM} experiments into space. 
 
The NASA Deep Space Atomic Clock (DSAC) mission has recently demonstrated a microwave trapped ion clock based on Hg$^+$ ions achieving a factor of $10$ improvement over previous space-based clocks \cite{Burt2021}. The such clock was proposed for the auto-navigation of spacecraft~\cite{2018Navigation}. Cold atom microwave clock was demonstrated in space in \cite{CAC2018}. The
ACES (Atomic Clock Ensemble in Space) mission \cite{ACES} is planned to perform an absolute measurement of the
red-shift effect between the microwave PHARAO clock on-board the International Space Station (ISS) and clocks
on Earth, to improve such limit by an order of magnitude.

The progress in the development of optical clocks has been extraordinary, with three orders of magnitude improvement in uncertainty over the last 15 years \cite{LudBoyYe15}. Several optical clocks have reached uncertainty at the 10$^{-18}$ level (see, e.g.,~\cite{PhysRevLett.123.033201}), with further improvements expected as there is no apparent technical limit.
Portable high-precision optical clocks were also demonstrated \cite{KelBurKal19}, which is a prerequisite for space deployment.
 Various clock-comparisons and clock-cavity comparison experiments are sensitive to $d_e^{(i)}$, $d_{m_e}^{(i)}$, and $d_{g}^{(i)}$. The applications of the different clock types of clocks and optical cavities for \ac{ULDM} searches were recently reviewed in \cite{2022UDM}.

Deployment of high-precision optical clocks in space will enable both practical and fundamental applications, including tests of general relativity \cite{FOCOS}, \ac{DM} searches \cite{Sun}, gravitational wave detection in new wavelength ranges \cite{Kolkowitz_2016,Fedderke:2021kuy}, relativistic geodesy \cite{Puetzfeld:2019kki}, linking Earth optical clocks \cite{Gozzard:2021wkf}, and others. The roadmap for cold atom technologies in space has been outlined in \cite{roadmap}.
 In the present work, we demonstrate another window of opportunity to study \ac{ULDM} with clocks in space, taking advantage of the space environments that are drastically different from that of the Earth. Being away from the Earth's surface allows us to test the quadratic models described above.
We used orbital parameters of proposal~\cite{FOCOS} as an example in Section~\ref{direction}. Ref.~\cite{FOCOS} describes a space mission concept that would place a state-of-the-art optical atomic clock in an eccentric orbit around Earth. The main mission goal is to test the gravitational red-shift, a classical test of general relativity, with sensitivity 30,000 times beyond current limits by comparing clocks on Earth and the spacecraft.
A high stability laser link between the Earthbound and space clocks is needed to connect the relative time, range, and velocity of the orbiting spacecraft to Earthbound stations. In general relativity, the tick rate of time slows in the presence of massive bodies, and locating one or more clocks in orbits around the Earth provides a low-noise environment for tests of gravity.
Sr high-precision optical atomic clock aboard an Earth-orbiting space station (OACESS) \cite{OACESS} was proposed for \ac{DM} searches, including static or quasi-static apparent variations of $m_e$ with changing height above
Earth's surface and transient changes in the apparent value of $\alpha$ due to the passage of a
relativistic scalar wave from an intense burst of extraterrestrial origin. The goal of this pathfinder mission is to compare
the space-based high-precision optical atomic clock (OAC) with one or more ultra-stable terrestrial OACs to search for
space-time-dependent signatures of dark scalar fields that manifest as anomalies in
the relative frequencies of station-based and ground-based clocks.
The OACESS will serve as a pathfinder for dedicated missions (e.g., FOCOS described above~\cite{FOCOS}) to establish high-precision OAC as space-time references in space.
We used the orbital parameters of the International Space Station ($\sim$ 400 km above the surface of the Earth) and FOCOS mission proposal~\cite{FOCOS} in 
Fig.~\ref{fig:dme_direction_phi2} and Fig.~\ref{fig:sat_dme_de}.

A version of the such proposal with a state-of-the-art cavity will enable test the quadratic models by also running a clock-cavity experiment as carried out in \cite{Kennedy:2020bac}, sensitive to $d_e^{(2)}$ coupling. We note that a time transfer link to Earth is not required for such an experiment. The clock-comparison experiment in space involving a molecular clock will also be sensitive to $d_{m_e}^{(2)}$. Molecular clocks are projected to reach $10^{-18}$ uncertainties \cite{HanKuzLun21}.

In Ref.~\cite{Sun}, a clock-comparison satellite mission with two clocks onboard to the inner reaches of the solar system was proposed to search for a \ac{DM} halo bound to the Sun and to look for the spatial variation of the \aclp{FC} associated with a change in the gravitation potential. Various clock combinations were considered to provide sensitivities to various couplings. This work showed that the projected sensitivity of space-based clocks for the detection of Sun-bound \ac{DM} halo exceeds the reach of Earth-based clocks by orders of magnitude. This mission in its proposed form can be used to test the quadratic coupling models. A \ac{DM} halo bound to the Sun can drastically improve the experimental reach due to much higher \ac{DM} densities.

\section{Theoretical Challenges of Models with Quadratic Couplings}\label{sec:Theoretical_Challenges_of_Models_with_Quadratic_Couplings}

In this section, we describe the theoretical issues related to theories with sizable quadratic couplings. The first is the \ac{EFT} expectation setting a hierarchy between the linear and the quadratic couplings of a generic theory. The second is the naturalness problem caused by the lightness of the scalar, as a desirable large scalar quadratic coupling is associated with high corrections to the scalar mass. In Section~\ref{sec:Examples_of_Technically-Natural_Models}, we present the symmetry principles that can give rise to a large hierarchy between linear and quadratic interactions, and in Section~\ref{sec:relaxed_relaxion_models} we present a model in which in addition the scalar is kept dynamically light even for detectable quadratic couplings.

\paragraph{Linear vs. quadratic \ac{EFT}:} 
Consider any naive dimension 4 (not necessarily of anomalous dimension greater than 4) \ac{SM} operator, $\mathcal{O}_{\text{SM}}$. Since the action of the theory is dimensionless, any coupling between the scalar \ac{DM} field and this operator has to be suppressed by some cutoff of the theory, denoted by $\Lambda$. For a linear coupling, we have only one power of $\Lambda$ suppression, $\frac{\phi}{\Lambda}\mathcal{O}_{\text{SM}}$, while for a quadratic coupling, we have two powers of $\Lambda$ suppression, $\frac{\phi^2}{\Lambda^2}\mathcal{O}_{\text{SM}}$. In the Wilsonian picture of \ac{RG}, $\Lambda$ is expected to be the largest scale of the described system. As a result, we naively expect a large hierarchy between the linear and quadratic interaction strength.

\begin{figure}[H]
	\centering
	\includegraphics[width=0.2\linewidth]{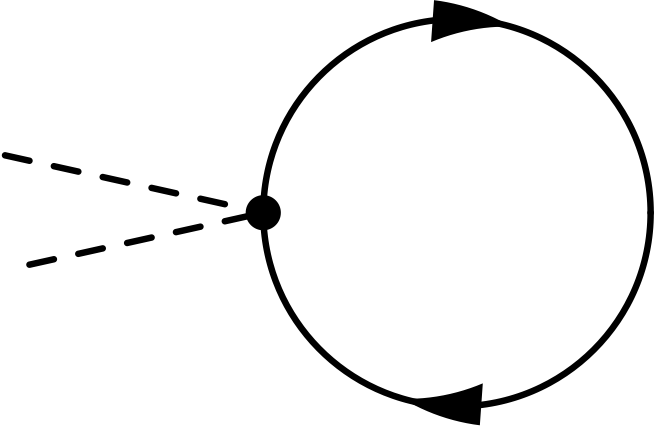}
	\caption{1-loop correction to the new scalar mass in the $\frac{d^{(2)}_{m_\psi}}{\Lambda^2}m_{\psi}\phi^2\psi\psi^c$ model.}
	\label{fig:phi_mass_correction_1loop}
\end{figure}

\paragraph{Naturalness:}~\label{sec:Naturalness}
Consider the following interaction between the \ac{DM} field and the \ac{SM} fermions,
\begin{equation}
 \L_{\text{int}} =\frac{d^{(2)}_{m_e}}{\Lambda^2}m_e\phi^2\psi_e\psi_e^c\,,\qquad \rho_A = \psi_e\psi_e^c\ \,, \label{eq:Stadnik_coupling}
\end{equation} 
where $\psi_e$ is the electron field. In a natural model, the quantum corrections to the mass of the light scalar, $\delta m_\phi^2$, are small compared to the classical bare mass parameter, $m_\phi^2$. The simplest 1-loop diagram that contributes to $\delta m_\phi^2$ is presented in Fig.~\ref{fig:phi_mass_correction_1loop}. 
Given the above electron coupling, the diagram gives the following mass correction
\begin{equation}
\delta (m^2_{\phi})^{\text{1-loop}} \sim \frac{4}{16\pi^2}\times \frac{d^{(2)}_{m_e}}{\Lambda^2}\left(m_e\Lambda^e_{\text{UV}}\right)^2\,,
\end{equation}
where $m_e$ is the electron mass, and $\Lambda^e_{\text{UV}}$ is the effective cut-off of the loop in Fig~\ref{fig:phi_mass_correction_1loop}, where new degrees of freedom are required to cancel the UV sensitivity of the scalar mass.
The requirement $ \delta m_{\phi}^2 \lesssim  m_{\phi}^2$ yields
\begin{equation}
\Lambda^e_{\text{UV}} \lesssim  \Lambda \,\frac{2\pi}{ \sqrt{d^{(2)}_{m_e}}}  \frac{ m_\phi}{m_e}\,. \label{eq:Naturalness_cutoff}
\end{equation}
For an ultra-light scalar, a measurable coupling of Eq.~\eqref{eq:Stadnik_coupling} requires a very small cutoff of the theory. 
For example, if we parameterize the experimental reach in terms of the effective temporal variation of the electron mass $\delta m_e/m_e$, induced by the \ac{ULDM} field $\phi$, we find that 
\begin{equation}\label{eq:Naturalness_cutoff_detectable}
\Lambda^e_{\text{UV}} \lesssim \frac{2\pi}{m_e}\sqrt{\frac{ 2 \rho_{\rm DM}}{\left(\delta m_e/m_e\right)^{\rm exp}} } \lesssim 20 \, \eV \left(\frac{10^{-18}}{\left(\delta m_e/m_e\right)^{\rm exp}}\right)^{1/2} \left(\frac{\rho_{\rm DM}}{\rho_\text{DM}^\odot}\right)^{1/2}\! , 
\end{equation}
where, $\rho_\text{DM}^\odot = 0.4\GeV/(\rm cm)^3$ is the mean galactic \ac{DM} energy density, similar to the one expected in our solar system~\cite{Salucci:2010qr}, and $\left(\delta m_e/m_e\right)^{\rm exp}$ is defined as the experiment sensitivity to the variations of the electron mass. 
The relatively low cutoff of Eq.~\eqref{eq:Naturalness_cutoff_detectable} is theoretically unfavorable, as it suggests that there exist some new fields with masses below $20\eV$ that are coupled to the \ac{SM} fermions. 
The same analysis can be done for \ac{DM} coupling to the photons, the quarks and the gluons as well. 
The quarks couplings yield the following bound
\begin{equation}\label{eq:Naturalness_cutoff_detectable_quarks}
\Lambda_{\text{UV}}^{\text{q}} \lesssim \frac{2\pi}{m_q}\sqrt{\frac{ \rho_{\rm DM}}{\left(\delta m_q/m_q\right)^{\rm exp}} } \lesssim  1 \, \keV \left(\frac{10^{-22}}{\left(\delta m_q/m_q\right)^{\rm exp}}\right)^{1/2} \left(\frac{\rho_{\rm DM}}{\rho_\text{DM}^\odot}\right)^{1/2}\! . 
\end{equation}
Due to the difference in the degree of divergence, the parametric form of the cut-offs for the \ac{DM} couplings to the photons and the gluons are different than that of Eqs.~\eqref{eq:Naturalness_cutoff_detectable}-\eqref{eq:Naturalness_cutoff_detectable_quarks}. For the gluon coupling, we get
\begin{equation}\label{eq:Naturalness_cutoff_detectable_gluons}
\Lambda_{\text{UV}}^{\text{g}} \lesssim \left(  8\pi^2\frac{ \rho_{\rm DM}}{\left(\delta \alpha_{s}/ \alpha_{s}\right)^{\rm exp}}\right)^{1/4}  \lesssim  40 \, \keV \left(\frac{10^{-22}}{\left(\delta \alpha_{s}/ \alpha_{s}\right)^{\rm exp}}\right)^{1/4} \left(\frac{\rho_{\rm DM}}{\rho_\text{DM}^\odot}\right)^{1/4}\! . 
\end{equation}

The cutoff can be raised if the on-Earth \ac{DM} density, which drives the oscillations observed in terrestrial \ac{DDM} searches, is enhanced compared to the mean galactic \ac{DM} density.\footnote{A much higher density is allowed if the \ac{DM} is forming a halo around the Earth~\cite{Banerjee:2019epw,Banerjee:2022wzk,haloformation}.} 
Note that while the parameterisation above can be easily applied to any \ac{DDM} experiment probing such oscillations, the effect of a \ac{DM} density enhancement on the experimental sensitivity is model-dependent and experiment-dependent. One should also note that the, existence of new physics at the keV scale is constrained by various astrophysical and cosmological considerations, which are subjected to recent critical investigations~\cite{Allahverdi:2020bys,DeRocco:2020xdt}.
A linear coupling between the \ac{DM} and the \ac{SM} may also possess a naturalness problem for CP-even scalars. However, it could be protected due to the existence of accidental/non-accidental symmetries, which are absent in the case of a quadratic coupling. In addition, a CP-odd linear theory can be embedded in an axion-like theory, thus protected from a fine-tuning problem, as explained in the following section.

\section{Examples of Technically-Natural Models of \ac{DM} with Quadratic Interactions}\label{sec:Examples_of_Technically-Natural_Models}

In this section, we survey a few models which yield a technically-natural hierarchy between the linear \ac{ULDM} coupling and the quadratic one, allowing the quadratic interactions to dominate the phenomenology of the \ac{ULDM}. 
We begin by addressing the symmetries protecting these couplings in the agnostic \ac{EFT} approach, identifying those that may retain the linear couplings small in subsection~\ref{subsec:agnostic EFT approach}. 
We then specify two models in which the linear interactions are absent - a \ac{pNGB} effective model in subsection~\ref{subsec:PNGB} and a UV-complete Higgs-portal model in subsection~\ref{subsec:HiggsPortal}. 
Finally, in subsection~\ref{subsec:Clockwork} we present two variations of the Clockwork framework in which the hierarchy between the linear and the quadratic couplings can be ameliorated in a technically-natural way. 
Although these models present theoretically-sound mechanisms for altering the naive hierarchy between the strength of the quadratic and the linear interactions, a naturally light scalar implies the quadratic interactions despite being dominant, are beyond the reach of current and near future \ac{DDM} searches. 
We present a possible solution to this issue in Section~\ref{sec:relaxed_relaxion_models}.

\subsection[EFT perspective of the linear vs the quadratic couplings]{\ac{EFT} perspective of the linear vs. the quadratic couplings}\label{subsec:agnostic EFT approach}
In this subsection, we analyze the effective interaction between the $\phi$ sector and the \ac{SM} without specifying a UV model. We consider the following three types of operators,
\begin{equation}
	\mathcal{L}_{\text{int}}\left(\phi\right)=c_{1}^{\text{lin}}\phi\,{\cal O}_{\text{SM}}+c_{2}^{\text{lin}}\partial_{\mu}\phi\, J_{\text{SM}}^{5\mu}+c_{1}^{\text{quad}}\phi^{2}\,{\cal O}_{\text{SM}}\, .
	\label{eq:agnostic_operators}
\end{equation}
Here we assume $\phi$ is a light \ac{pNGB} and thus a pseudo-scalar, while ${\cal O}_{\text{SM}}$ is taken to be a dimension four, CP-even operator, consist of \ac{SM} fields only, such as ${\cal O}_{\text{SM}}= F_{\mu\nu}F^{\mu\nu},\; m_\psi \psi \psi^c \;$etc, $J_{\text{SM}}^{5\mu}=\bar{\psi} \gamma^\mu \gamma^5\psi$ is the \ac{SM} axial current. 
At this point, we ignore possible self coupling of $\phi$ and the possible interactions with some hidden sectors.

In general, we expect the linear $\phi$ couplings to dominate over the quadratic couplings unless the quadratic $\phi^{2}$ couplings are protected by symmetry. In this context, we consider three types of accidental symmetries that act non-trivially on the field $\phi$
\begin{equation}
	G_{\text{global}}\supset U(1)_\Phi \times CP\times\mathbb{Z}_{2}^{\phi}\, .
\end{equation}
The $U(1)_\Phi$ symmetry is a non-linearly realized
global $U\left(1\right)$ group, which acts as a constant shift for a
\ac{pNGB}, thus protecting the \ac{pNGB} from
acquiring a mass. Thus, the smallness of the \ac{DM} mass $\phi$ is protected
by a shift symmetry,
\begin{equation}
	U(1)_\Phi:\quad\phi\mapsto\phi+\alpha\,.
\end{equation}	
Therefore, unless the \ac{SM} is charged under the $U(1)_\Phi$ group, both $c_{1}^{\text{lin}}\phi\,{\cal O}_{\text{SM}}$ and
$c_{1}^{\text{quad}}\phi^{2}\,{\cal O}_{\text{SM}}$ breaks this symmetry. Moreover, we consider both $CP$ and $\mathbb{Z}_{2}^{\phi}$
to act similarly on the field $\phi$. $CP$ is an external symmetry with well-defined transformation rules for the \ac{SM} fields, while the internal $\mathbb{Z}_{2}^{\phi}$ can be taken to affect only $\phi$. 
For example, we consider the action of the discrete groups as follows
\begin{equation}
	\mathbb{Z}_{2}^{\phi}:\quad\phi\mapsto-\phi \, ,
\end{equation}
while a general CP transformation in an arbitrary basis can be written as
\begin{align}
	CP:  \qquad \nonumber \phi\left(t,\vec{x}\right)  &\mapsto-\phi\left(t,-\vec{x}\right)\label{eq:CP_scalar}\\ \!\!\!\!\!\!\!\!
	{\cal O}_{\text{SM}}\left(t,\vec{x}\right) &\mapsto {\cal O}_{\text{SM}}\left(t,-\vec{x}\right) \\  \!\!\!\!\!\! 
	\partial_\mu J^{5\mu}_{\text{SM}}\left(t,\vec{x}\right) &\mapsto -\partial_\mu J^{5\mu}_{\text{SM}}\left(t,-\vec{x}\right)\,.
\end{align}

Each operator in Eq.~\eqref{eq:agnostic_operators} may break one or more of the global symmetries. 
We summarize the symmetry breaking pattern in Table~\ref{tab:cheching_symmetries}. 
\begin{table}[H]
	\begin{center}
		\begin{tabular}{c|ccc}
			& $U\left(1\right)_{\text{shift}}$ & $CP$ & $\mathbb{Z}_{2}^{\phi}$\tabularnewline
			\hline 
			$c_{1}^{\text{lin}}\phi\,{\cal O}_{\text{SM}}$ & x & x & x\tabularnewline
			$c_{2}^{\text{lin}}\partial_{\mu}\phi\, J_{\text{SM}}^{5\mu}$ & $\checked$ & $\checked$ & x\tabularnewline
			$c_{1}^{\text{quad}}\phi^{2}\,{\cal O}_{\text{SM}}$ & x & $\checked$ & $\checked$ \tabularnewline
		\end{tabular}
		\par\end{center}
	\caption{Symmetry preserving and symmetry breaking operators.}
	\label{tab:cheching_symmetries}
\end{table}
In principle, each symmetry can be broken by a different sector. 
Thus, the naive expectation that the highly irrelevant operators such as $\phi^{2}\,{\cal O}_{\text{SM}}$ are less relevant than the linear ones does not hold. 
Moreover, the $\mathbb{Z}_{2}^{\phi}$ is considered a good approximate symmetry if the internal $\mathbb{Z}_{2}^{\phi}$ is highly protected. 
Thus, the quadratic coupling i.e. $c_{1}^{\text{quad}}\phi^{2}\,{\cal O}_{\text{SM}}$, can have the leading effect on the violation of \ac{EP} and/or oscillations of the \aclp{FC}. 
The difference between the quadratic theory and a linear is emphasized through the analytic expressions of the \ac{EP} constraints, as shown in Eq.~\eqref{eq:etq_EP_quadratic_Stadnik_background} and Eq.~\eqref{eq:etq_EP_linear_Stadnik_background}. 
As discussed below Eq.~\eqref{eq:AP-quad_electrons}, a quadratically coupled \ac{DM} induces \aclp{FC} oscillations at an angular frequency of $\omega= 2m_\phi$ compare to that of the a linear coupling such as $c_{1}^{\text{lin}}\phi{\cal O}_{\text{SM}}$  ($c_{2}^{\text{lin}}\partial_{\mu}\phi J_{\text{SM}}^{5\mu}$) which yields \acl{FC} oscillations (spin precision) at an angular frequency of $\omega=m_\phi$. 
In order to be able to probe both the quadratic and the CP-even linear interaction in \ac{DDM} experiments, we must require that both the couplings are not suppressed compared to one another. 
This criteria posses an apparent tuning problem from the perspective of a naive \ac{EFT} analysis. which can be ameliorated as shown in Subsection~\ref{subsec:Clockwork}. 

\subsection[A Simple pseudo-Goldstone model as a naturally ultra-light scalar DM candidate]{A Simple pseudo-Goldstone model as a naturally ultra-light scalar DM candidate}
\label{subsec:PNGB}
There are several ways to avoid the naturalness requirement of a small effective cutoff. One of the most appealing solutions is to consider $\phi$ as a \ac{pNGB} of some \ac{SSB}. 
In the non-linear sigma model description of the Goldstone interactions, one must add an appropriate linear $\phi$ coupling, as well as other polynomial powers of $\phi$ interactions.
The low energy theory of a spontaneous broken $U(1)_\Phi$ symmetry can be described by
\begin{align}
	\L_{\text{EFT}} = &\, \L_{\text{SM}} +\frac{f^2}{2}\partial_{\mu}U^\dagger\partial^{\mu}U-m_\psi U\psi\psi^c 
	+\frac{1}{\Lambda^2}\partial_{\mu}U^\dagger\partial^{\mu}Um_\psi U\psi\psi^c +...+\text{h.c.}\,, \label{eq:non-linear_sigma_EFT}
\end{align}
where $U=e^{i\frac{\phi}{f}}$, $\psi$ is some \ac{SM} fermion with mass $m_\psi$, $f$ is the scale at which $U(1)_\Phi$ is broken spontaneously, $\Lambda\gtrsim f$ is the cutoff of the theory, and the ellipsis represents higher derivatives and/or higher dimensional irrelevant operators in the Lagrangian. 
As shown in Eq.~\eqref{eq:L_axion_phi2}, the last term of Eq.~\eqref{eq:non-linear_sigma_EFT} gives rise to a quadratic interaction between the \ac{SM} fermions and $\phi$. 
This derivative term arises naturally from the low-energy physics and does not require an ad-hoc source in the UV. For example, consider a complex scalar field with a $U(1)_\Phi$ preserving potential
\begin{equation}
	V(\Phi^\dagger \Phi) = \lambda_{\Phi}  \left( \Phi^\dagger\Phi- \frac{f^2}{2} \right)^2.
	\label{eq:potential}
\end{equation} 
This potential results in the \ac{SSB} of the $U(1)_\Phi$ symmetry. Therefore, at low energies, expanding around the true vacuum of the theory, one finds a mass less Goldstone boson, $\phi$, appearing as the phase of the complex scalar as
\begin{equation}
	\Phi=\frac{f+\rho}{\sqrt{2}}e^{i\frac{\phi}{f}}\,,~\label{eq:Phi}
\end{equation}
where $\rho$ is the radial mode of $\Phi$ with mass $m_\rho\sim f$. 
The above parameterization manifestly provides an interaction between $\rho$ and $\phi$, which arises from the kinetic term of $\Phi$ as
\begin{align}
	\partial^{\mu}\Phi^{\dagger}\partial_{\mu}\Phi&\simeq\partial^{\mu}\left(\frac{f+\rho}{\sqrt{2}}e^{-i\frac{\phi}{f}}\right)\partial_{\mu}\left(\frac{f+\rho}{\sqrt{2}}e^{i\frac{\phi}{f}}\right)
	\nonumber \\
	&=\frac{1}{2}\partial_{\mu}\rho\partial^{\mu}\rho+\frac{1}{2}\partial_{\mu}\phi\partial^{\mu}\phi+\frac{1}{2f^{2}}\left(f+\rho\right)^{2}\partial_{\mu}\phi\partial^{\mu}\phi\,. \label{eq:rho_dphidphi}
\end{align}
Thus, Eq.~\eqref{eq:rho_dphidphi} suggests that upon integrating out the radial mode, a coupling between $\rho$ and the \ac{SM} will result in a low energy effective coupling between $\partial_{\mu}\phi\partial^{\mu}\phi$ and the \ac{SM}.
Going back to Eq.~\eqref{eq:non-linear_sigma_EFT} and expanding the exponent in terms of the small fluctuations of $\phi/f$, one finds 
\begin{align}
	\L_{\text{EFT}} \supset   &\; \L_{\text{SM}} +\frac{1}{2}\partial_{\mu}\phi^\dagger\partial^{\mu}\phi-m_\psi\left(1+i\frac{\phi}{f}-\frac{\phi^2}{2f^2}+...\right)\psi\psi^c 
	\nonumber \\ &
	+\frac{1}{m_\rho^2f^2}\partial_{\mu}\phi^\dagger\partial^{\mu}\phi m_\psi\left(1+i\frac{\phi}{f}-\frac{\phi^2}{2f^2}+...\right)\psi\psi^c +...+\text{h.c.}\,.
\end{align}
The corrections to $\delta m^2_{\phi}$ from the quadratic $\phi^2\psi\psi^c $ are exactly canceled by the corrections from the other interactions. 
As already mentioned, this cancellation is guaranteed by the shift symmetry of $\phi$, which is a non-linear realization of the original $U(1)_\Phi$ symmetry. This non-linearly $U(1)_\Phi$ symmetry also forbids any potential of $\phi$ as well which is manifested in a different basis. To see how it works, one can do an axial field redefinition as 
\begin{equation}
	\psi\longrightarrow e^{i\frac{\phi}{2f}}\psi\,,\qquad  \psi^c\longrightarrow e^{i\frac{\phi}{2f}}\psi^c\,,
\end{equation} 
which yields the following effective Lagrangian
\begin{equation} \label{eq:ALP_EFT}
	\L_{\text{EFT}}' \supset \L_{\text{SM}} +\frac{1}{2}\partial_{\mu}\phi^\dagger\partial^{\mu}\phi-m_\psi\psi\psi^c -\frac{1}{f}\partial_{\mu}\phi J^{\mu}_A+\frac{1}{m_\rho^2 f^2}	\partial_{\mu}\phi\partial^{\mu}\phi m_\psi\psi\psi^c +...+\text{h.c.}\,,
\end{equation}
where $J^{\mu}_A= \bar{\psi}\bar{\sigma}^\mu\partial_\mu\psi+\bar{\psi^c} \bar{\sigma}^\mu\partial_\mu\psi^c$ is the axial current. 
Note that we ignored any anomalies as these can be eliminated by an appropriate choice for the $U(1)_\Phi$ charges of the other fermions. Even in the absence of anomalies, the shift symmetry would be explicitly broken by a soft mass term for $\phi$, related to its nature as a \ac{DM} candidate. 
Therefore, by using the \ac{EOM} for $\phi$, one can replace, to leading order, $\partial_\mu\phi\partial^\mu\phi\rightarrow m_\phi^2\phi^2$, yielding an interaction term similar to the one in Eq.~\eqref{eq:Stadnik_coupling}
\begin{align}\label{eq:L_axion_phi2}
	\L^\text{nat}_{\text{int}} = \frac{1}{ \Lambda^2f^2}m_\psi\partial_\mu\phi\partial^\mu\phi\psi\psi^c \longrightarrow \frac{m^2_\phi}{ \Lambda^2f^2}m_\psi\phi^2\psi\psi^c \,.
\end{align}
As expected, this implies that a natural $d^{(2)}_{m_i}$ coupling would be proportional to $m_\phi^2$, protecting $m_\phi$ against radiative corrections. 
\\
The model presented above is usually discussed in the context of \acp{ALP}. 
As we are interested in the (pseudo) scalar-electrons coupling, we note that it is strongly constrained by stellar evolution consideration, as those couplings provide alternative channels for stellar energy loss processes~\cite{Raffelt:2012sp,Redondo:2013lna,Hardy:2016kme}. For instance, the most stringent bound
on the pseudo-scalar electron-Yukawa coupling, $\L\supset i g^p_e \phi \bar{\psi_e}\gamma^5\psi_e$, is $ g_e^p \lesssim 3\times 10^{-13}\, $, obtained from the evolution of red giants~\cite{Raffelt:2012sp}. 
This can be translated to a bound on the \ac{ALP} decay constant as $f \gtrsim 2\times 10^{9}\,\GeV\,$. 
The temporal oscillation of the mass of electron for an \ac{ALP} decay constant allowed by cosmological consideration can be written as
\begin{equation}
	\frac{\delta m_e}{m_e} \simeq \frac{\rho_{\rm DM}}{f^4}\lesssim 2 \times 10^{-79}\, \leri{\frac{2\times 10^9\GeV}{f}}^4\left(\frac{\rho_{\rm DM}}{\rho_\text{DM}^\odot}\right) \,, \label{eq:Ridiculous_Sensitivity}
\end{equation}
which for detection, requires unrealistically high precision from the current and the near future \ac{DDM} searches~\cite{Safronova:2017xyt}.

Below, we present different types of light scalar models which do not generate a linear axion-like interaction of the form $\frac{1}{f}\partial_{\mu}\phi J^{\mu}_A$, or in which such linear interaction is suppressed, and are thus significantly less constrained. 
In these models, $\phi$ mainly interacts with the \ac{SM} via its quadratic derivatives. These interactions can be naturally obtained from the kinetic mixing between $\phi$ and its radial mode, as shown in Eq~\eqref{eq:rho_dphidphi}.

One example of such realization which leads to quadratic $\phi^2$ couplings without linear couplings is to consider the $\mathbb{Z}_{2}^{\phi}$ even coupling of a complex scalar field $\Phi$ as
\begin{equation}\label{phi_SM}
	\L_{\Phi-\rm SM} = \L_{\rm SM}+ \partial^{\mu}\Phi^{\dagger}\partial_{\mu}\Phi + V(\Phi^{\dagger}\Phi) + 16\pi^2 \frac{\Phi^\dagger\Phi}{\Lambda} \psi\psi^c\,,  
\end{equation}
where $\psi,\psi^c$ are some \ac{SM} fermionic fields with mass $m_\psi$, $\Lambda$ is the cut-off scale of the effective interaction. As explained in Section~\ref{sec:Naturalness}, imposing a $U(1)_\Phi$ symmetry that acts solely on $\Phi$ ensures that $\Phi$ may only couple to the \ac{SM} through powers of $\Phi^\dagger\Phi \,\mathcal{O}^s_\text{SM}$ or higher derivative powers, where $\mathcal{O}^s_\text{SM}$ is an operator contains the \ac{SM} fields which is a singlet of all \ac{SM} symmetries.
\\
Using Eq.~\eqref{eq:rho_dphidphi} and the last term of Eq.~\eqref{phi_SM}, after integrating out the radial mode $\rho$ at energies below $m_{\rho}$, we get an effective interaction between $\partial_\mu\phi\partial^\mu\phi$ and the \ac{SM} fermion $\psi$ as, 
\begin{equation}\label{eq:dphidphi_ee}
	\L_{\rm EFT} \supset\frac{16\pi^2}{ \Lambda m_{\rho}^2}\,  \partial_\mu\phi\partial^\mu\phi\, \psi\psi^c \, .
\end{equation}
We require that the effective cutoff would be the largest scale of the \ac{EFT}, and thus
\begin{equation}\label{eq:cutoff_natural_bound}
	\Lambda \gtrsim {\rm Max}\left[4\pi f,m_\psi\right] \quad \Rightarrow \quad  \Lambda  \gtrsim {\rm Max}\left[ m_\rho, m_\psi\right]\,.
\end{equation} 
Note that the interaction term between the $\Phi$ sector and the \ac{SM} explicitly breaks the chiral symmetry and generates a correction to the \ac{SM} fermion mass at the tree-level. 
Assuming no fine cancellation against other contributions to the fermion mass, we require that the correction to $m_\psi$ from Eq.~\eqref{phi_SM} is smaller than the physical value found in experiments. Thus,
\begin{equation}
\label{me_naturalness}
\delta m_\psi =  \frac{16\pi^2}{\Lambda}\frac{f^2}{2 }\lesssim m_\psi 
\quad \Rightarrow \quad \Lambda \gtrsim  \frac{16\pi^2\, f^2}{2\, m_\psi}\,. 
\end{equation}
By the consistency of the Goldstone theory, where $m_{\rho}\lesssim 4\pi f$, we obtain
\begin{equation}
\label{eq:Lambda_me_nat}
\Lambda\gtrsim  \frac{m_{\rho}^2}{2\, m_\psi} \,\, .
\end{equation}
We note that if the radial mode $\rho$ is lighter than $\psi$, we expect Eq.~\eqref{eq:cutoff_natural_bound} to give a stronger lower bound on $\Lambda$ than Eq.~\eqref{eq:Lambda_me_nat} as, 
\begin{equation}\label{eq:Lambda_lower_bound}
	\Lambda \gtrsim  {\rm Max}\left[  \frac{m_{\rho}^2}{2\, m_\psi},\,\, m_\rho\,\, ,m_\psi \right] \gtrsim m_\psi
	\,.
\end{equation}
However, as the strength of the $\phi^2$ interactions is inversely proportional to $\Lambda$, raising the cut-off requires a higher experimental sensitivity to detect the temporal variation of $m_\psi$. 
If we parameterise the sensitivity of a \ac{DDM} search experiment in terms of the variation of the electron mass $(\delta m_e/m_e)$, the maximal $\Lambda$ such an experiment can probe is
\begin{equation}\label{eq:exp_bound}
	\Lambda \lesssim 1 \,\MeV
	\left(\frac{10^{-18}}{\leri{\delta m_e/m_e}^{\rm exp}}\right) \left(\frac{30\, \eV}{m_\rho}\right)^2\left(\frac{\rho_{\rm DM}}{\rho_\text{DM}^\odot}\right)\,.
\end{equation}
This translate to a bound on the mass of the radial mode, which has to satisfy
\begin{align}
	\label{eq:m_{rho}}
	m_\rho\lesssim 45 \eV
	\left(\frac{10^{-18}}{\leri{\delta m_e/m_e}^{\rm exp}}\right)^{1/2}\left(\frac{\rho_{\rm DM}}{\rho_\text{DM}^\odot}\right)^{1/2}\!.
\end{align}
The characteristic sensitivity of $\leri{\delta m_e/m_e}^{\rm exp}\sim 10^{-18}$, would allow probing models with $\rho$ as heavy as $m_\rho\sim 45\, \eV$, saturating the requirement above and a corresponding maximal cutoff of the same order.
We note that an enhanced local \ac{DM} density would allow \ac{DDM} searches to probe models with higher cut-offs and heavier radial modes accordingly.

\subsection{The Higgs-portal model - an example of a UV complete theory with no linear DM couplings}~\label{subsec:HiggsPortal}
In this section, we provide an example of a specific renormalizable UV model that could result in the effective low energy \ac{SM} with additional interactions of the form of Eq.~\eqref{eq:Stadnik_Lint_quad}. 
We allow other even orders of derivatives of $\phi$, such as $(\partial_\mu\phi\partial^\mu\phi)^2$ or $\square \phi \square\phi$, to be coupled to the \ac{SM}, but forbid linear derivative couplings. 
To achieve the desirable low energy \ac{EFT}, we extend the \ac{SM} field content by introducing a new complex scalar field $\Phi$, which is a singlet of the \ac{SM} gauge group. We impose an additional $U(1)_\Phi$ global symmetry, acting only on the $\Phi$ field, under which the \ac{SM} fields are neutral.

\begin{figure}
	\centering
	\includegraphics[width=0.4\linewidth]{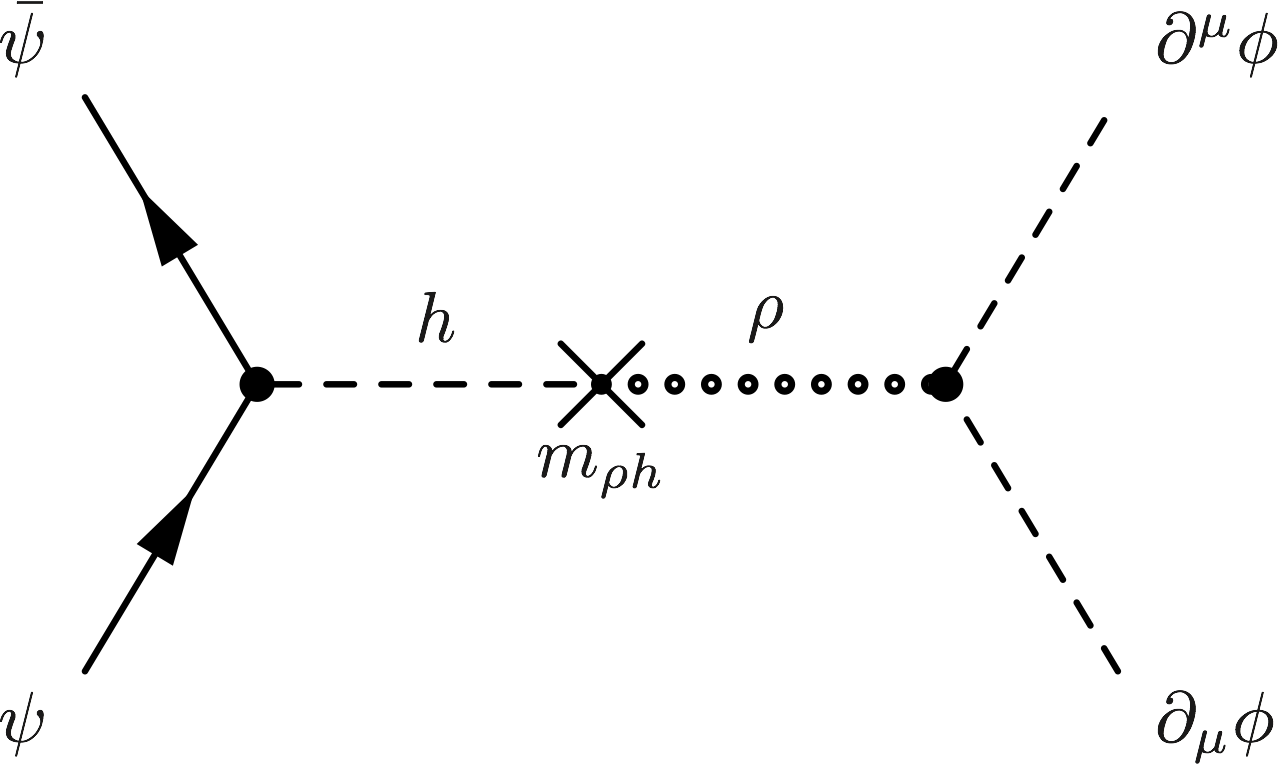}
	\: $\rightarrow$ \:
	\includegraphics[width=0.4\linewidth]{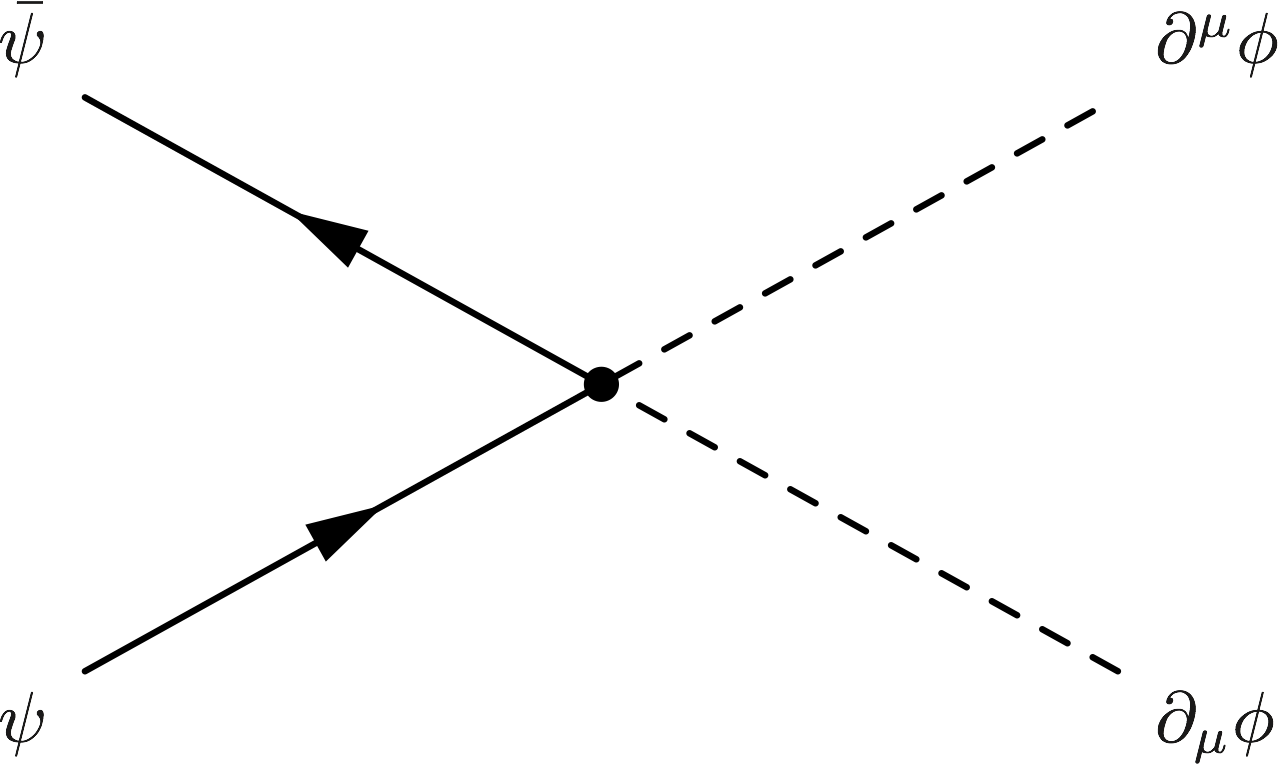}
	\caption{A diagrammatic description of the effective operator in the low energy theory. }
	\label{fig:EFT_generated_diagram1}
\end{figure}

The most general renormalizable model can be written as
\begin{align}
	\L_{\text{UV}}= &\: \L_{\text{SM}}+\partial_{\mu}\Phi^\dagger\partial^{\mu}\Phi - \lambda_{\Phi H}\Phi^\dagger\Phi  H^\dagger H -V(\Phi^\dagger \Phi)\,,
\end{align}
where $\L_{\text{SM}}$ denotes the usual \ac{SM} Lagrangian, $V(\Phi^\dagger \Phi)$ is the potential described in Eq.~\eqref{eq:potential} and $H$ is the \ac{SM} Higgs doublet. 
We assume that the potential of $\Phi$ induces a \ac{SSB} of the $U(1)_\Phi$ symmetry, upon which the low energy description of the theory is given in terms of the radial and compact modes of $\Phi$, presented in Eq.~\eqref{eq:Phi}. After the electroweak symmetry breaking, the Higgs portal coupling induces an interaction between the $\Phi$ sector and the \ac{SM} fermions through a mixing of the radial mode $\rho$ with the physical Higgs singlet $h$ as,
\begin{align}
	\L_{\text{H-portal}}\simeq\   - \lambda_{\Phi H}\left(\frac{f+\rho}{\sqrt{2}}\right)^2 \left(\frac{v_H+h}{\sqrt{2}} \right)^2\,.
\end{align}
After diagonalizing the mass matrix $M_{\rho h}$, one obtains a $h-\rho$ mixing angle of 
\begin{align}
	\sin\theta_{h\rho} \approx \frac{\lambda_{\Phi H}\,v_H\,f}{m^2_h-m^2_\rho}\,.
\end{align}
After integrating out $\rho$ at energy below $m_\rho$, we obtain interactions between $\partial_\mu\phi\partial^\mu\phi$ and the light \ac{SM} fermions as discussed before. 
In the limit of small mixing, we obtain the following effective operator
\begin{align}
	\L_{\text{EFT}} &\simeq \frac{\sin\theta_{h\rho}\,m_\psi}{f m^2_\rho\, v_H}\partial_\mu\phi\partial^\mu\phi\psi\psi^c\label{eq:higgs_portal_EFT}=\frac{\lambda_{\Phi H}}{\leri{m^2_h-m^2_\rho}}\frac{m_\psi}{m^2_\rho }\partial_\mu\phi\partial^\mu\phi\psi\psi^c\,, 
\end{align}
which is shown diagrammatically in Fig.~\ref{fig:EFT_generated_diagram1}. 

While this model is UV complete, it sets a very low upper bound on the mass of $\rho$. 
In order for such a model to be detectable considering the current experimental sensitivity, the mass of the radial mode would have to be $m_\rho\lesssim 10^{-5} \eV$. 
While such a light invisible scalar that couples to the Higgs makes this model less appealing, we consider it to be more of a proof of concept rather than a conclusive case study. 
In the next subsection, we would take a more agnostic approach, studying the low-energy behavior of the natural quadratic coupling without specifying a UV completion.

\subsection{The clockwork mechanism - an example of a tunable hierarchy between linear and quadratic couplings }\label{subsec:Clockwork}

In the previous Section~\ref{subsec:agnostic EFT approach}, we introduced a general EFT approach to linear and quadratic couplings between the \ac{DM} $\phi$ and some CP-even \ac{SM} operators such as ${\cal O}_{\text{SM}}= F_{\mu\nu}F^{\mu\nu},\; m_\psi \psi \psi^c\;$ etc. 
We argued that different operators might break/preserve different approximate symmetries. Therefore, there could be, in principle, a large hierarchy between the dimensionless coefficient of different types of operators. For example, in some models, we expect the linear $\phi$ couplings to be suppressed and the quadratic couplings of $\phi^2$ to the \ac{SM} to dominate the physical effects on the induced forces and potential, even though their naive dimension is higher than the linear couplings.

In this section, we provide two examples where the suppression of the linear couplings is based on symmetry principles. 
The examples are based on Clockwork framework~\cite{Choi:2015fiu, Kaplan:2015fuy}, and the suppression is due to the existence of a large hierarchy between the effective periodicity, $F$, compare to the smaller periodicity $f$, which is the dynamical scale of a spontaneous symmetry breaking. A detailed description of the Clockwork model can be found in Appendix~\ref{app:Clockwork}.

In the Clockwork model, the remaining $U(1)_{\text{shift}}$ symmetry, which keeps the \ac{DM} mass small, is identified with the remaining $U\!\left(1\right)_{\text{clock}}$ of the N+1 Clockwork model, as shown in Eq.~\eqref{eq:clockwork_charges} where $N$ is the number of Clockwork sites. 
Note that in the limit of exact $U(1)_{\text{shift}}$ symmetry, $\phi$ is mass-less. 
In order to achieve suppression of the linear $\phi$ couplings, we assign small charges to the \ac{SM} fermions, i.e., $[\psi]\sim \mathcal{O}(1)$, such that the leading symmetry preserving interaction is on the last Clockwork site,
\begin{equation}\label{eq:OurClock_last_site}
	\L_{\text{int}}^{\text{leading}} \supset c_N\frac{\Phi_{N}}{\Lambda} LHE^{c}+V_{\text{mixing}}\left(\Phi_N,H\right)+ \text{h.c.}\,,
\end{equation}   
where $L$ are the left-handed $SU(2)$ doublet \ac{SM} Lepton fields, while $E^{c}$ are the left-handed $SU(2)$ singlet \ac{SM} Lepton fields.
We assume that $V_{\text{mixing}}\left(\Phi_N,H\right)$ breaks CP spontaneously by some parameter $\theta_{\text{CP}}\ll1$, which is defined as the CP mixing angle between the Clockwork and the Higgs CP eigenstates:
\begin{align}
	\phi = & \cos\theta_{\text{CP}} \hat{\phi}+\sin\theta_{\text{CP}}\hat{h}\,, \\
	h = & -\sin\theta_{\text{CP}}\hat{\phi}+\cos\theta_{\text{CP}} \hat{h}\,.
\end{align}
$V_{\text{mixing}}\left(\Phi_N,H\right)$ can give rise to both CP-even and CP-odd interactions between $\phi$ and the \ac{SM} fields.

At low energies, we can integrate out the heavy modes of the Clockwork and obtain an effective Lagrangian of the \ac{pNGB} which is identified as the \ac{ULDM} field,
\begin{equation}
	\L_{\text{EFT- int}}^{\text{leading}} \supset c_N\frac{f\,e^{i\frac{\phi}{F}}}{\Lambda} h\psi\psi^c +V_{\text{mixing}}\left(\frac{\phi}{F},H\right)+ \text{h.c.}\,,
\end{equation}
where $c_N$ is some $\mathcal{O}(1)$ coefficient. 
As a result, the derivative coupling of $\phi$, of the form $\frac{f}{\Lambda}\frac{\partial_{\mu}\phi}{F} m_\psi \left[ \bar{\psi}\bar{\sigma}^\mu\psi-\bar{\psi}^c\bar{\sigma}^\mu\psi^c \right]$, is highly suppressed by  $F=3^N f$.

The higher dimensional operator
\begin{equation}\label{eq:OurClock_first_site}
	\L_{\text{int}}^{\text{NL}} \supset c_0\frac{\Phi_{0}^\dagger\Phi_{0}}{\Lambda^2} m_\psi\psi\psi^c + \text{h.c.} \, ,
\end{equation}  
gives rise to a quadratic coupling that is only suppressed by $f$, not $F$. $c_0$ is some $\mathcal{O}(1)$ coefficient. 
After integrating out the heavy Clockwork radial modes, from Eq.~\eqref{eq:OurClock_first_site} we obtain, 
\begin{equation}
	\L_{\text{EFT- int}} \supset c_0\frac{1}{2f^2\Lambda^2} \partial_{\mu}\phi\partial^{\mu}\phi \,m_\psi\psi\psi^c \, .
\end{equation}
Since $F=3^N f$, is just an artifact scale of the Clockwork model, it can be larger than the cutoff scale, $\Lambda\gtrsim f$. 
Thus, we achieve a hierarchy between the linear and the quadratic (dimensionless) couplings. 

Moreover, adding the both the interactions of Eq.~\eqref{eq:OurClock_last_site} and Eq.~\eqref{eq:OurClock_first_site} leads to a collective breaking of the $U\!\left(1\right)_{\text{clock}}$ symmetry. 
As a result, one can write the effective potential generated for this \ac{pNGB}.  
The 1-loop \ac{CW} effective potential of $\phi$ can be written as, 
\begin{equation}
	\!\!\!\! V_{\text{CW}}\left(\phi\right)=\frac{\left(-1\right)^{\mathcal F}}{64\pi^{2}}Tr\left[2M^{\dagger}\left(\phi\right)M\left(\phi\right)\Lambda_{c}^{2}+\left(M^{\dagger}\left(\phi\right)M\left(\phi\right)\right)^{2}\ln\frac{M^{\dagger}\left(\phi\right)M\left(\phi\right)}{\tilde{\Lambda}_{c}^{2}}\right]\!.\!\!
	\label{eq:CW_poetntial}
\end{equation}
In the equation above, Tr is performed over all field degrees of freedom, $(-1)^{\mathcal F}$ is $+1$ for bosons and $-1$ for fermions. 
The momentum cutoff $\Lambda_{c}$, is taken to be of the order of the radial mode mass, which is of $\mathcal{O}(f)$. 
Thus, at the leading order the effective potential of $\phi$ can be approximately written as 
\begin{equation}\label{eq:DM_Clockwork_mass}
	V(\phi) \simeq -c_0c_N\frac{m_\psi^2f^3}{16\pi^2\Lambda}\cos\left(\frac{\phi}{F} \right).
\end{equation}
The induced \ac{DM} potential of Eq.~\eqref{eq:DM_Clockwork_mass} generates a contribution to the \ac{DM} mass of
\begin{equation}
	m_{\phi}^2 \simeq c_0c_N\frac{m_\psi^2f^3}{16\pi^2\Lambda} \frac{1}{F^2}\,.
\end{equation}
Therefore, if Eq.~\eqref{eq:DM_Clockwork_mass} is the only source of the \ac{DM} mass, the quadratic interaction will also be suppressed by $1/F^2$, and similar to the linear interaction. Therefore, we require that there is an additional source of the \ac{DM} mass, $m_{\phi}^2$ which is not suppressed by $1/F^2$.

Another way to employ the Clockwork mechanism is one where the shift symmetry is broken at the two ends of the Clockwork chain i.e. at the first site, at $\Phi_0$, and at the last site, at $\Phi_N$. 
We assume two different $Z_2$ symmetries are conserved in each site, such that the Lagrangian of these $U(1)_{\text{clock}}$ breaking sectors can be written as,
\begin{equation}\label{eq:Clockwork_two_Z2_breakings}
	\L_{\text{int}}^{\text{leading}} \supset  c_N\frac{\Phi_N - \Phi_N^\dagger}{\Lambda} m_\psi \psi \psi^c + c_0 \frac{\Phi_0 + \Phi_0^\dagger}{\Lambda} m_\psi \psi \psi^c\,.
\end{equation}
Therefore, the effective Lagrangian of the \ac{pNGB} takes the form of
\begin{equation}\label{eq:Clockwork_two_Z2_breakings_EFT}
	\L_{\text{effective}}^{\text{leading}} \simeq  c_N\frac{f\sin\left(\frac{\phi}{F} \right) }{\Lambda} m_\psi \psi \psi^c + c_0 \frac{f\cos\left(\frac{\phi}{f} \right)}{\Lambda} m_\psi \psi \psi^c\,.
\end{equation}
Assuming the clockwork potential is dominated by a backreaction potential which has a minimum near $\phi\simeq 0$, one can expand the trigonometric functions to achieve the desired hierarchy between the linear coupling and the quadratic coupling, relatively suppressed by $f/F\ll1$, as 
\begin{equation}\label{eq:Clockwork_two_Z2_breakings_EFT2}
	\L_{\text{effective}}^{\text{leading}} \simeq  \frac{c_N\,f}{\Lambda F} \phi m_\psi \psi \psi^c +  \frac{c_0}{\Lambda f} \phi^2  m_\psi \psi \psi^c\,.
\end{equation}

\section{Sensible Models of Light Scalars with Large Quadratic Couplings} \label{sec:relaxed_relaxion_models}

As mentioned in previous sections, we are interested in keeping the scalar naturally light, while also maintaining its quadratic interactions within experimental reach. This should be achieved in conjunction to ensuring that the linear scalar interactions are suppressed.
In this section we shall consider two main constructions that realize such a scenario. 
We begin by briefly reviewing the idea presented in~\cite{Hook:2018jle}, which involves a QCD axion with a $\mathbb Z_N$ symmetry acting on $N$ copies of the \ac{SM}. We then move to describe a realization of the relaxed-relaxion idea~\cite{Banerjee:2020kww}, that shows that relaxion models may yield naively unnaturally-large couplings for a light relaxion. 

\subsection{Quadratic interactions of the naturally light $\mathbb Z_N$ QCD-axion}
It is interesting to note that there is a class of models that naturally leads to sizeable quadratic interactions between the \ac{ULDM} and the \ac{SM} field, with no corresponding linear scalar coupling.
This is precisely the effective theory of a certain kind of a naturally light QCD-axion model, where the axion mass is suppressed relative to the conventional models due to the presence of $N$ copies of the \ac{SM}, which furnish a $\mathbb Z_N$ symmetry~\cite{Hook:2018jle,DiLuzio:2021pxd}\,. 
In this class of models, while the axion couplings to the \ac{SM} fields (our \ac{SM}) follow the conventional models, the axion mass is suppressed by an additional factor of $z^N$, with $z\equiv m_u/m_d\sim 1/2$ given by the up to down quark mass ratio.
Furthermore, as was demonstrated in~\cite{Kim:2022ype}, while at linear order, the axion couplings to the \ac{SM} fields are only to pseudo-scalar operators (thus preserving parity), the axion also possesses quadratic couplings to the hadrons.
This leads to exciting phenomenology as the QCD axion can be searched for in experiments that consider the variation of coupling constants instead of the conventional QCD-axion searches (see {\it e.g.}~\cite{Graham:2013gfa}).
For more information, we refer the reader to~\cite{Kim:2022ype}.

\subsection{Quadratic interactions of a light relaxed-relaxion}
Another possibility to ameliorate the naturalness bound for a light scalar is by relaxing its mass in a dynamical way~\cite{Banerjee:2020kww,Banerjee:2022wzk}. 
Dynamical relaxation of a light scalar is discussed in the context of the relaxion mechanism~\cite{Graham:2015cka}, where the light scalar field scans the Higgs mass parameter starting from some high cut-off down to its measured value. 
As discussed~\cite{Banerjee:2020kww}, due to the small incremental change of the Higgs \ac{VEV} as a function of the scalar field value, the scalar stops at a shallow part of its potential and the stopping point (in the field space) is very close to $\phi/f\sim \pi/2$. 
As a result of the shallowness of the scalar's potential, its mass is suppressed compare to the naive expected value, however the interaction strength with the Higgs/\ac{SM} is not. Thus for low energy observers, the scalar appears to be unnatural although the relaxion mechanism is constructed in a technically-natural way. 

Below we use a similar idea in order to suppress the mass of the \ac{ULDM} field, while maintaining its quadratic interactions observable and suppressing its linear interactions with the \ac{SM}. 
To achieve that we relax the mass of a hidden sector Higgs, $\hat H$, from some cut-off $\Lambda$ to its true \ac{VEV} $\hat v$, and invoke an interaction of the scalar with the \ac{SM} as given in Eq.~\eqref{eq:new_coupling}. 

\subsubsection{Interaction with the SM}

\begin{figure}
	\centering
	\hspace*{0.8 cm}
	\includegraphics[scale=0.3]{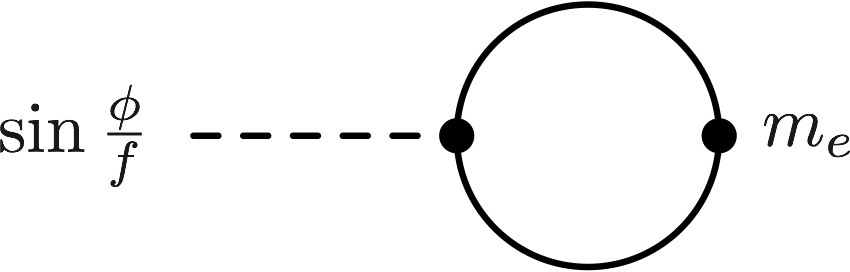}
	\vspace{0.2 cm}
	\textbf{ \huge{}+ }
	\includegraphics[scale=0.3]{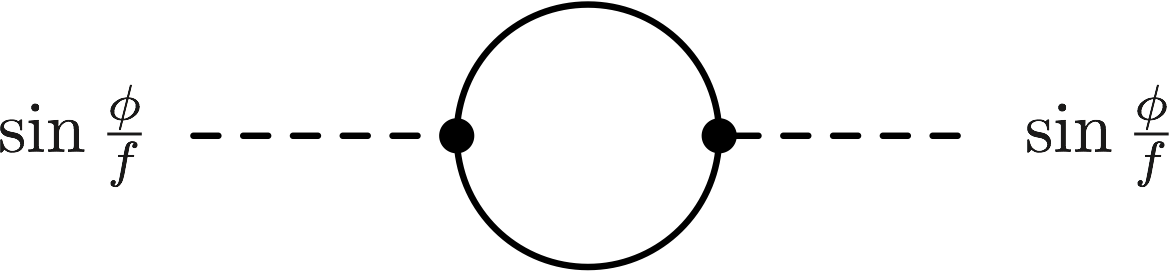}
	\caption{Feynman diagrams for generating Coleman Weinberg for $\phi$ at 1-loop order coming from Eq.~\ref{eq:new_coupling}.}
	\label{fig:coleman_weinberg}
\end{figure}

We assume that the field $\phi$ is coupled to the light \ac{SM} matter as\footnote{In principle there could be some renormalizable portal interaction between the \ac{SM} Higgs and $\hat H$. Thus through the mixing with the hidden Higgs, $\phi$ will also have some interaction with the \ac{SM}. However we consider the strength of such portal interaction to be small and thus this contribution is negligible. }
\begin{equation}\label{eq:new_coupling}
\L_{\text{int }}\supset -\sin\left(\frac{\phi}{f}\right)\left[\sum_{\psi=e,u,d}g_{\phi \psi}m_{\psi}\psi\psi^{c} +\frac{g_{\phi \gamma}}{2} F_{\mu\nu} F^{\mu\nu}  +\frac{g_{\phi g}}{2}G_{\mu\nu}G^{\mu\nu}+\text{h.c.}\right]\,,
\end{equation}
which can be obtained by demanding that the UV completion of this sector includes a linear complex field $\Phi=\leri{\rho+f} \exp\leri{i\frac{\phi}{f}}$, and respects a version of charge conjugation under which $\Phi\rightarrow -\Phi^*$.
At leading order, Eq.~\eqref{eq:new_coupling} generates a quadratic coupling to the light matter in a technically natural way. 
We can calculate the 1-loop potential of $\phi$, generated by the diagrams presented in Fig.~\ref{fig:coleman_weinberg}, similarly to Eq.\eqref{eq:CW_poetntial}, as
\begin{equation}
 V\left(\phi\right)\simeq \frac{1}{16\pi^{2}}\left[\!\sum_{\psi=e,u,d}\!\!\!2\, g_{\phi \psi}\, m_{\psi}^{2}\Lambda_{\rm cut}^2\! \left[\sin\left(\frac{\phi}{f}\right)\!+\!g_{\phi \psi}\sin^{2}\left(\frac{\phi}{f}\right)\right]\!\!+\!\!\sum_{A=\gamma,g}\!\! g_{\phi A}\Lambda_{\rm cut}^4\sin\left(\frac{\phi}{f}\right)\!\right]\!,
\label{eq:coleman_weinberg}
\end{equation}
where $\Lambda_{\rm cut} \gtrsim m_e$ is the momentum cut-off of the loop. 
The overall potential for $\phi$ can be written as,
\begin{align}
V(\phi,\hat{H}) &=  -g\Lambda^3\phi +\leri{\Lambda^2-g\Lambda\phi-\mu_{\rm br}^2\cos\frac{\phi}{f}} |\hat{H}|^2 +|\hat{H}|^4 \nonumber\\
& +\frac{1}{8\pi^{2}}\, \sum_{\psi=e,u,d}g_{\phi \psi} m^2_{\psi}\Lambda_{\rm cut}^2\, \sin\frac{\phi}{f} +\frac{1}{16\pi^{2}}\sum_{A=\gamma,g} g_{\phi A}\Lambda_{\rm cut}^4\sin\frac{\phi}{f} \, ,
\label{eq:relaxion_potential}
\end{align}
$\hat{H}$ belongs to a new hidden sector whose mass is being relaxed from some cut-off $\Lambda$ to its true vacuum expectation value $\hat{v} = \left<\hat{H}\right>$, $g$ is small theory parameter and $\mu_{\rm br}\lesssim \hat{v}$ is the scale of the cosine potential (for more details of this kind of construction see {\it e.g.}~\cite{Graham:2015cka,Banerjee:2020kww}). 

To achieve a successful relaxation of the mass parameter of $\hat{H}$, we require 
\begin{equation}
\mu_{\rm br}^2 \hat{v}^2 \gtrsim \frac{1}{8\pi^{2}}\, g_{\phi \psi}\, m_{\psi}^{2}\, \Lambda_{\rm cut}^2
\label{eq:h_indep_wiggles}\,.
\end{equation} 
Minimizing Eq.~\eqref{eq:relaxion_potential} with respect to both $\hat{H}$ and $\phi$, and using the above constraint, we find the field stopping point as
\begin{equation}\label{eq:stopping_point}
\frac{\phi_0}{f} \simeq \frac{\pi}{2} - \frac{\mu_{\rm br}^2}{ 2\hat{v}^2}-\frac{g_{\phi \psi}\, m_{\psi}^{2}\, \Lambda_{\rm cut}^2}{8\pi^2 \mu_{br}^2 \hat{v}^2}\pm \frac{\mu_{\rm br}}{\Lambda}\equiv \frac{\pi}{2} -\delta_{\pi/2}\,,
\end{equation}
where $\delta_{\pi/2}\ll 1$ is defined as the deviation from $\pi/2$. 
As explained in~\cite{Banerjee:2020kww}, we obtain the mass of the light scalar $\phi$ as, 
\bea
\label{eq:mass_relaxed_relaxion}
m_\phi^2 \approx \frac{\mu_{\rm br}^2 \hat{v}^2}{f^2}\frac{\mu_{\rm br}}{\Lambda} = \frac{\mu_{\rm br}^2 \hat{v}^2}{f^2}\delta \,, 
\eea
which is suppressed compare to the naive expected value of $m_{\rm naive}^2\sim \mu_{\rm br}^2\hat{v}^2/f^2$ by a small parameter defined as $\delta =\mu_{\rm br}/\Lambda\ll 1$. This suppression is a result of the flatness of the effective potential of $\phi$, which characterizes the first point at which the derivative of the backreaction potential, mainly controlled by a periodic function of $\phi$ with a slowly rising amplitude, balances out the constant derivative of the UV term.

Expanding Eq.~\eqref{eq:new_coupling} around the stopping point of $\phi$ gives rise to linear and quadratic couplings of $\phi$ to the \ac{SM} as
\begin{align}
\L_{\rm int} \supset &- \sum_{\psi=e,u,d}g_{\phi \psi}m_{\psi} \psi \psi^c \left(1 - \frac{\delta_{\pi/2}^2}{2}- \frac{\phi^2}{2 f^2} - \delta_{\pi/2} \frac{\phi}{f} \right)
\nonumber \\ 
&  +\left[\frac{g_{\phi \gamma}}{2} F_{\mu\nu} F^{\mu\nu} + \frac{g_{\phi g}}{2}G_{\mu\nu}G^{\mu\nu}\right]\left(1 - \frac{\delta_{\pi/2}^2}{2}- \frac{\phi^2}{2 f^2} - \delta_{\pi/2} \frac{\phi}{f} \right)
\,. 
\label{eq:relaxion_quad-coupling}
\end{align}
Moreover, for a valid weakly coupled theory, we require
\begin{align}
& \delta m_{\psi}\lesssim m_{\psi} \qquad \text{ i.e} \: g_{\phi \psi}\lesssim \mathcal{O}(1)\,,
\\
& \delta \alpha_{(s)}\lesssim\alpha_{(s)} \qquad  \!\! \text{i.e. } \: g_{\phi \gamma (g)}\lesssim \mathcal{O}(1)\, .
\label{eq:e_mass}
\end{align}
Combining Eq.~\eqref{eq:h_indep_wiggles} and Eq.~\eqref{eq:e_mass} we get
\begin{eqnarray}
g_{\phi \psi} \lesssim {\rm Min}\left[1,8 \pi^2\frac{\mu_{\rm br}^2\hat{v}^2}{m_\psi^2\, \Lambda^2}\right]\,,
\\
g_{\phi \gamma (g)} \lesssim  {\rm Min}\left[1,4 \pi^2\frac{\mu_{\rm br}^2\hat{v}^2}{ \Lambda^4}\right]\, .
\label{eq:g_lower_bound}
\end{eqnarray}

\subsubsection{Relaxation of the cutoff}
Consider a model where the field stopping point is not near $\phi_0/f \sim \frac{\pi}{2}$ as in Eq.~\eqref{eq:stopping_point}, but a naive order one stopping point, 
\begin{equation}\label{eq:stopping_point_naive}
	\left( \frac{\phi_0}{f}\right)_{\text{Naive}} \sim \mathcal{O}(1)\,.
\end{equation}
For simplicity, let us focus on the quadratic coupling of $\phi$ to the electron
\begin{equation}
	\L \supset  g_{\phi e} \frac{\phi^2}{f^2}\, m_e \, e e^c\, .
\end{equation}
Given the naive stopping point, this interaction leads to a quadratic contribution to the scalar mass as
\begin{equation}
	\left( \delta m_{\phi}^2\right)_{\text{Naive}}  \simeq g_{\phi e} \,m_e^2\, \frac{\Lambda_{e}^2}{8 \pi^2 f^2}\,,
\end{equation}
where $\Lambda_{e}$ is the cut-off of the electron loop. 
In addition, if the scalar field $\phi$ accounts for the \ac{DM} in the present universe, it acquires a time-dependent background value. Thus, the same coupling would induce temporal variations of the mass of the electron as
\begin{equation}
	\frac{\delta m_e}{m_e}(t) \simeq g_{\phi e} \frac{\rho_{\rm DM}}{f^2 m_{\phi}^2}\,.
\end{equation}
We then obtain a simple relation between the variation of the \acl{FC} and the correction to the mass of $\phi$
\begin{equation}
	\left( \delta m_{\phi}^2\right)_{\text{Naive}} = \left(\frac{\delta m_e}{m_e}\right)(t)  \times \frac{m_{e}^2\, m_{\phi}^2\,  \Lambda_{e}^2}{\rho_{\rm DM}\, 8\pi^2}\,. 
\end{equation}
Without the relaxed relaxion dynamics, a theory involving such interactions can only be natural if 
\begin{equation}
	\label{eq:nat_criteria}
		\left( \delta m_{\phi}^2\right)_{\text{Naive}}\lesssim m_{\phi}^2 \Rightarrow  \frac{\delta m_e}{m_e}(t) \times \frac{\Lambda_{e}^2}{8 \pi^2} \lesssim \frac{\rho_{\rm DM}}{m_{e}^2}\,.\\
\end{equation}
Thus, naively, theories of natural light scalars generating observable variations of constants of nature would imply new physics should appear at relatively low scales as
\begin{align}
     \Lambda_{e} \lesssim 30 \eV\, \left(\frac{\rho_{\rm DM}}{\rho_{\rm DM}^{\odot}}\right)^{1/2} \leri{\frac{10^{-18}}{ \leri{\delta m_e/m_e}^{\rm exp}}}^{1/2}.
     \label{eq:naive_cutoff}
\end{align}
In light of this consideration, we can determine whether a theory described by Eq.~\eqref{eq:relaxion_potential} is natural or unnatural, by quantifying by how much the cut-off of the theory is being relaxed away from the naive estimation above.

In the relaxed scenario, the 1-loop order corrections to the mass of $\phi$ are already taken into account via the effective potential in Eq.~\eqref{eq:relaxion_potential}, yielding the stopping point given in Eq.~\eqref{eq:stopping_point}, and the suppressed mass in Eq.~\eqref{eq:mass_relaxed_relaxion}. Therefore, naturalness does not require that these 1-loop corrections are kept smaller than $m_\phi$. However, for the relaxation to be successful (both of the mass of $\hat{H}$ and of the mass of $\phi$), we must require that the coupling of $\phi$ to the \ac{SM} does not overcome the derivative of the $\hat{H}$-dependent potential of $\phi$ at the first stopping point. This requirement is expressed in Eq.~\eqref{eq:h_indep_wiggles}, and can be re-expressed as
\begin{align}
    \frac{{g}_{\phi e}\, m_e^2\, \Lambda_e^2}{ 8 \pi^2 f^2 }\lesssim \frac{\mu_{\rm br}^2\hat{v}^2}{f^2}\,,
\end{align}
or
\begin{align}
    \leri{\delta m_{\phi}^2}_{\rm{Naive}}\lesssim \frac{m_\phi^2}{\delta}\,,
\end{align}
which is less restrictive than the naive requirement in Eq.~\eqref{eq:nat_criteria} since $\delta\lesssim 1$.
Consequently, the phenomenological effects associated with the coupling of $\phi$ to the \ac{SM} may be enhanced, without significantly altering the mass of the scalar. Namely, the cut-off scale of such a theory $\Lambda_e$, including a light scalar and observable quadratic interactions with the \ac{SM}, is relaxed by a factor of $1/\sqrt{\delta}$ with respect to the naive prediction of Eq.~\eqref{eq:naive_cutoff}.

\subsubsection{A large hierarchy between the linear and quadratic coupling due to the Relaxed Relaxion mechanism}\label{sec:Relaxed_Relaxion_hierarchis_lin_vs_quad}
A large hierarchy between the effective linear and quadratic couplings of $\phi$ can be simply parameterize in terms of the deviation of the scalar's stopping point from $\frac{\pi}{2}$, denoted by $\delta_{\pi/2}$. The different scaling between the dimensionless linear and quadratic couplings is
\begin{equation}
	c_{\text{linear}} \sim  g_{\phi (\psi,A)}\times\delta_{\pi/2},\qquad \text{while} \qquad c_{\text{quadratic}}  \sim  g_{\phi (\psi,A)}\,.
\end{equation}
In our natural theory we consider $g_{\phi (\psi,A)}\sim \mathcal{O}\left( 1\right) $. Moreover, we wish to have natural values of the scaled couplings, which gives new bounds that are absent if only either the linear or the quadratic coupling were to be turned on. For example, in order to match the observed bound of $d_{m_{e}}^{\left(2\right)}\sim10^{14}$, one must require that
\begin{equation}
	\mathcal{O}(1)\sim c_{m_{e}}^{\left(2\right)}=d_{m_{e}}^{\left(2\right)}\times\frac{f^{2}}{M_{\text{pl}}^{2}}\simeq\frac{f^{2}}{\text{GeV}^{2}}\times10^{-23}\;\Rightarrow\; f\simeq3\times 10^{11}\,\text{GeV}\,,
\end{equation}
in order to get $\mathcal{O}(1)$ coupling. 
This, in turn, sets the value of the relaxion dynamical scale, $f$.
For the linear coupling, matching the observed bound on $d_{m_e}^{\left(1\right)}\sim10^{-3}$ translates to
\begin{equation}
	c_{m_{e}}^{\left(1\right)}=d_{m_{e}}^{\left(1\right)}\times\frac{f}{M_{\text{pl}}}\simeq 10^{-10} \sim \delta_{\pi/2}
\end{equation}
which sets the value of $\delta_{\pi/2}$.
Assuming no cancellation between the different terms contributing to $\delta_{\pi/2}$, we get the following constraints:
\begin{equation}\label{eq:deltaPi/2_constraints}
	\begin{cases}
		\frac{\mu_{{\rm br}}^{2}}{\hat{v}^{2}}\lesssim 10^{-10}\,\\
		\frac{\mu_{{\rm br}}}{\Lambda}\lesssim 10^{-10}\,\\
		\frac{g_{\phi\psi}\,m_{\psi}^{2}\,\Lambda_{\rm cut}^{2}}{8\pi^{2}\mu_{br}^{2}\hat{v}^{2}}\lesssim 10^{-10}\,.
	\end{cases}
\end{equation}
The first two constraints from Eq.~\eqref{eq:deltaPi/2_constraints} can be easily satisfied by constructing $\mu_{{\rm br}} \ll \hat{v} , \Lambda$. The last constraint from Eq.~\eqref{eq:deltaPi/2_constraints} can also be achieved. However, it suggests that it is more likely to see an effect of both linear and quadratic couplings to light \ac{SM} matter.

\begin{figure}
	\centering
	\includegraphics[width=1\linewidth]{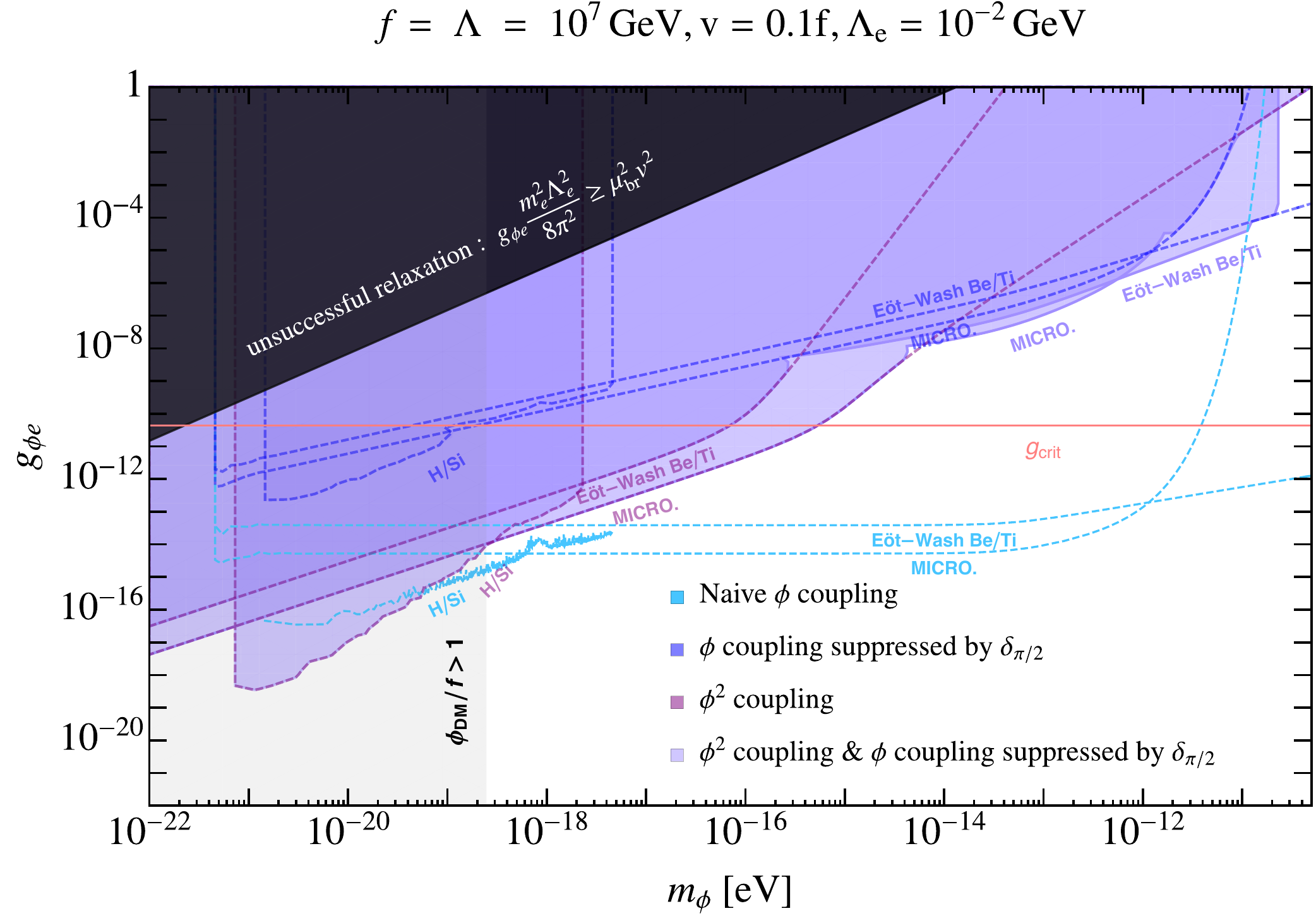}
	\caption{The allowed parameter space of an \ac{ULDM} relaxed-relaxion inspired model with $f=\Lambda=10\,\hat{v}=10^7\text{\,GeV}$, and $\Lambda_{e}=10\,\MeV$. Blue - bounds on a model with only the suppressed linear coupling given, purple - bounds on a model with only the quadratic coupling, light blue - bounds on a model with a naive linear coupling, lilac - bounds given by the combination of the quadratic coupling and the suppressed linear coupling. Pink - the critical coupling for the quadratic interactions. Black - the region in which there is no relaxation of the mass parameter of $\hat{H}$, gray - region in which the \ac{ULDM} amplitude is greater than $f$.}
	\label{fig:f105gevlabdaeme}
\end{figure}

An example of the allowed parameter space for this model, including the theoretical considerations above as well as the experimental bounds, is presented in Fig~\ref{fig:f105gevlabdaeme} for a fixed $f=10\,\hat{v}=10^7\,\text{GeV}$ and a cutoff of $\Lambda_{e}=10\, \text{MeV}$. We plot both the bounds on the quadratic coupling and on the linear coupling (for a generic $\mathcal{O}\leri{1}$ stopping point and for the relaxed-relaxion mechanism) separately, however, since they are not independent in this model, the bound should be drawn from satisfying these constraints simultaneously (shaded lilac region of Fig.~\ref{fig:f105gevlabdaeme}). This yields non-trivial constraints both due to the interdependence of these couplings in this model, making it impossible to turn off the quadratic coupling without turning off the linear coupling as well, and also taking into account the non-linear effects discussed in subsection~\ref{subsec:lin_and_quad_model}. Note that the \ac{EP} constraints involving the quadratic coupling strongly depend on the radius of the source $R_C$ and on the distance from the center of the source $r$. Our bounds are calculated assuming $R_C=r=R_\oplus$ for E\"ot-Wash Be/Ti~\cite{Schlamminger:2007ht} and $R_C=R_\oplus  \,,r=R_\oplus +700$\, km for MICROSCOPE~\cite{PhysRevLett.129.121102}, with $R_\oplus$ being the radius of the Earth. Although we have focused on the electron coupling $g_{\phi e}$, the same analysis can be easily extended to other couplings between $\phi$ and the \ac{SM} fields, such as $g_{\phi\gamma},g_{\phi u(d)}$ and $g_{\phi g}$. 

\section{Conclusions}\label{sec:conclusion}
\acresetall
In this work, we considered a special class of spin-0 \ac{ULDM} models. In this class, the dominant operators describing the interactions between the \ac{ULDM} field $\phi$ and the \ac{SM} are quadratic in the \ac{ULDM} field, namely, of the form $\phi^2\mathcal{O}_{\rm SM}$ where $\mathcal{O}_{\rm SM}$ composed of \ac{SM} fields. 

A dominant quadratic interaction between the \ac{ULDM} field and the \ac{SM} poses two problems to a quantum field theory. First, naive \ac{EFT} power counting suggests that linear \ac{ULDM} couplings would dominate over the quadratic couplings, as the former are associated with effective operators of lower mass-dimensions. Second, generic quadratic interactions generate a large additive contribution to the mass of the \ac{ULDM}, thus resulting in a naturalness problem. 
We discussed theoretical constructs in which these challenges are ameliorated. First, we presented technically-natural models where the dominant interaction between the \ac{ULDM} and the \ac{SM} is in fact quadratic with the \ac{ULDM} field, and the linear (scalar) coupling is either absent or suppressed. Second, with a $\mathbb Z_N$ QCD axion model and a relaxed-relaxion inspired framework, we demonstrated that a large quadratic coupling and a parametrically smaller mass can be achieved naturally.

We have also studied the phenomenology of this class of models, considering various terrestrial and space-based existing and near future experimental probes. Finally, we considered the interplay between indirect searches for the \ac{ULDM} field, related to tests of the violation of the \ac{EP} and the existence of the fifth-force, and the direct searches associated with temporal oscillations of the fundamental constants of nature.

\section*{Acknowledgements}
We thank Kfir Blum, Dmitry Budker, Joshua Eby, Huyngjin Kim, Eric Madge and Yevgeny Stadnik for useful discussions. 
AB, GP, and MS would like to thank the Aspen Center for Physics for its hospitality and support where the part of the work was performed.
The work of AB is supported by the Azrieli foundation. 
The work of GP is supported by grants from BSF-NSF, Friedrich Wilhelm Bessel research award, GIF, ISF, Minerva, SABRA Yeda-Sela WRC Program, the Estate of Emile Mimran, and the Maurice and Vivienne Wohl Endowment. 
The work of MS was supported by the NSF QLCI Award OMA - 2016244, NSF Grant PHY-2012068,  and  European Research Council (ERC) under the European Union’s Horizon 2020 research and innovation program (Grant Number 856415). IS is supported by a fellowship from the Ariane de Rothschild Women Doctoral Program.

\newpage
\appendix

\section{The \ac{EOM} of \ac{DM} for Linearly Coupled model }\label{App:The_solution of the Linear model and a Quadratic}
Let us consider a linearly coupled CP-even scalar \ac{DM} with the \ac{SM} described by Eq.~\eqref{eq:Stadnik_Lint_linear}. 
The \ac{EOM} for the \ac{DM} field $\phi$ can be written as, 
\begin{equation}
\left(\frac{\partial^{2}}{\partial t^{2}}-\nabla\cdot\nabla+m_{\phi}^{2}\right)\phi= J_{\text{source}}\left(r\right)\, ,
\end{equation}
where $J_{\text{source}}\left(r\right) \equiv -\frac{Q^C_id_i^{\left(1\right)}}{M_{\text{Pl}}}\rho_{C}\left(r\right)$ is coming because of the presence of a source term with density $\rho_C$ and dilatonic $Q_i^C$ are the dilatonic charges for the body C, corresponding to the i-th fundamental constant (and corresponding coupling). 
We consider a homogeneous spherically symmetric source satisfies 
\begin{equation}
\rho_{C}\left(r\right)=\begin{cases}
\frac{3M_{C}}{4\pi R_{C}^{3}} & r\leq R_{C}\\
0 & r>R_{C}\,,
\end{cases}\label{eq:stadnik_source}
\end{equation}
where $M_C$ and $R_C$ are the mass and the radius of the body C. 
The solution to the \ac{EOM} for the scalar field $\phi$, in the presence of a homogeneous spherical source, is 
\begin{equation}\label{eq:E.O.M_linear}
\phi\left(t,x\right)=\phi_{0}\cos\left(m_{\phi}t+\delta\right)-\frac{Q^C_id_i^{\left(1\right)}}{M_{\text{Pl}}}\,I\left( m_{\phi} R_C\right) \frac{M_C}{r}\,e^{-m_\phi r}\,,
\end{equation}
where the function $I\left(x\right)$ is given by
\begin{equation}
I\left(x\right) = 3 \dfrac{x \cosh x - \sinh x}{x^3} \: =\begin{cases}
1 & x\ll1\\
\frac{e^{x}}{2x} & x\gg1\,,
\end{cases}
\end{equation}
and $\phi_0=\sqrt{2 \rho_{\rm DM}}/m_\phi$, and $\delta$ is a random phase of the \ac{DM} background. $\rho_{\rm DM}$ is the local \ac{DM} density.  
The sensitivity of \ac{EP} tests can be written as (Eq.~\eqref{eq:etq_EP_linear_Stadnik_background} in the main text)
\begin{align} \label{eq:Sensitivity_EP_linear}
\eta^\text{EP}_{\rm lin}= 2 \frac{|\vec a_{A,C}-\vec a_{B,C}|}{|\vec a_{A,C}+\vec a_{B,C}|}& \approx\frac{|\leri{Q_i^A d^{(1)}_i-Q_i^B d^{(1)}_i}\vec\nabla V_C\leri{\phi}|}{\frac{G M_C}{r^2}}\nonumber\\ 
& \simeq Q^C_j d^{(1)}_j {\left( Q_i^A d^{(1)}_i-Q_i^B d^{(1)}_i\right) }e^{-m_\phi r}\,.
\end{align}

Having a time-varying background field, as in Eq.~\eqref{eq:E.O.M_linear}, induces a small temporal dependence of \aclp{FC} which can be probed by \ac{DDM} searches.  
In \ac{DDM} searches, we compare two systems (say A and B) with sensitivity coefficients $\kappa^{A/B}_i$ corresponding to the i-th fundamental constant. 
The sensitivity of the \ac{DDM} searches can be written as (Eq.~\eqref{eq:etq_DDM_linear_Stadnik_background} in the main text)
\begin{align} \label{eq:Sensitivity_AP_linear}
\eta^\text{DDM}_{\rm lin}=\frac{|\delta Y(t) |}{Y}\approx  \frac{\Delta \kappa_i d_{i}^{(1)}}{M_{\text{Pl}}} \phi_0\, \cos\left(m_{\phi}t+\delta\right) \,,
\end{align}
where $\Delta\kappa_i=\kappa^{A}_i-\kappa^{B}_i$ is the difference of the sensitivity coefficients of a specific transition. 

\section{The EOM of a Quadratically Coupled DM Model }\label{sec:Screening of the Quadratic potential}
The \ac{EOM} for the \ac{DM} field in the quadratic model, described by Eq.~\eqref{eq:Stadnik_Lint_quad}, is
\begin{equation}\label{eq:EOM_quadratic}
\left(\frac{\partial^{2}}{\partial t^{2}}-\nabla\cdot\nabla+\tilde{m}_{\phi}^{2}\left(r\right)\right)\phi=0\, ,
\end{equation}
where we define $\tilde{m}_{\phi}^{2}\left(r\right) \equiv m_{\phi}^{2}+\frac{Q_i^C d_i^{(2)}}{M_{\text{Pl}}^{2}}\rho_{C}\left(r\right)$. We also replace the \ac{SM} matter fields, with the background source density, $\rho_{C}\left(r\right)$, in the presence of a spherically homogeneous source as described in Eq.~\eqref{eq:stadnik_source}. 

The \ac{EOM} yields a solution outside the source body of the form
\begin{equation}\label{eq:phi_quadratic_solution}
\phi\left(t,x\right)=\phi_{0}\cos\left(m_{\phi}t+\delta\right)\left[1-s_{C}^{\left(2\right)}[d^{(2)}_i]\,\frac{GM_{C}}{r}\right]\,.
\end{equation}
In the above equation, $C$ stands for the source body with 
\begin{align}
s^{(2)}_{C}[d^{(2)}_i]= Q^C_i d^{(2)}_i J_{\textrm{sign}[d^{(2)}_i]}\leri{\sqrt{3 Q^C_i d^{(2)}_i \frac{GM_{C}}{R_C}}}\,,
\end{align}
where the function $J_{\textrm{sign}[d^{(2)}_i]}\leri{x}$ is defined in~\cite{Hees:2018fpg} as
\begin{align}
J_+(x) = 3\frac{x-\tanh x}{x^3}\,\,\,{\rm and}\,\,\,J_-(x) = 3\frac{\tan x- x}{x^3}\,.
\end{align}
Note that, we consider the case where all the couplings are positive i.e. $\textrm{sign}[d^{(2)}_i]\geq0$. Also, $G$ and $\Mpl$ can be used interchangeably with $8\pi G=1/\Mpls$. 

\subsection{The criticality of quadratic couplings} \label{subsection:criticality}

As seen from the solution of Eq.~\eqref{eq:phi_quadratic_solution}, for very large positive values of the coupling coefficient, $d^{(2)}_{i}Q^C_i\frac{GM_C}{R_C} \gg 1$, $J_+(x)$ can be approximated as
\begin{equation}
\lim_{x\gg 1} J_+(x) = \frac{3}{x^2} -\frac{3}{x^3}+\mathcal{O}\left(\frac{1}{x^4}\right).
\end{equation}
In the large coupling limit, $s^{(2)}_C[d^{(2)}_i]$ can be written as
\begin{equation}
s^{(2)}_{C}[d^{(2)}_i]= \frac{R_C}{GM_{C}}\left(1-\sqrt{\frac{d_i^{\rm crit}}{d^{(2)}_i}}\right)\,,
\label{eq:scdcrit}
\end{equation}
where we define $d_i^{\rm crit} = R_C/(3 Q^C_i GM_C)$. 
For sub-critical values of the coupling coefficient $d^{(2)}_{i}Q_i^C\frac{GM_c}{R_c} \ll 1$, $J_+(x)\approx 1+ \mathcal{O}\left(x^2\right)$, and $s_{C}[d^{(2)}_i]=Q^C_i d^{(2)}_i$.  
Therefore, the solution of the \ac{EOM}, Eq.~\eqref{eq:phi_quadratic_solution}, can be written as
\begin{equation}
\phi\left(t,x\right)\simeq\phi_{0}\cos\left(m_{\phi}t+\delta\right)	\begin{cases}
\left[1-\frac{R_C}{r}\left(1-\sqrt{\frac{d_i^{\rm crit}}{d^{(2)}_i}}\right)\right]\,{\rm for}\,\,\, d^{(2)}_{i}\gg d_i^{\rm crit} \\
\left[1-\frac{R_C}{r}\frac{d^{(2)}_i}{3\,d_i^{\rm crit}}\right]\,\,{\rm for}\,\,\, d^{(2)}_{i} \ll d_i^{\rm crit}\,.
\end{cases}
\label{eq:phi2sol}
\end{equation} 

The motion of a test mass can be derived from the geodesic path~\cite{Hees:2018fpg}:
\begin{align}\label{eq:worldine_action}
S_T = & \; \int d\tau \, m(\phi)\sqrt{-g(v_\gamma(\tau),v_\gamma(\tau))} \nonumber
\\
= &\;  \int d\tau \,m(\phi)\sqrt{-g_{\mu\nu} \frac{d x^\mu}{d\tau}\frac{d x^\nu}{d\tau}  } \, ,
\end{align}
where $T$ stands for a test-particle. From Eq.~\eqref{eq:worldine_action} we can derive the geodesic path of a test particle as,
\begin{equation}
\left| \vec{a}_T \right|  \simeq  \left| \vec{a}_{\text{Gravity}} \right| - \frac{Q_i^C d^{(2)}_{i}}{2\Mpls} \,\phi \left[ \nabla \phi + \vec{v}\dot{\phi}\right]\,. 
\end{equation}
Let us denote two different test bodies by A and B. The \ac{EP} test, measure the contribution to the violation of the universal free fall from purely gravity theory.
The sensitivity of \ac{EP} tests, which essentially gives the upper bound on the couplings, is then given by (Eq.~\eqref{eq:etq_EP_quadratic_Stadnik_background} in the main text)
\begin{align} \label{eq:Sensitivity_EP_quadratic}
\!\!\!\!
\eta^\text{EP} =2 \frac{|\vec\nabla V_{A,C}-\vec\nabla V_{B,C}|}{|\vec\nabla V_{A,C}+\vec\nabla V_{B,C}|}\simeq s_{C}^{\left(2\right)}[d^{(2)}_i]\leri{\Delta Q}_j^{AB} d^{(2)}_j \frac{\phi_0^2}{2M^2_{\text{Pl}}}\left[1-s_{C}^{\left(2\right)}\frac{GM_{C}}{r}\right],\!\!
\end{align}
where $\leri{\Delta Q}_j^{AB}=\leri{Q_j^A -Q_j^B}$.

The \ac{DDM} sensitivity in terms of the quadratic couplings is given by
\begin{align} \label{eq:Sensitivity_AP_quadratic}
\eta^\text{DDM}=\frac{|\delta Y(t) |}{Y}\simeq  \frac{\Delta \kappa_i d_{i}^{(2)}}{M_{\text{Pl}}^2} \phi_0^2\, \left[1-s_{C}^{\left(2\right)}\frac{GM_{C}}{r}\right]^2\times\css\,,
\end{align}
where $\Delta\kappa_i\equiv \kappa^A_i -\kappa^B_i$ is the difference of the sensitivity coefficients of a specific transition (see {\it e.g.}~\cite{Safronova:2017xyt} and refs. therein), and we have defined $\css=\cos^2\left(m_{\phi}t+\delta\right)$ .
\\
Using Eq.~\eqref{eq:phi2sol}, we can write the \ac{DDM} sensitivity at the surface of the central body, in the small coupling limit $d^{(2)}_{i} \ll d_i^{\rm crit}$, as
\begin{align}
\left. \eta^\text{DDM} \right|_{\left( d^{(2)}_{i} \ll d_i^{\rm crit} ,  r\simeq R_c\right) } & \approx  \frac{\Delta \kappa_i d_{i}^{(2)}}{M_{\text{Pl}}^2} \phi_0^2\, 	\left[1-\frac{R_C}{3\,r}\frac{d^{(2)}_i}{d_i^{\rm crit}}\right]^2\times\css,
\end{align}
and in the large coupling limit, $d^{(2)}_{i} \gg d_i^{\rm crit}$, as 
\begin{align} \label{eq:Sensitivity_AP_quadratic_above_citical}
\left. \eta^\text{DDM} \right|_{\left( d^{(2)}_{i} \gg d_i^{\rm crit} ,  r\simeq R_c\right) }  & \simeq  \frac{\Delta \kappa_i d_{i}^{(2)}}{M_{\text{Pl}}^2} \phi_0^2\, 	\left[1-\frac{R_C}{r}\left(1-\sqrt{\frac{d_i^{\rm crit}}{d^{(2)}_i}}\right)\right]^2\times\css\nonumber
\\ & \simeq
\frac{\Delta \kappa_i} {M_{\text{Pl}}^2} \phi_0^2\, \left[d_i^{\rm crit} +\frac{\Delta r}{R_C}d_{i}^{(2)}\sqrt{\frac{d_i^{\rm crit}}{d^{(2)}_i}} +\left( \frac{\Delta r}{R_C}\right)^2 d_{i}^{(2)} \right]\times\css.
\end{align}
Here we have defined $\Delta r = r-R_c$. 
On the surface of the spherical body, where $\frac{\Delta r}{r} \ll 1$, the first term of Eq.~\eqref{eq:Sensitivity_AP_quadratic_above_citical} does not depend on the coupling $d^{(2)}_i$.
Therefore, above criticality, $d^{(2)}_{i} \gg d_i^{\rm crit}$, the change in \aclp{FC} becomes independent on $d^{(2)}_{i}$ and cannot be translated to bound on the couplings. 

\section{EP bounds from the Classical solution vs Quantum corrections in a quadratic theory}\label{App:Classical_background_of_phi2}

The background value of the quadratically coupled \ac{DM} field, which was discussed in Section~\ref{sec:Screening of the Quadratic potential} and in the main text, yields some variations in the acceleration of a test particle. 
The changes in the acceleration can be approximated by
\begin{align}
\nonumber \left|\Delta a^{\text{Classical}}\right| & \simeq \left| \frac{d^{(2)}_{i}Q_i^A}{\Mpls} \,\phi \left[ \nabla \phi + \vec{v}\dot{\phi}\right] \right| 
\\
&\simeq d^{(2)}_i Q_i^A \frac{\phi_0^2}{M^2_{\text{Pl}}} s_{C}^{\left(2\right)}\frac{GM_{C}}{r^2} \left[1-s_{C}^{\left(2\right)}\frac{GM_{C}}{r}\right]\,,
\label{eq:acceleration_Stadnik1}
\end{align}
where $Q_i^A$ is the dilatonic charge of the test body. 
As discussed before, for relatively large couplings, $d^{(2)}_i \gg d_i^{\rm crit}$, Eq.~\eqref{eq:acceleration_Stadnik1} can be approximated by,
\begin{equation}
\left|\Delta a^{\text{Classical}}\right| \simeq  d^{(2)}_i Q_i^A \frac{\phi_0^2}{M^2_{\text{Pl}}} \frac{R_C}{r^2} \left[\frac{\Delta r}{R_c} + \epsilon\right] \, , \qquad  \text{with}\;\;\; \Delta r \equiv r-R_c \, .\label{eq:acceleration_Stadnik2}
\end{equation}
Thus, for distances $r\simeq R_c$ there exists a suppression of the classics potential. 
$\epsilon\ll1$ is higher order the sub-leading corrections of $\mathcal{O}((\Delta r)^2,\sqrt{d_i^{\rm crit}/d^{(2)}_i})$. 
If we consider an electron as the test body, the above formula can be simplified to, 
\begin{equation}
\left|\Delta a^{\text{Classical}}\right| \simeq  d^{(2)}_{m_e} \frac{\phi_0^2}{M^2_{\text{Pl}}} \frac{R_C}{r^2} \left[\frac{\Delta r}{R_c} + \epsilon\right] \,.\label{eq:acceleration_Stadnik_3}
\end{equation}

\begin{figure}[h!]
	\centering
	\hspace*{0.8 cm}
	\includegraphics[scale=0.3]{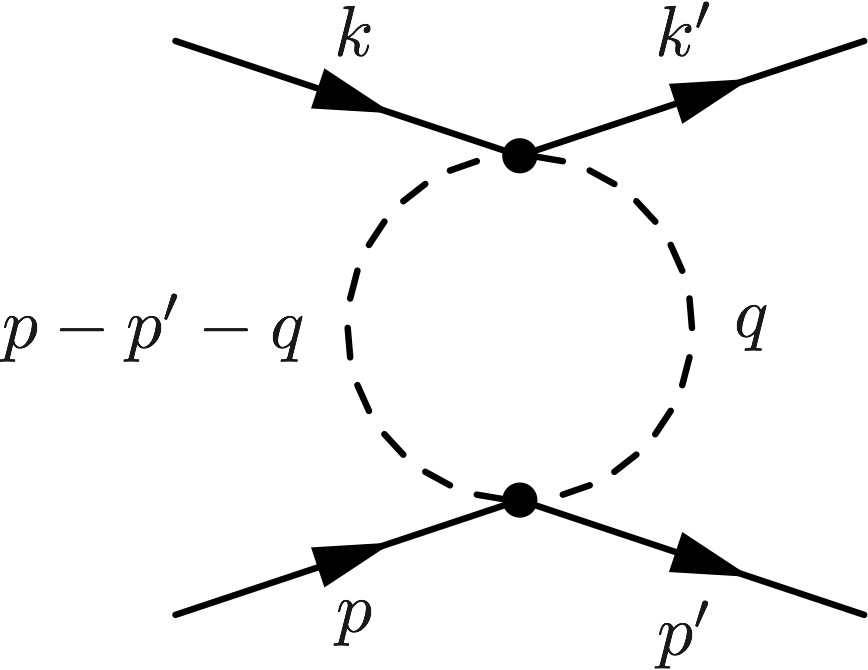}
	\vspace{0.2 cm}
	\textbf{\qquad \huge{+} \;}
	\includegraphics[scale=0.3]{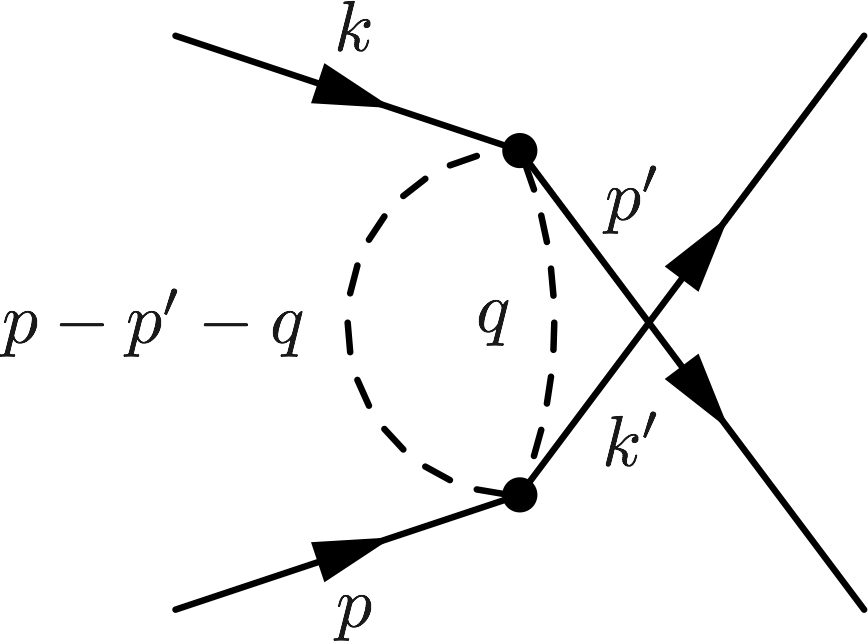}
	\caption{1-loop Feynman diagrams that generates the non relativistic potential}
	\label{fig:1-loop_diagrams_for_potential}
\end{figure}

We can compare the classical effect to the quantum corrections from a $\phi^2$ exchange. The quantum potential calculated in the non relativistic limit, $M_\text{body}\gg |\textbf{q}|$, where $M_\text{body}$ is the mass of the test particle, while $|\textbf{q}|$ is the momentum transfer by the interaction/potential. In the Born approximation, the scattering amplitude in the non-relativistic limit, can be mapped to the potential by
\begin{equation}
i{\cal M}\equiv\left\langle p'|iT|p\right\rangle =-i\tilde{V}\left(\mathbf{q}\right)2\pi\delta\left(E_{\mathbf{p}}-E_{\mathbf{p'}}\right)\, , 	
\end{equation} 
where $i{\cal M}$ is the scattering amplitude from a state with momentum $p$ to a state with momentum $p'$, and $\tilde{V}\left(\mathbf{q}\right)$ is the non-relativistic potential in 3d-momentum space. Assuming conservation of spin and other quantum numbers, the non-relativistic amplitude can be approximated as
\begin{equation}
i{\cal M} = -4m^2G\left(p-p'\right)
\end{equation}
where $G\left(p-p'\right)$ is the value of amputated diagram. In our $\phi^2$ coupling scenario, the leading diagrams are 1-loop
order, as in Fig.~\ref{fig:1-loop_diagrams_for_potential}.
\\
Performing a contour integral to evaluate the integral over $|\textbf{q}|$ in the complex plane yields the potential between an electron and a heavy central body, 
\begin{align}
V\left(\mathbf{r}\right)=-\int\frac{d^{3}\mathbf{q}}{\left(2\pi\right)^{3}}G_{1-loop}\left(\mathbf{q}\right)e^{i\mathbf{q}\cdot\mathbf{x}}
=
\begin{cases}
-\frac{(d^{(2)}_{m_e})^2Q^C_{m_e} m_{e}M_C}{(2\Mpls)^264\pi^{3}r^{2}}\frac{\sqrt{m_{\phi}}e^{-2m_{\phi}r}}{\sqrt{r}} & m_{\phi}r\gg1\\
-\frac{(d^{(2)}_{m_e})^2Q^C_{m_e}m_e M_C}{(2\Mpls)^2\,64\pi^{3}r^{3}} & m_{\phi}r\ll1\,.
\end{cases}
\end{align}
In the limit $m_{\phi}r\gg1$, the potential is exponentially suppressed by $e^{-2m_{\phi}r}$ and $\frac{\sqrt{m_\phi}}{r^5/2}$, and thus negligible. 
We can also show that in the other limit i.e. $m_{\phi}r\ll1$, the quantum potential is also negligible compare to the classical one. 
To see this, one can compare the accelerations from quantum and classical effects. The acceleration from quantum effect can be written as,  
\begin{equation}
\lim_{m_{\phi}r\ll1 }\left|\Delta a^{\text{Quantum}}\right| \simeq
\left|\frac{3 (d^{(2)}_{m_e})^2Q^C_{m_e} M_C}{256\pi^3\Mpl^4 r^{4}}\right|\, . \label{eq:quantum_acceleration}
\end{equation}
Comparing Eq.~\eqref{eq:acceleration_Stadnik_3} to Eq.~\eqref{eq:quantum_acceleration} and using $\phi_0=\sqrt{2\rho_{\rm DM}}/m_\phi$ we find
\bea
\lim_{m_{\phi}r\ll1 }\left|\dfrac{\Delta a^{\text{Classical}}}{\Delta a^{\text{Quantum}}}\right| &\simeq & \frac{512\, \pi^{3}}{3\,d^{(2)}_{m_e} Q^C_{m_e}}\frac{\rho_{\text{DM}}\Mpls}{m_{\phi}^{2}} \frac{R_C}{M_C} r^{2} \left[\frac{\Delta r}{R_c}\right]  \nonumber\\ 
&\simeq & 10^{50}\left[\frac{\Delta r}{R_c}\right]\left(\frac{10^{-14}\,\text{eV}}{m_{\phi}}\right)^{2}\left(\frac{\rho_{\rm DM}}{\rho_\text{DM}^\odot}\right)
\,,
\eea
where, we take $r\simeq R_C$, and use the Earth as the source body with $Q^C_{m_e}\simeq 2.7\times 10^{-4}$. 
For the coupling, we use the current bound of $d^{(2)}_{m_e}\sim 10^{14}$ at the considered mass as shown in Fig.~\ref{fig:dme_AP_EP}.  
Thus, even if one is very close to the surface of the central body (very close to the surface of the Earth), the classical effects are still dominating compare to the quantum ones. This is expected from the classical behavior of the scalar field, with a large occupation number.

\section{The Clockwork model}\label{app:Clockwork}
Consider $N+1$ complex scalar fields, $\Phi_{0}$ to $\Phi_{N}$, where each field represents a site on a lattice in the theory (field) space. The potential of these uncoupled sites is
\begin{equation}
	V_{\text{clock}}\left(\left\{ \Phi_{j}\right\} \right)=\sum_{i=0}^{N}\left(-m_{i}^{2}\Phi_{i}^{\dagger}\Phi_{i}+\lambda_{i}\left|\Phi_{i}^{\dagger}\Phi_{i}\right|^{2}\right)\,.
\end{equation}
For simplicity, we consider the case where all couplings at different sites scale the same, i.e.
\begin{equation}
	\forall_{i}:\;\; m_{i}^{2}\simeq m^{2}>0\,,\qquad\lambda_{i}\simeq\lambda\,.
\end{equation}
Small deviation from the above alignment will have a negligible effect
on the following analysis.

The clockwork potential has a $U\!\left(1\right)^{N+1}$ global symmetry
which is broken spontaneously by $N+1$ \acp{VEV} of the complex scalar
fields, $\left\langle \Phi_{i}\right\rangle \equiv\frac{f}{\sqrt{2}}=\sqrt{\frac{m^{2}}{2\lambda}}$.
In addition to the above potential, we explicitly break the global
$U\!\left(1\right)^{N+1}$ symmetry to a single $U\!\left(1\right)$,
by connecting neighboring sites through a small parameter, $\left|\epsilon\right|<1$,
\begin{equation}
	\Delta V_{\text{clock}}\left(\left\{ \Phi_{j}\right\} \right)=-\sum_{j=0}^{N-1}\left(\epsilon\Phi_{j}^{\dagger}\Phi_{j+1}^{3}+\text{h.c.}\right)\,.\label{eq:deltaV_clock}
\end{equation}
Expanding all the scalars around their \acp{VEV},
\[
\Phi_{j}=\frac{1}{\sqrt{2}}(f+\rho_{j})U_{j}\,,\qquad U_{j}=\text{e}^{i\pi_{j}/f}\,,
\]
and taking the limit $\epsilon\ll\lambda\sim1$, such that the radial
modes can be decoupled, we get the following potential for the compact
degrees of freedom,
\[
\Delta V_{\text{clock}}\left(\left\{ \pi_{j}\right\} \right)=-\frac{\epsilon f^{4}}{2}\sum_{j=0}^{N-1}\cos\left(\frac{3\pi_{j+1}-\pi_{j}}{f}\right)\,.
\]
The above potential respects a $U\!\left(1\right)$ symmetry under
which the fields, $\Phi_{0},\Phi_{1}...\Phi_{N}$, have charges $Q=1,\frac{1}{3}...,\frac{1}{3^{N}}$.
The potential induces a mass matrix of the $N+1$ \ac{pNGB}s,
\[
M_{\left(\pi\right)}^{2}=\epsilon\frac{f^{2}}{2}\begin{pmatrix}1 & -3 & 0 & \cdots &  & 0\\
	-3 & 10 & -3 & \cdots &  & 0\\
	0 & -3 & 10 & \cdots &  & 0\\
	\vdots & \vdots & \vdots & \ddots &  & \vdots\\
	&  &  &  & 10 & -3\\
	0 & 0 & 0 & \cdots & -3 & 3^{2}
\end{pmatrix}\,.
\]
The matrix $M_{\left(\pi\right)}^{2}$ becomes diagonal in the field
basis $\phi_{j}$ ($j=0,\,...,N$), related to the basis $\pi_{j}$
by a real $(N+1)\times(N+1)$ orthogonal matrix $O$,
\[
\pi=O\phi\,,~~~~~~O^{T}M^{2}O={\rm diag}\left(m_{\phi_{0}}^{2},\dots,m_{\phi_{N}}^{2}\right)\, .
\]
The eigenvalues of the mass matrix are given by
\begin{equation}
	m_{\phi_{0}}^{2}=0\,,~~~m_{\phi_{k}}^{2}=\lambda_{k}\,\epsilon\frac{f^{2}}{2}\,,~~~~~k=1,..,N\,,\label{eq:clockwork_mass}
\end{equation}
\begin{equation}
	\lambda_{k}\equiv10-6\cos\frac{k\pi}{N\!+\!1}\,.
\end{equation}
The projection of the \ac{pNGB} $\pi_{j}$, into the scalars in the mass
basis $\phi_{k}$, are given by the elements of the rotation matrix,
\[
O_{j0}=\frac{{\cal N}_{0}}{3^{j}}\,,~~~O_{jk}={\cal N}_{k}\left[3\sin\frac{jk\pi}{N\!+\!1}-\sin\frac{\left(j+1\right)k\pi}{N\!+\!1}\right]\,,~~~j=0,..,N\,;~~k=1,..,N
\]
\begin{equation}
	{\cal N}_{0}\equiv\sqrt{\frac{9-1}{9-3^{-2N}}}\:,~~~~{\cal N}_{k}\equiv\sqrt{\frac{2}{(N\!+\!1)\lambda_{k}}}\:.
\end{equation}
As expected by the breaking of $\left[U\left(1\right)\right]^{N+1}\rightarrow U\left(1\right)$
, there are $N$ massive scalars $\phi_{k},\;\left(k=1,...,N\right)$,
and a single mass-less Goldstone whose wave function is exponentially peaked on the first site,
\begin{equation}
	\phi\equiv\phi_{0}={\cal N}_{0}\sum_{j=0}^{N}\frac{1}{3^{j}}\pi_{j}\,.
\end{equation}
For $N\gg1$ the normalization constant is ${\cal N}_{0}(N)\approx\sqrt{8/9}$.
The overlap between the mass-less eigenstate $\phi$ and the site $j$
is $\langle\pi_{j}|\phi\rangle\approx1/3^{j}$. Under the spontaneously
broken $U\!\left(1\right)_{\text{clock}}$ symmetry, $\phi$ and the
other compact fields transform as follows,
\begin{align}
	U\!\left(1\right)_{\text{clock}}:\qquad & \pi_{j}\mapsto\pi_{j}+\frac{1}{3^{j}}f\alpha\ , 
	\nonumber \\
	& \phi\mapsto\phi+f\alpha\ ,\label{eq:clockwork_charges}
\end{align}
where $\alpha\in\left[0,2\pi\times3^{N}\right]$. Introducing an explicit
breaking of the global $U\!\left(1\right)_{\text{clock}}$ at the
site $j$ would generate a potential for $\phi$ with periodicity
of order $3^{j}f$. Therefore, by placing the backreaction sector
at the $0^{\text{th}}$ site and the rolling sector
at the $N^{\text{th}}$ site, we obtain the desired hierarchy between
the periodicities, $F/f\approx3^{N}$.

\bibliographystyle{JHEP}
\bibliography{phi2}

\providecommand{\href}[2]{#2}\begingroup\raggedright\begin{thebibliography}{100}

\bibitem{Catena:2009mf}
R.~Catena and P.~Ullio, \emph{{A novel determination of the local dark matter
  density}}, \href{https://doi.org/10.1088/1475-7516/2010/08/004}{\emph{JCAP}
  {\bfseries 08} (2010) 004} [\href{https://arxiv.org/abs/0907.0018}{{\ttfamily
  0907.0018}}].

\bibitem{Graham:2013gfa}
P.~W. Graham and S.~Rajendran, \emph{{New Observables for Direct Detection of
  Axion Dark Matter}},
  \href{https://doi.org/10.1103/PhysRevD.88.035023}{\emph{Phys. Rev. D}
  {\bfseries 88} (2013) 035023}
  [\href{https://arxiv.org/abs/1306.6088}{{\ttfamily 1306.6088}}].

\bibitem{Arvanitaki:2014faa}
A.~Arvanitaki, J.~Huang and K.~Van~Tilburg, \emph{{Searching for dilaton dark
  matter with atomic clocks}},
  \href{https://doi.org/10.1103/PhysRevD.91.015015}{\emph{Phys. Rev.}
  {\bfseries D91} (2015) 015015}
  [\href{https://arxiv.org/abs/1405.2925}{{\ttfamily 1405.2925}}].

\bibitem{Graham:2015ifn}
P.~W. Graham, D.~E. Kaplan, J.~Mardon, S.~Rajendran and W.~A. Terrano,
  \emph{{Dark Matter Direct Detection with Accelerometers}},
  \href{https://doi.org/10.1103/PhysRevD.93.075029}{\emph{Phys. Rev.}
  {\bfseries D93} (2016) 075029}
  [\href{https://arxiv.org/abs/1512.06165}{{\ttfamily 1512.06165}}].

\bibitem{Banerjee:2018xmn}
A.~Banerjee, H.~Kim and G.~Perez, \emph{{Coherent relaxion dark matter}},
  \href{https://doi.org/10.1103/PhysRevD.100.115026}{\emph{Phys. Rev.}
  {\bfseries D100} (2019) 115026}
  [\href{https://arxiv.org/abs/1810.01889}{{\ttfamily 1810.01889}}].

\bibitem{Kolb:1990vq}
E.~W. Kolb and M.~S. Turner, \emph{{The Early Universe}}, vol.~69. 1990,
  \href{https://doi.org/10.1201/9780429492860}{10.1201/9780429492860}.

\bibitem{Bar:2018acw}
N.~Bar, D.~Blas, K.~Blum and S.~Sibiryakov, \emph{{Galactic rotation curves
  versus ultralight dark matter: Implications of the soliton-host halo
  relation}}, \href{https://doi.org/10.1103/PhysRevD.98.083027}{\emph{Phys.
  Rev. D} {\bfseries 98} (2018) 083027}
  [\href{https://arxiv.org/abs/1805.00122}{{\ttfamily 1805.00122}}].

\bibitem{Bar:2019bqz}
N.~Bar, K.~Blum, J.~Eby and R.~Sato, \emph{{Ultralight dark matter in disk
  galaxies}}, \href{https://doi.org/10.1103/PhysRevD.99.103020}{\emph{Phys.
  Rev. D} {\bfseries 99} (2019) 103020}
  [\href{https://arxiv.org/abs/1903.03402}{{\ttfamily 1903.03402}}].

\bibitem{Arvanitaki:2014wva}
A.~Arvanitaki, M.~Baryakhtar and X.~Huang, \emph{{Discovering the QCD Axion
  with Black Holes and Gravitational Waves}},
  \href{https://doi.org/10.1103/PhysRevD.91.084011}{\emph{Phys. Rev. D}
  {\bfseries 91} (2015) 084011}
  [\href{https://arxiv.org/abs/1411.2263}{{\ttfamily 1411.2263}}].

\bibitem{Dine:2022mjw}
M.~Dine, \emph{{The Problem of Axion Quality: A Low Energy Effective Action
  Perspective}},  \href{https://arxiv.org/abs/2207.01068}{{\ttfamily
  2207.01068}}.

\bibitem{Banerjee:2022wzk}
A.~Banerjee, J.~Eby and G.~Perez, \emph{{From axion quality and naturalness
  problems to a high-quality ZN QCD relaxion}},
  \href{https://doi.org/10.1103/PhysRevD.107.115011}{\emph{Phys. Rev. D}
  {\bfseries 107} (2023) 115011}
  [\href{https://arxiv.org/abs/2210.05690}{{\ttfamily 2210.05690}}].

\bibitem{Kamionkowski:1992mf}
M.~Kamionkowski and J.~March-Russell, \emph{{Planck scale physics and the
  Peccei-Quinn mechanism}},
  \href{https://doi.org/10.1016/0370-2693(92)90492-M}{\emph{Phys. Lett.}
  {\bfseries B282} (1992) 137}
  [\href{https://arxiv.org/abs/hep-th/9202003}{{\ttfamily hep-th/9202003}}].

\bibitem{Barr:1992qq}
S.~M. Barr and D.~Seckel, \emph{{Planck scale corrections to axion models}},
  \href{https://doi.org/10.1103/PhysRevD.46.539}{\emph{Phys. Rev.} {\bfseries
  D46} (1992) 539}.

\bibitem{Davidi:2017gir}
O.~Davidi, R.~S. Gupta, G.~Perez, D.~Redigolo and A.~Shalit, \emph{{Nelson-Barr
  relaxion}}, \href{https://doi.org/10.1103/PhysRevD.99.035014}{\emph{Phys.
  Rev.} {\bfseries D99} (2019) 035014}
  [\href{https://arxiv.org/abs/1711.00858}{{\ttfamily 1711.00858}}].

\bibitem{Davidi:2018sii}
O.~Davidi, R.~S. Gupta, G.~Perez, D.~Redigolo and A.~Shalit, \emph{{The
  hierarchion, a relaxion addressing the Standard Model?s hierarchies}},
  \href{https://doi.org/10.1007/JHEP08(2018)153}{\emph{JHEP} {\bfseries 08}
  (2018) 153} [\href{https://arxiv.org/abs/1806.08791}{{\ttfamily
  1806.08791}}].

\bibitem{Calmet:2021iid}
X.~Calmet and F.~Kuipers, \emph{{Implications of quantum gravity for dark
  matter}}, \href{https://doi.org/10.1142/S0218271821420049}{\emph{Int. J. Mod.
  Phys. D} {\bfseries 30} (2021) 2142004}
  [\href{https://arxiv.org/abs/2107.13529}{{\ttfamily 2107.13529}}].

\bibitem{Perez:2020dbw}
G.~Perez and A.~Shalit, \emph{{High quality Nelson-Barr solution to the strong
  CP problem with $\theta=\pi$}},
  \href{https://doi.org/10.1007/JHEP02(2021)118}{\emph{JHEP} {\bfseries 02}
  (2021) 118} [\href{https://arxiv.org/abs/2010.02891}{{\ttfamily
  2010.02891}}].

\bibitem{Choi:1985cb}
K.~Choi and J.~E. Kim, \emph{{DYNAMICAL AXION}},
  \href{https://doi.org/10.1103/PhysRevD.32.1828}{\emph{Phys. Rev. D}
  {\bfseries 32} (1985) 1828}.

\bibitem{Kim:1984pt}
J.~E. Kim, \emph{{A COMPOSITE INVISIBLE AXION}},
  \href{https://doi.org/10.1103/PhysRevD.31.1733}{\emph{Phys. Rev. D}
  {\bfseries 31} (1985) 1733}.

\bibitem{Contino:2021ayn}
R.~Contino, A.~Podo and F.~Revello, \emph{{Chiral models of composite axions
  and accidental Peccei-Quinn symmetry}},
  \href{https://doi.org/10.1007/JHEP04(2022)180}{\emph{JHEP} {\bfseries 04}
  (2022) 180} [\href{https://arxiv.org/abs/2112.09635}{{\ttfamily
  2112.09635}}].

\bibitem{Dine:2012sla}
M.~Dine, T.~Banks and S.~Sachdev, eds., \emph{{Proceedings, Theoretical
  Advanced Study Institute in Elementary Particle Physics (TASI 2010). String
  Theory and Its Applications: From meV to the Planck Scale}: {Boulder,
  Colorado, USA, June 1-25, 2010}}, (Singapore), World Scientific, 2012.
\newblock 10.1142/8153.

\bibitem{Hook:2018dlk}
A.~Hook, \emph{{TASI Lectures on the Strong CP Problem and Axions}},
  {\emph{PoS} {\bfseries TASI2018} (2019) 004}
  [\href{https://arxiv.org/abs/1812.02669}{{\ttfamily 1812.02669}}].

\bibitem{Oswald:2021vtc}
R.~Oswald et~al., \emph{{Search for Dark-Matter-Induced Oscillations of
  Fundamental Constants Using Molecular Spectroscopy}},
  \href{https://doi.org/10.1103/PhysRevLett.129.031302}{\emph{Phys. Rev. Lett.}
  {\bfseries 129} (2022) 031302}
  [\href{https://arxiv.org/abs/2111.06883}{{\ttfamily 2111.06883}}].

\bibitem{WRESL2}
O.~Tretiak, X.~Zhang, N.~L. Figueroa, D.~Antypas, A.~Brogna, A.~Banerjee
  et~al., \emph{{Improved Bounds on Ultralight Scalar Dark Matter in the
  Radio-Frequency Range}},
  \href{https://doi.org/10.1103/PhysRevLett.129.031301}{\emph{Phys. Rev. Lett.}
  {\bfseries 129} (2022) 031301}
  [\href{https://arxiv.org/abs/2201.02042}{{\ttfamily 2201.02042}}].

\bibitem{Froggatt:1978nt}
C.~D. Froggatt and H.~B. Nielsen, \emph{{Hierarchy of Quark Masses, Cabibbo
  Angles and CP Violation}},
  \href{https://doi.org/10.1016/0550-3213(79)90316-X}{\emph{Nucl. Phys.}
  {\bfseries B147} (1979) 277}.

\bibitem{Gelmini:1980re}
G.~B. Gelmini and M.~Roncadelli, \emph{{Left-Handed Neutrino Mass Scale and
  Spontaneously Broken Lepton Number}},
  \href{https://doi.org/10.1016/0370-2693(81)90559-1}{\emph{Phys. Lett. B}
  {\bfseries 99} (1981) 411}.

\bibitem{Kim:1979if}
J.~E. Kim, \emph{{Weak Interaction Singlet and Strong CP Invariance}},
  \href{https://doi.org/10.1103/PhysRevLett.43.103}{\emph{Phys. Rev. Lett.}
  {\bfseries 43} (1979) 103}.

\bibitem{Shifman:1979if}
M.~A. Shifman, A.~I. Vainshtein and V.~I. Zakharov, \emph{{Can Confinement
  Ensure Natural CP Invariance of Strong Interactions?}},
  \href{https://doi.org/10.1016/0550-3213(80)90209-6}{\emph{Nucl. Phys. B}
  {\bfseries 166} (1980) 493}.

\bibitem{Zhitnitsky:1980tq}
A.~R. Zhitnitsky, \emph{{On Possible Suppression of the Axion Hadron
  Interactions. (In Russian)}}, {\emph{Sov. J. Nucl. Phys.} {\bfseries 31}
  (1980) 260}.

\bibitem{Dine:1981rt}
M.~Dine, W.~Fischler and M.~Srednicki, \emph{{A Simple Solution to the Strong
  CP Problem with a Harmless Axion}},
  \href{https://doi.org/10.1016/0370-2693(81)90590-6}{\emph{Phys. Lett. B}
  {\bfseries 104} (1981) 199}.

\bibitem{Graham:2015cka}
P.~W. Graham, D.~E. Kaplan and S.~Rajendran, \emph{{Cosmological Relaxation of
  the Electroweak Scale}},
  \href{https://doi.org/10.1103/PhysRevLett.115.221801}{\emph{Phys. Rev. Lett.}
  {\bfseries 115} (2015) 221801}
  [\href{https://arxiv.org/abs/1504.07551}{{\ttfamily 1504.07551}}].

\bibitem{Ema:2016ops}
Y.~Ema, K.~Hamaguchi, T.~Moroi and K.~Nakayama, \emph{{Flaxion: a minimal
  extension to solve puzzles in the standard model}},
  \href{https://doi.org/10.1007/JHEP01(2017)096}{\emph{JHEP} {\bfseries 01}
  (2017) 096} [\href{https://arxiv.org/abs/1612.05492}{{\ttfamily
  1612.05492}}].

\bibitem{Brzeminski:2020uhm}
D.~Brzeminski, Z.~Chacko, A.~Dev and A.~Hook, \emph{{Time-varying fine
  structure constant from naturally ultralight dark matter}},
  \href{https://doi.org/10.1103/PhysRevD.104.075019}{\emph{Phys. Rev. D}
  {\bfseries 104} (2021) 075019}
  [\href{https://arxiv.org/abs/2012.02787}{{\ttfamily 2012.02787}}].

\bibitem{Flacke:2016szy}
T.~Flacke, C.~Frugiuele, E.~Fuchs, R.~S. Gupta and G.~Perez,
  \emph{{Phenomenology of relaxion-Higgs mixing}},
  \href{https://doi.org/10.1007/JHEP06(2017)050}{\emph{JHEP} {\bfseries 06}
  (2017) 050} [\href{https://arxiv.org/abs/1610.02025}{{\ttfamily
  1610.02025}}].

\bibitem{Choi:2016luu}
K.~Choi and S.~H. Im, \emph{{Constraints on Relaxion Windows}},
  \href{https://doi.org/10.1007/JHEP12(2016)093}{\emph{JHEP} {\bfseries 12}
  (2016) 093} [\href{https://arxiv.org/abs/1610.00680}{{\ttfamily
  1610.00680}}].

\bibitem{Banerjee:2020kww}
A.~Banerjee, H.~Kim, O.~Matsedonskyi, G.~Perez and M.~S. Safronova,
  \emph{{Probing the Relaxed Relaxion at the Luminosity and Precision
  Frontiers}}, \href{https://doi.org/10.1007/JHEP07(2020)153}{\emph{JHEP}
  {\bfseries 07} (2020) 153}
  [\href{https://arxiv.org/abs/2004.02899}{{\ttfamily 2004.02899}}].

\bibitem{Hees:2018fpg}
A.~Hees, O.~Minazzoli, E.~Savalle, Y.~V. Stadnik and P.~Wolf, \emph{{Violation
  of the equivalence principle from light scalar dark matter}},
  \href{https://doi.org/10.1103/PhysRevD.98.064051}{\emph{Phys. Rev. D}
  {\bfseries 98} (2018) 064051}
  [\href{https://arxiv.org/abs/1807.04512}{{\ttfamily 1807.04512}}].

\bibitem{Lin:2019uvt}
T.~Lin, \emph{{Dark matter models and direct detection}},
  \href{https://doi.org/10.22323/1.333.0009}{\emph{PoS} {\bfseries 333} (2019)
  009} [\href{https://arxiv.org/abs/1904.07915}{{\ttfamily 1904.07915}}].

\bibitem{Alexander:2016aln}
J.~Alexander et~al., \emph{{Dark Sectors 2016 Workshop: Community Report}},  8,
  2016, \href{https://arxiv.org/abs/1608.08632}{{\ttfamily 1608.08632}}.

\bibitem{Hook:2018jle}
A.~Hook, \emph{{Solving the Hierarchy Problem Discretely}},
  \href{https://doi.org/10.1103/PhysRevLett.120.261802}{\emph{Phys. Rev. Lett.}
  {\bfseries 120} (2018) 261802}
  [\href{https://arxiv.org/abs/1802.10093}{{\ttfamily 1802.10093}}].

\bibitem{DiLuzio:2021pxd}
L.~Di~Luzio, B.~Gavela, P.~Quilez and A.~Ringwald, \emph{{An even lighter QCD
  axion}}, \href{https://doi.org/10.1007/JHEP05(2021)184}{\emph{JHEP}
  {\bfseries 05} (2021) 184}
  [\href{https://arxiv.org/abs/2102.00012}{{\ttfamily 2102.00012}}].

\bibitem{Kim:2022ype}
H.~Kim and G.~Perez, \emph{{Oscillations of atomic energy levels induced by QCD
  axion dark matter}},  \href{https://arxiv.org/abs/2205.12988}{{\ttfamily
  2205.12988}}.

\bibitem{Safronova:2017xyt}
M.~S. Safronova, D.~Budker, D.~DeMille, D.~F.~J. Kimball, A.~Derevianko and
  C.~W. Clark, \emph{{Search for New Physics with Atoms and Molecules}},
  \href{https://doi.org/10.1103/RevModPhys.90.025008}{\emph{Rev. Mod. Phys.}
  {\bfseries 90} (2018) 025008}
  [\href{https://arxiv.org/abs/1710.01833}{{\ttfamily 1710.01833}}].

\bibitem{Antypas:2019yvv}
D.~Antypas, D.~Budker, V.~V. Flambaum, M.~G. Kozlov, G.~Perez and J.~Ye,
  \emph{{Fast apparent oscillations of fundamental constants}},
  \href{https://doi.org/10.1002/andp.201900566}{\emph{Annalen Phys.} {\bfseries
  532} (2020) 1900566} [\href{https://arxiv.org/abs/1912.01335}{{\ttfamily
  1912.01335}}].

\bibitem{Fischbach:1996eq}
E.~Fischbach and C.~Talmadge, \emph{{Ten years of the fifth force}},  in
  \emph{{31st Rencontres de Moriond: Dark Matter and Cosmology, Quantum
  Measurements and Experimental Gravitation}}, pp.~443--451, 1996,
  \href{https://arxiv.org/abs/hep-ph/9606249}{{\ttfamily hep-ph/9606249}}.

\bibitem{Damour:2010rp}
T.~Damour and J.~F. Donoghue, \emph{{Equivalence Principle Violations and
  Couplings of a Light Dilaton}},
  \href{https://doi.org/10.1103/PhysRevD.82.084033}{\emph{Phys. Rev. D}
  {\bfseries 82} (2010) 084033}
  [\href{https://arxiv.org/abs/1007.2792}{{\ttfamily 1007.2792}}].

\bibitem{Stadnik:2014tta}
Y.~V. Stadnik and V.~V. Flambaum, \emph{{Searching for dark matter and
  variation of fundamental constants with laser and maser interferometry}},
  \href{https://doi.org/10.1103/PhysRevLett.114.161301}{\emph{Phys. Rev. Lett.}
  {\bfseries 114} (2015) 161301}
  [\href{https://arxiv.org/abs/1412.7801}{{\ttfamily 1412.7801}}].

\bibitem{Stadnik:2015kia}
Y.~V. Stadnik and V.~V. Flambaum, \emph{{Can dark matter induce cosmological
  evolution of the fundamental constants of Nature?}},
  \href{https://doi.org/10.1103/PhysRevLett.115.201301}{\emph{Phys. Rev. Lett.}
  {\bfseries 115} (2015) 201301}
  [\href{https://arxiv.org/abs/1503.08540}{{\ttfamily 1503.08540}}].

\bibitem{Masia-Roig:2022net}
H.~Masia-Roig et~al., \emph{{Intensity interferometry for ultralight bosonic
  dark matter detection}},
  \href{https://doi.org/10.1103/PhysRevD.108.015003}{\emph{Phys. Rev. D}
  {\bfseries 108} (2023) 015003}
  [\href{https://arxiv.org/abs/2202.02645}{{\ttfamily 2202.02645}}].

\bibitem{Banks:2020gpu}
H.~Banks and M.~Mccullough, \emph{{Charting the Fifth Force Landscape}},
  \href{https://doi.org/10.1103/PhysRevD.103.075018}{\emph{Phys. Rev. D}
  {\bfseries 103} (2021) 075018}
  [\href{https://arxiv.org/abs/2009.12399}{{\ttfamily 2009.12399}}].

\bibitem{Budnik:2020nwz}
R.~Budnik, H.~Kim, O.~Matsedonskyi, G.~Perez and Y.~Soreq, \emph{{Probing the
  relaxed relaxion and Higgs portal scenarios with XENON1T scintillation and
  ionization data}},
  \href{https://doi.org/10.1103/PhysRevD.104.015012}{\emph{Phys. Rev. D}
  {\bfseries 104} (2021) 015012}
  [\href{https://arxiv.org/abs/2006.14568}{{\ttfamily 2006.14568}}].

\bibitem{Balkin:2021zfd}
R.~Balkin, J.~Serra, K.~Springmann, S.~Stelzl and A.~Weiler, \emph{{Density
  induced vacuum instability}},
  \href{https://doi.org/10.21468/SciPostPhys.14.4.071}{\emph{SciPost Phys.}
  {\bfseries 14} (2023) 071}
  [\href{https://arxiv.org/abs/2105.13354}{{\ttfamily 2105.13354}}].

\bibitem{Bar:2021kti}
N.~Bar, K.~Blum and C.~Sun, \emph{{Galactic rotation curves versus ultralight
  dark matter: A systematic comparison with SPARC data}},
  \href{https://doi.org/10.1103/PhysRevD.105.083015}{\emph{Phys. Rev. D}
  {\bfseries 105} (2022) 083015}
  [\href{https://arxiv.org/abs/2111.03070}{{\ttfamily 2111.03070}}].

\bibitem{Dailey_2020}
C.~Dailey, C.~Bradley, D.~F.~J. Kimball, I.~A. Sulai, S.~Pustelny,
  A.~Wickenbrock et~al., \emph{Quantum sensor networks as exotic field
  telescopes for multi-messenger astronomy},
  \href{https://doi.org/10.1038/s41550-020-01242-7}{\emph{Nature Astronomy}
  {\bfseries 5} (2020) 150}.

\bibitem{PhysRevD.105.063032}
H.~Kim and A.~Lenoci, \emph{Gravitational focusing of wave dark matter},
  \href{https://doi.org/10.1103/PhysRevD.105.063032}{\emph{Phys. Rev. D}
  {\bfseries 105} (2022) 063032}.

\bibitem{haloformation}
D.~Budker, J.~Eby, M.~Gorghetto, M.~Jiang and G.~Perez, \emph{{A Generic
  Formation Mechanism of Ultralight Dark Matter Solar Halos}},
  \href{https://arxiv.org/abs/2306.12477}{{\ttfamily 2306.12477}}.

\bibitem{Hook:2019pbh}
A.~Hook and J.~Huang, \emph{{Searches for other vacua. Part I. Bubbles in our
  universe}}, \href{https://doi.org/10.1007/JHEP08(2019)148}{\emph{JHEP}
  {\bfseries 08} (2019) 148}
  [\href{https://arxiv.org/abs/1904.00020}{{\ttfamily 1904.00020}}].

\bibitem{Salucci:2010qr}
P.~Salucci, F.~Nesti, G.~Gentile and C.~Martins, \emph{{The dark matter density
  at the Sun's location}},
  \href{https://doi.org/10.1051/0004-6361/201014385}{\emph{Astron. Astrophys.}
  {\bfseries 523} (2010) A83}
  [\href{https://arxiv.org/abs/1003.3101}{{\ttfamily 1003.3101}}].

\bibitem{Pitjev_2013}
N.~P. Pitjev and E.~V. Pitjeva, \emph{Constraints on dark matter in the solar
  system}, \href{https://doi.org/10.1134/s1063773713020060}{\emph{Astronomy
  Letters} {\bfseries 39} (2013) 141}.

\bibitem{Tsai:2022jnv}
Y.-D. Tsai, J.~Eby, J.~Arakawa, D.~Farnocchia and M.~S. Safronova, \emph{{New
  Constraints on Dark Matter and Cosmic Neutrino Profiles through Gravity}},
  \href{https://arxiv.org/abs/2210.03749}{{\ttfamily 2210.03749}}.

\bibitem{Schlamminger:2007ht}
S.~Schlamminger, K.~Y. Choi, T.~A. Wagner, J.~H. Gundlach and E.~G. Adelberger,
  \emph{{Test of the equivalence principle using a rotating torsion balance}},
  \href{https://doi.org/10.1103/PhysRevLett.100.041101}{\emph{Phys. Rev. Lett.}
  {\bfseries 100} (2008) 041101}
  [\href{https://arxiv.org/abs/0712.0607}{{\ttfamily 0712.0607}}].

\bibitem{PhysRevD.61.022001}
G.~L. Smith, C.~D. Hoyle, J.~H. Gundlach, E.~G. Adelberger, B.~R. Heckel and
  H.~E. Swanson, \emph{Short-range tests of the equivalence principle},
  \href{https://doi.org/10.1103/PhysRevD.61.022001}{\emph{Phys. Rev. D}
  {\bfseries 61} (1999) 022001}.

\bibitem{PhysRevLett.129.121102}
{\scshape MICROSCOPE Collaboration} collaboration, \emph{$microscope$ mission:
  Final results of the test of the equivalence principle},
  \href{https://doi.org/10.1103/PhysRevLett.129.121102}{\emph{Phys. Rev. Lett.}
  {\bfseries 129} (2022) 121102}.

\bibitem{Kennedy:2020bac}
C.~J. Kennedy, E.~Oelker, J.~M. Robinson, T.~Bothwell, D.~Kedar, W.~R. Milner
  et~al., \emph{{Precision Metrology Meets Cosmology: Improved Constraints on
  Ultralight Dark Matter from Atom-Cavity Frequency Comparisons}},
  \href{https://doi.org/10.1103/PhysRevLett.125.201302}{\emph{Phys. Rev. Lett.}
  {\bfseries 125} (2020) 201302}
  [\href{https://arxiv.org/abs/2008.08773}{{\ttfamily 2008.08773}}].

\bibitem{Aharony:2019iad}
S.~Aharony, N.~Akerman, R.~Ozeri, G.~Perez, I.~Savoray and R.~Shaniv,
  \emph{{Constraining Rapidly Oscillating Scalar Dark Matter Using Dynamic
  Decoupling}}, \href{https://doi.org/10.1103/PhysRevD.103.075017}{\emph{Phys.
  Rev. D} {\bfseries 103} (2021) 075017}
  [\href{https://arxiv.org/abs/1902.02788}{{\ttfamily 1902.02788}}].

\bibitem{Abe_2021}
M.~Abe, P.~Adamson, M.~Borcean, D.~Bortoletto, K.~Bridges, S.~P. Carman et~al.,
  \emph{Matter-wave atomic gradiometer interferometric sensor ({MAGIS}-100)},
  \href{https://doi.org/10.1088/2058-9565/abf719}{\emph{Quantum Science and
  Technology} {\bfseries 6} (2021) 044003}.

\bibitem{Hees:2016gop}
A.~Hees, J.~Gu\'ena, M.~Abgrall, S.~Bize and P.~Wolf, \emph{{Searching for an
  oscillating massive scalar field as a dark matter candidate using atomic
  hyperfine frequency comparisons}},
  \href{https://doi.org/10.1103/PhysRevLett.117.061301}{\emph{Phys. Rev. Lett.}
  {\bfseries 117} (2016) 061301}
  [\href{https://arxiv.org/abs/1604.08514}{{\ttfamily 1604.08514}}].

\bibitem{Anastassopoulos:2017ftl}
{\scshape CAST} collaboration, \emph{{New CAST Limit on the Axion-Photon
  Interaction}}, \href{https://doi.org/10.1038/nphys4109}{\emph{Nature Phys.}
  {\bfseries 13} (2017) 584}
  [\href{https://arxiv.org/abs/1705.02290}{{\ttfamily 1705.02290}}].

\bibitem{Capozzi:2020cbu}
F.~Capozzi and G.~Raffelt, \emph{{Axion and neutrino bounds improved with new
  calibrations of the tip of the red-giant branch using geometric distance
  determinations}},
  \href{https://doi.org/10.1103/PhysRevD.102.083007}{\emph{Phys. Rev. D}
  {\bfseries 102} 083007} [\href{https://arxiv.org/abs/2007.03694}{{\ttfamily
  2007.03694}}].

\bibitem{Raffelt:2006cw}
G.~G. Raffelt, \emph{{Astrophysical axion bounds}},
  \href{https://doi.org/10.1007/978-3-540-73518-2_3}{\emph{Lect. Notes Phys.}
  {\bfseries 741} (2008) 51}
  [\href{https://arxiv.org/abs/hep-ph/0611350}{{\ttfamily hep-ph/0611350}}].

\bibitem{Buschmann:2021juv}
M.~Buschmann, C.~Dessert, J.~W. Foster, A.~J. Long and B.~R. Safdi,
  \emph{{Upper Limit on the QCD Axion Mass from Isolated Neutron Star
  Cooling}}, \href{https://doi.org/10.1103/PhysRevLett.128.091102}{\emph{Phys.
  Rev. Lett.} {\bfseries 128} (2022) 091102}
  [\href{https://arxiv.org/abs/2111.09892}{{\ttfamily 2111.09892}}].

\bibitem{Beznogov:2018fda}
M.~V. Beznogov, E.~Rrapaj, D.~Page and S.~Reddy, \emph{{Constraints on
  Axion-like Particles and Nucleon Pairing in Dense Matter from the Hot Neutron
  Star in HESS J1731-347}},
  \href{https://doi.org/10.1103/PhysRevC.98.035802}{\emph{Phys. Rev. C}
  {\bfseries 98} (2018) 035802}
  [\href{https://arxiv.org/abs/1806.07991}{{\ttfamily 1806.07991}}].

\bibitem{Reynolds:2019uqt}
C.~S. Reynolds, M.~C.~D. Marsh, H.~R. Russell, A.~C. Fabian, R.~Smith,
  F.~Tombesi et~al., \emph{{Astrophysical limits on very light axion-like
  particles from Chandra grating spectroscopy of NGC 1275}},
  \href{https://doi.org/10.3847/1538-4357/ab6a0c}{\emph{Astrophys. J.}
  {\bfseries 890} (2020) 59}
  [\href{https://arxiv.org/abs/1907.05475}{{\ttfamily 1907.05475}}].

\bibitem{Abel:2017rtm}
C.~Abel et~al., \emph{{Search for Axionlike Dark Matter through Nuclear Spin
  Precession in Electric and Magnetic Fields}},
  \href{https://doi.org/10.1103/PhysRevX.7.041034}{\emph{Phys. Rev. X}
  {\bfseries 7} (2017) 041034}
  [\href{https://arxiv.org/abs/1708.06367}{{\ttfamily 1708.06367}}].

\bibitem{Banerjee:2019epw}
A.~Banerjee, D.~Budker, J.~Eby, H.~Kim and G.~Perez, \emph{{Relaxion Stars and
  their detection via Atomic Physics}},
  \href{https://doi.org/10.1038/s42005-019-0260-3}{\emph{Commun. Phys.}
  {\bfseries 3} (2020) 1} [\href{https://arxiv.org/abs/1902.08212}{{\ttfamily
  1902.08212}}].

\bibitem{Banerjee:2019xuy}
A.~Banerjee, D.~Budker, J.~Eby, V.~V. Flambaum, H.~Kim, O.~Matsedonskyi et~al.,
  \emph{{Searching for Earth/Solar Axion Halos}},
  \href{https://doi.org/10.1007/JHEP09(2020)004}{\emph{JHEP} {\bfseries 09}
  (2020) 004} [\href{https://arxiv.org/abs/1912.04295}{{\ttfamily
  1912.04295}}].

\bibitem{Touboul:2017grn}
P.~Touboul et~al., \emph{{MICROSCOPE Mission: First Results of a Space Test of
  the Equivalence Principle}},
  \href{https://doi.org/10.1103/PhysRevLett.119.231101}{\emph{Phys. Rev. Lett.}
  {\bfseries 119} (2017) 231101}
  [\href{https://arxiv.org/abs/1712.01176}{{\ttfamily 1712.01176}}].

\bibitem{Berge:2017ovy}
J.~Berg\'e, P.~Brax, G.~M\'etris, M.~Pernot-Borr\`as, P.~Touboul and J.-P.
  Uzan, \emph{{MICROSCOPE Mission: First Constraints on the Violation of the
  Weak Equivalence Principle by a Light Scalar Dilaton}},
  \href{https://doi.org/10.1103/PhysRevLett.120.141101}{\emph{Phys. Rev. Lett.}
  {\bfseries 120} (2018) 141101}
  [\href{https://arxiv.org/abs/1712.00483}{{\ttfamily 1712.00483}}].

\bibitem{Safronova:2019lex}
M.~S. Safronova, \emph{{The Search for Variation of Fundamental Constants with
  Clocks}}, \href{https://doi.org/10.1002/andp.201800364}{\emph{Annalen Phys.}
  {\bfseries 531} (2019) 1800364}.

\bibitem{Antypas:2020rtg}
D.~Antypas, O.~Tretiak, K.~Zhang, A.~Garcon, G.~Perez, M.~G. Kozlov et~al.,
  \emph{{Probing fast oscillating scalar dark matter with atoms and
  molecules}}, \href{https://doi.org/10.1088/2058-9565/abe472}{\emph{Quantum
  Sci. Technol.} {\bfseries 6} (2021) 034001}
  [\href{https://arxiv.org/abs/2012.01519}{{\ttfamily 2012.01519}}].

\bibitem{Wagner:2012ui}
T.~A. Wagner, S.~Schlamminger, J.~H. Gundlach and E.~G. Adelberger,
  \emph{{Torsion-balance tests of the weak equivalence principle}},
  \href{https://doi.org/10.1088/0264-9381/29/18/184002}{\emph{Class. Quant.
  Grav.} {\bfseries 29} (2012) 184002}
  [\href{https://arxiv.org/abs/1207.2442}{{\ttfamily 1207.2442}}].

\bibitem{Su:1994gu}
Y.~Su, B.~R. Heckel, E.~G. Adelberger, J.~H. Gundlach, M.~Harris, G.~L. Smith
  et~al., \emph{{New tests of the universality of free fall}},
  \href{https://doi.org/10.1103/PhysRevD.50.3614}{\emph{Phys. Rev. D}
  {\bfseries 50} (1994) 3614}.

\bibitem{FOCOS}
A.~{Derevianko}, K.~{Gibble}, L.~{Hollberg}, N.~R. {Newbury}, C.~{Oates}, M.~S.
  {Safronova} et~al., \emph{{Fundamental physics with a state-of-the-art
  optical clock in space}},
  \href{https://doi.org/10.1088/2058-9565/ac7df9}{\emph{Quantum Science and
  Technology} {\bfseries 7} (2022) 044002}.

\bibitem{Dymarsky:2013pqa}
A.~Dymarsky, Z.~Komargodski, A.~Schwimmer and S.~Theisen, \emph{{On Scale and
  Conformal Invariance in Four Dimensions}},
  \href{https://doi.org/10.1007/JHEP10(2015)171}{\emph{JHEP} {\bfseries 10}
  (2015) 171} [\href{https://arxiv.org/abs/1309.2921}{{\ttfamily 1309.2921}}].

\bibitem{Coradeschi:2013gda}
F.~Coradeschi, P.~Lodone, D.~Pappadopulo, R.~Rattazzi and L.~Vitale, \emph{{A
  naturally light dilaton}},
  \href{https://doi.org/10.1007/JHEP11(2013)057}{\emph{JHEP} {\bfseries 11}
  (2013) 057} [\href{https://arxiv.org/abs/1306.4601}{{\ttfamily 1306.4601}}].

\bibitem{Taylor:1988nw}
T.~R. Taylor and G.~Veneziano, \emph{{Dilaton Couplings at Large Distances}},
  \href{https://doi.org/10.1016/0370-2693(88)91290-7}{\emph{Phys. Lett. B}
  {\bfseries 213} (1988) 450}.

\bibitem{Kaplan:2000hh}
D.~B. Kaplan and M.~B. Wise, \emph{{Couplings of a light dilaton and violations
  of the equivalence principle}},
  \href{https://doi.org/10.1088/1126-6708/2000/08/037}{\emph{JHEP} {\bfseries
  08} 037} [\href{https://arxiv.org/abs/hep-ph/0008116}{{\ttfamily
  hep-ph/0008116}}].

\bibitem{Lee:2020zjt}
J.~Lee, E.~Adelberger, T.~Cook, S.~Fleischer and B.~Heckel, \emph{{New Test of
  the Gravitational $1/r^2$ Law at Separations down to 52 $\mu$m}},
  \href{https://doi.org/10.1103/PhysRevLett.124.101101}{\emph{Phys. Rev. Lett.}
  {\bfseries 124} (2020) 101101}
  [\href{https://arxiv.org/abs/2002.11761}{{\ttfamily 2002.11761}}].

\bibitem{Tan:2020vpf}
W.-H. Tan et~al., \emph{{Improvement for Testing the Gravitational
  Inverse-Square Law at the Submillimeter Range}},
  \href{https://doi.org/10.1103/PhysRevLett.124.051301}{\emph{Phys. Rev. Lett.}
  {\bfseries 124} (2020) 051301}.

\bibitem{Olive:2007aj}
K.~A. Olive and M.~Pospelov, \emph{{Environmental dependence of masses and
  coupling constants}},
  \href{https://doi.org/10.1103/PhysRevD.77.043524}{\emph{Phys. Rev. D}
  {\bfseries 77} (2008) 043524}
  [\href{https://arxiv.org/abs/0709.3825}{{\ttfamily 0709.3825}}].

\bibitem{Burt2021}
E.~Burt, J.~Prestage, R.~Tjoelker, D.~Enzer, D.~Kuang, D.~Murphy et~al.,
  \emph{Demonstration of a trapped-ion atomic clock in space},
  \href{https://doi.org/10.1038/s41586-021-03571-7}{\emph{Nature} {\bfseries
  595} (2021) 43}.

\bibitem{2018Navigation}
T.~A. Ely, E.~A. Burt, J.~D. Prestage, J.~M. Seubert and R.~L. Tjoelker,
  \emph{Using the deep space atomic clock for navigation and science},
  \href{https://doi.org/10.1109/TUFFC.2018.2808269}{\emph{IEEE Transactions on
  Ultrasonics, Ferroelectrics, and Frequency Control} {\bfseries 65} (2018)
  950}.

\bibitem{CAC2018}
L.~Liu, D.-S. L{\"u}, W.~biao Chen, T.~Li, Q.~Qu, B.~Wang et~al.,
  \emph{In-orbit operation of an atomic clock based on laser-cooled 87rb
  atoms}, {\emph{Nature Communications} {\bfseries 2760} (2018) }.

\bibitem{ACES}
L.~{Cacciapuoti}, M.~{Armano}, R.~{Much}, O.~{Sy}, A.~{Helm}, M.~P. {Hess}
  et~al., \emph{{Testing gravity with cold-atom clocks in space. The ACES
  mission}}, \href{https://doi.org/10.1140/epjd/e2020-10167-7}{\emph{European
  Physical Journal D} {\bfseries 74} (2020) 164}.

\bibitem{LudBoyYe15}
A.~D. {Ludlow}, M.~M. {Boyd}, J.~{Ye}, E.~{Peik} and P.~O. {Schmidt},
  \emph{{Optical atomic clocks}},
  \href{https://doi.org/10.1103/RevModPhys.87.637}{\emph{Reviews of Modern
  Physics} {\bfseries 87} (2015) 637}
  [\href{https://arxiv.org/abs/1407.3493}{{\ttfamily 1407.3493}}].

\bibitem{PhysRevLett.123.033201}
S.~M. Brewer, J.-S. Chen, A.~M. Hankin, E.~R. Clements, C.~W. Chou, D.~J.
  Wineland et~al., \emph{$^{27}{\mathrm{al}}^{+}$ quantum-logic clock with a
  systematic uncertainty below ${10}^{\ensuremath{-}18}$},
  \href{https://doi.org/10.1103/PhysRevLett.123.033201}{\emph{Phys. Rev. Lett.}
  {\bfseries 123} (2019) 033201}.

\bibitem{KelBurKal19}
J.~Stuhler and other, \emph{Opticlock: Transportable and easy-to-operate
  optical single-ion clock},
  \href{https://doi.org/10.1016/j.measen.2021.100264}{\emph{Measurement:
  Sensors} {\bfseries 18} (2021) 100264}.

\bibitem{2022UDM}
D.~Antypas et~al., \emph{New horizons: Scalar and vector ultralight dark
  matter}, \href{https://doi.org/10.48550/ARXIV.2203.14915}{\emph{arXiv:
  2203.14915} (2022) }.

\bibitem{Sun}
Y.-D. Tsai, J.~Eby and M.~S. Safronova, \emph{{Direct detection of ultralight
  dark matter bound to the Sun with space quantum sensors}},
  \href{https://doi.org/10.1038/s41550-022-01833-6}{\emph{Nature Astron.}
  {\bfseries 7} (2023) 113} [\href{https://arxiv.org/abs/2112.07674}{{\ttfamily
  2112.07674}}].

\bibitem{Kolkowitz_2016}
S.~Kolkowitz, I.~Pikovski, N.~Langellier, M.~Lukin, R.~Walsworth and J.~Ye,
  \emph{Gravitational wave detection with optical lattice atomic clocks},
  \href{https://doi.org/10.1103/physrevd.94.124043}{\emph{Physical Review D}
  {\bfseries 94} (2016) }.

\bibitem{Fedderke:2021kuy}
M.~A. Fedderke, P.~W. Graham and S.~Rajendran, \emph{{Asteroids for
  \ensuremath{\mu}Hz gravitational-wave detection}},
  \href{https://doi.org/10.1103/PhysRevD.105.103018}{\emph{Phys. Rev. D}
  {\bfseries 105} (2022) 103018}
  [\href{https://arxiv.org/abs/2112.11431}{{\ttfamily 2112.11431}}].

\bibitem{Puetzfeld:2019kki}
D.~Puetzfeld and C.~L\"ammerzahl, eds., \emph{{Relativistic Geodesy}},
  vol.~196. Springer, 2019,
  \href{https://doi.org/10.1007/978-3-030-11500-5}{10.1007/978-3-030-11500-5}.

\bibitem{Gozzard:2021wkf}
D.~R. Gozzard, L.~A. Howard, B.~P. Dix-Matthews, S.~Karpathakis, C.~Gravestock
  and S.~W. Schediwy, \emph{{Ultrastable Free-Space Laser Links for a Global
  Network of Optical Atomic Clocks}},
  \href{https://doi.org/10.1103/PhysRevLett.128.020801}{\emph{Phys. Rev. Lett.}
  {\bfseries 128} (2022) 020801}
  [\href{https://arxiv.org/abs/2103.12909}{{\ttfamily 2103.12909}}].

\bibitem{roadmap}
I.~Alonso et~al., \emph{{Cold atoms in space: community workshop summary and
  proposed road-map}},
  \href{https://doi.org/10.1140/epjqt/s40507-022-00147-w}{\emph{EPJ Quant.
  Technol.} {\bfseries 9} (2022) 30}
  [\href{https://arxiv.org/abs/2201.07789}{{\ttfamily 2201.07789}}].

\bibitem{OACESS}
V.~Schkolnik et~al., \emph{{Optical atomic clock aboard an Earth-orbiting space
  station (OACESS): enhancing searches for physics beyond the standard model in
  space}}, \href{https://doi.org/10.1088/2058-9565/ac9f2b}{\emph{Quantum Sci.
  Technol.} {\bfseries 8} (2023) 014003}
  [\href{https://arxiv.org/abs/2204.09611}{{\ttfamily 2204.09611}}].

\bibitem{HanKuzLun21}
D.~{Hanneke}, B.~{Kuzhan} and A.~{Lunstad}, \emph{{Optical clocks based on
  molecular vibrations as probes of variation of the proton-to-electron mass
  ratio}}, \href{https://doi.org/10.1088/2058-9565/abc863}{\emph{Quantum
  Science and Technology} {\bfseries 6} (2021) 014005}
  [\href{https://arxiv.org/abs/2007.15750}{{\ttfamily 2007.15750}}].

\bibitem{Allahverdi:2020bys}
R.~Allahverdi et~al., \emph{{The First Three Seconds: a Review of Possible
  Expansion Histories of the Early Universe}},
  \href{https://arxiv.org/abs/2006.16182}{{\ttfamily 2006.16182}}.

\bibitem{DeRocco:2020xdt}
W.~DeRocco, P.~W. Graham and S.~Rajendran, \emph{{Exploring the robustness of
  stellar cooling constraints on light particles}},
  \href{https://doi.org/10.1103/PhysRevD.102.075015}{\emph{Phys. Rev. D}
  {\bfseries 102} (2020) 075015}
  [\href{https://arxiv.org/abs/2006.15112}{{\ttfamily 2006.15112}}].

\bibitem{Raffelt:2012sp}
G.~Raffelt, \emph{{Limits on a CP-violating scalar axion-nucleon interaction}},
  \href{https://doi.org/10.1103/PhysRevD.86.015001}{\emph{Phys. Rev.}
  {\bfseries D86} (2012) 015001}
  [\href{https://arxiv.org/abs/1205.1776}{{\ttfamily 1205.1776}}].

\bibitem{Redondo:2013lna}
J.~Redondo and G.~Raffelt, \emph{{Solar constraints on hidden photons
  re-visited}},
  \href{https://doi.org/10.1088/1475-7516/2013/08/034}{\emph{JCAP} {\bfseries
  1308} (2013) 034} [\href{https://arxiv.org/abs/1305.2920}{{\ttfamily
  1305.2920}}].

\bibitem{Hardy:2016kme}
E.~Hardy and R.~Lasenby, \emph{{Stellar cooling bounds on new light particles:
  plasma mixing effects}},
  \href{https://doi.org/10.1007/JHEP02(2017)033}{\emph{JHEP} {\bfseries 02}
  (2017) 033} [\href{https://arxiv.org/abs/1611.05852}{{\ttfamily
  1611.05852}}].

\bibitem{Choi:2015fiu}
K.~Choi and S.~H. Im, \emph{{Realizing the relaxion from multiple axions and
  its UV completion with high scale supersymmetry}},
  \href{https://doi.org/10.1007/JHEP01(2016)149}{\emph{JHEP} {\bfseries 01}
  (2016) 149} [\href{https://arxiv.org/abs/1511.00132}{{\ttfamily
  1511.00132}}].

\bibitem{Kaplan:2015fuy}
D.~E. Kaplan and R.~Rattazzi, \emph{{Large field excursions and approximate
  discrete symmetries from a clockwork axion}},
  \href{https://doi.org/10.1103/PhysRevD.93.085007}{\emph{Phys. Rev.}
  {\bfseries D93} (2016) 085007}
  [\href{https://arxiv.org/abs/1511.01827}{{\ttfamily 1511.01827}}].

\end{thebibliography}\endgroup

\end{document}